\documentclass[12pt,oneandhalf,chaparabic,phys,ms,eng]{metu}
\pdfoutput=1
\usepackage[utf8]{inputenc}
\usepackage{appendix} 
\usepackage{graphicx} 
\usepackage{amsmath,amssymb,amsfonts} 
\usepackage{stackengine} 
\usepackage{tabularx}
\usepackage{mathtools} 
\usepackage{yhmath} 
\usepackage{bm} 
\usepackage{tikz} 
\usepackage{cite} 
\allowdisplaybreaks[4] 

\newcommand{\sfigure}[4]{
	\centering
	\includegraphics[scale=#2]{#1}
	\caption[#3]{#4}
	\label{#1}
	}

\newcommand{\figref}[1]{Fig.~\ref{#1}}
\newcommand{\secref}[1]{Section~\ref{#1}}
\newcommand{\charef}[1]{Chapter~\ref{#1}}
\newcommand{\tabref}[1]{Table~\ref{#1}}
\newcommand{\appref}[1]{Appendix~\ref{#1}}
\newcommand{\equref}[1]{Eqn.~(\ref{#1})}

\def\vok{\mathrel{\rlap{\lower0pt\hbox{\hskip1pt{$k$}}}
    \raise6pt\hbox{$\neg$}}}
\def\voc{\mathrel{\rlap{\lower0pt\hbox{\hskip1pt{$c$}}}
    \raise4pt\hbox{$\neg$}}}
\author{Soner Albayrak}
\title{A Lorentz Violating Theory: Its Nonminimal Extension in the Photon Sector}
\turkishtitle{Lorentz Kıran Bir Model: Bu Modelin Foton Sektöründeki Uzantısı}
\date{February 2016}
\director[prof]{Gülbin Dural Ünver}
\headofdept[prof]{Mehmet Tevfik Zeyrek}
\supervisor[assocprof]{İsmail Turan}
\departmentofsupervisor{Physics Department, METU}
\committeememberi[prof]{Ali Ulvi Yılmazer}
\affiliationi{Physics Engineering Department, Ankara University}
\committeememberii[assocprof]{İsmail Turan}
\affiliationii{Physics Department, METU}
\committeememberiii[prof]{Tahmasib Aliyev}
\affiliationiii{Physics Department, METU}
\committeememberiv[prof]{Altuğ Özpineci}
\affiliationiv{Physics Department, METU}
\committeememberv[assocprof]{Seçkin Kürkçüoğlu}
\affiliationv{Physics Department, METU}
\keywords{Lorentz violation, CPT violation, Standard Model Extension, nonminimal SME, nonrenormalizable photon sector, vacuum-orthogonal models}
\anahtarklm{Lorentz kırılımı, CPT kırılımı, Genişletilmiş Standard Model, minimal olmayan SME, yeniden normalize olmayan Foton sektör, vakuma dik modeller}
\abstract{The relentless efforts of the physics community has not yet availed us the solution of how to unify the Quantum Mechanics with General Relativity, a puzzle that has engaged the minds of the physicists for almost a century. The insufficiency of today's and foreseeable future's technology for a direct reach into the Planck energies at which the fundamental theory, the Quantum Theory of Gravity, lies has lead to the search of the low energy effects of that fundamental Planck level theory irregardless of the details of it. In this thesis, one of the leading candidates of such an exotic effect, that is the violation of Lorentz and CPT symmetries is analyzed. The action level effective field theoretical framework for such an analysis called  \emph{Standard Model Extension} has already been in the literature for the last two decades; here, the nonminimal photon sector of such a framework is examined from a quantum field theoretical point of view. All possible polarization vectors for different kinds of CPT violations, the generic forms of the dispersion relations that these polarization vectors satisfy, and the corresponding propagators of the photon field are explicitly calculated. Special models of Lorentz violations are introduced, and a particular one called vacuum orthogonal model is analyzed extensively. It is found that this particular form of Lorentz violation cannot induce any effects that is detectable in the vacuum propagation of the photon. Isotropic and leading order cases of the Lorentz violation is studied and this found result is explicitly shown.}
\oz{Son yüzyıl boyunca fizikçileri meşgul eden Kuantum Mekaniğinin Genel Görelilik ile birleştirilememesi problemi, fizik camiasının bütün çabalarına rağmen halen çözülebilmiş değil. Ne bugünün ne de yakın geleceğin teknolojisinin doğrudan Planck enerjisinde bir gözlem yapmak için yeterli olması, Fizikçileri bu enerjide geçerli olan temel kuramın --kuramın kendisinin ne olduğuyla ilgilenmeden-- düşük enerjilerdeki sıradışı etkilerinin araştırılmasına yönlendirmiş durumda. Bu tezde de, böyle sıradışı olası etkilerin önde gelen adayı olan Lorentz ve CPT simetrilerinin kırılımı inceleniyor. Böyle bir analizi eylem seviyesinde etken alan kuramı ile inceleyen bir sistem, \emph{Genişletilmiş Standard Model}, son yirmi yıldır literatürde mevcut durumda; bu sistemin minimal olmayan foton sektörünün kuantum alan kuramsal analizi de burada yapılmaktadır. Farklı çeşit Lorentz kırılımları için olası bütün polarizasyon vektörleri, bu polarizasyon vektörlerinin uyduğu enerji momentum ilişkilerinin genel halleri, ve bu foton alanlarına karşılık gelen yayıcılar açık olarak hesaplanmıştır. Lorentz kırılımlarının özel modelleri tanıtılmış, ve vakuma dik denilen bir model ayrıca kapsamlı bir şekilde çalışılmıştır. Bu özel modelin, fotonun vakumdaki yayılımında gözlemlenebilecek bir etki oluşturamayacağı bulunmuş; sadece eşyönlü katsayılardan ve sadece öncül seviyedeki katsayılardan oluşan Lorentz kırılımı özel durumları için bu bulgu bariz bir şekilde gösterilmiştir.}
\dedication{To Tutku Ünver, without whom I would not have discovered the scientist within me}
\acknowledgments{I would like to thank my supervisor, Assoc.~Prof. İsmail Turan, for his endless support, incredible encouragements, and valuable time; and most importantly, for the role model he provided me with: I consider it to be my luck and honor to be able to have worked with him.

I would like to thank my colleague and friend Ufuk Taşdan for his kind interest and encouraging support toward my work. Also, I would like to thank my friend Musa Burak Erdihan for his mere presence in my life for the last 11 years. Additionally, I would like to give my special thanks to Taylan Nurlu, who help me forget a theoretical physicists inevitable doom of being never understood hence never sincerely appreciated by their friends and family, as he has always been there to listen and try to understand what I do throughout this thesis. We shared more than a few discussions.

Moreover, I thank TÜBİTAK for financial support through \emph{2210 - National Scholarship Program for MS Students} during my master education.

Last but not least, I thank my friends and my family for being there whenever I needed them.}
\usepackage{hyperref}
\usepackage[all]{hypcap}
\usepackage{todonotes}
\usepackage{graphicx}
\graphicspath{ {./images/} }
\usepackage{rotating}
\usepackage{xy} 
\usepackage{booktabs}
\usepackage{pifont}
\usepackage{color}
\usepackage{listings}
\usepackage{pdfpages}
\usepackage{array}
\usepackage{float}
\restylefloat{figure}

\DeclareGraphicsExtensions{.pdf,.png,.jpg}

\begin{document}
\begin{preliminaries}
\begin{theglossary}{{\it nm}SME}
\item[LV] Lorentz Violation / Lorentz Violating
\item[SM] Standard Model			
\item[LVT] Lorentz Violating Terms	
\item[VSR] Very Special Relativity		
\item[DSR] Deformed Special Relativity
\item[EFT] Effective Field Theory / Effective Field Theoretical
\item[CPT] Charge Conjugation, Parity Transformation, and Time Reversal
\item[SME] Standard Model Extension												
\item[RMS] Robertson-Mansouri-Sexl Model
\item[EOM] Equations of Motion
\item[VOM] Vacuum Orthogonal Model
\item[CPTV] CPT Violation / CPT Violating				
\item[{\it m}SME] Minimal Extension of Standard Model
\item[{\it nm}SME] Non-minimal Extension of Standard Model
\end{theglossary}

\end{preliminaries}
%
%
%

\chapter{Introduction}
\label{CHAPTER:INTRODUCTION}
One of the important quests in Physics for the last century, if not the most, has been the unification of Quantum Mechanics (QM) with General Relativity (GR), in the search for the ultimate theory of all known physics. The mathematical incompatibility of these two theories, along with the incline toward the incompleteness of \emph{Standard Model} (SM), has kindled the desire to build another theory, so-called Quantum Theory of Gravity, from which the SM and GR emerge under proper limits. Several different approaches have been developed as to be candidates for the Quantum Theory of Gravity, among which string theories, loop quantum gravity, noncommutative field theories, space-time foam, geometrodynamics, and many others  can be named. However, in contrast to the abundance of theoretical possibilities, a clue regarding what the ultimate theory would be among those eludes us, as that would be experimentally extracted in the Planck scales only. Therefore, that our technology for such an experiment is impossible both today and in the foreseeable future leaves us in the dark if the experimental results of those candidates are directly sought.

It has been realized for the last couple of decades that the progress can be turned upside down: Instead of starting from the theoretical candidates and searching for their effects, the exotic effects in the attainable energies can be searched from where the high energy theory is reached. Indeed, it is much more practical to search for such effects which can be treated as the low energy effects of the quantum gravity. Exotic effects, in this sense, can be any phenomena that can be explained by neither SM nor GR. Among such exotic effects, the violation of Lorentz and CPT symmetries is one of the leading candidates. That is mostly because almost all current approaches to the quantum gravity naturally allow the violation of Lorentz symmetry\cite{39.683,359.545,hep-ph/9501341,gr-qc/9809038,gr-qc/0411101,hep-th/0105082,2.604,144.349,hep-th/9805095,hep-th/9912169,hep-th/0312032,hep-ph/0610216}; however, we can also bluntly see why Lorentz symmetry is at least to be modified in the Planck scales with a simple example. In an almost flat part of the spacetime in which the Special Relativity holds, one needs to be able to boost to another frame for which the length of a stick is arbitrarily small due to the length contraction. However, it is assumed that the Planck length is the shortest possible length, which indicates that the Special Relativity as it stands cannot hold for the Planck level.

Once we accept that Lorentz symmetry can be broken, the validity of CPT symmetry becomes weaker; because, CPT symmetry is dictated by so-called the CPT theorem which states that any Lorentz invariant local quantum field theory with a Hermitian Hamiltonian must have CPT symmetry\cite{PCT}. Hence, the \emph{Lorentz Violation} (LV) may or may not induce \emph{CPT Violation} (CPTV), whereas breaking of the CPT symmetry directly induces breaking of the Lorentz symmetry as long as positivity of energy (Hermitian Hamiltonian) and causality (locality of the fields) are preserved. There are some theories in the literature, dealing with nonlocal models for the sake of CPTV without LV\cite{699.177}; however, it is generally assumed that locality is a much more stringent bound, which should be respected, than the Lorentz symmetry whose violation is already allowed in the leading quantum gravity candidates.

The breaking of Lorentz symmetry can be either spontaneous or explicit, which have quite different implications. An explicit breaking of Lorentz symmetry means that the fundamental theory at the Planck level does not have the Lorentz symmetry at all; that means, the principle of relativity, the heart of Lorentz symmetry, is invalid at all energies. Such a breaking is indeed a radical one, and results in not only physical consequences, but also philosophical implications as the universe should be a biased one about a particular position or velocity at all energies. There are some papers in the literature dealing with such a breaking\cite{97.021601}; however, it has been shown in an \emph{Effective Field Theoretical} (EFT) approach that the usual Riemann geometry cannot be maintained for gravity under an explicit breaking\cite{hep-th/0312310}. Alternative geometries like Riemann-Finsler can overcome this problem\cite{1104.5488}, yet there is really not that much motivation to consider explicit breaking of Lorentz symmetry by paying the price of abandoning the usual Riemann geometry.

For the case of the spontaneous breaking of the Lorentz symmetry, in contrast to the explicit one, the fundamental theory in the Planck scale does respect the principal of Relativity, and does have some form of Lorentz symmetry, probably an accordingly modified one as we do not know the exact form of the quantum gravity either. The key point is that this symmetry breaks for the vacuum solution, just like the Higgs mechanism, hence is not reflected in the solution of the theory, that is the universe we observe. In that sense, the spontaneous breaking attracts much more appeal than the explicit one as there is no philosophical problems about how biased the universe is: There is no specific reason for its particular bias, it would have as well been biased in a completely different form as the occurrence is spontaneous. This is like the direction that a pencil falls when it is let after being initially hold vertically: There is nothing special relating to that direction, and any other direction would be as well; however, a direction had to be picked as pencil went from high energy (vertical position) to low energy (horizontal position). In the real world, the direction that the pencil takes is surely a result of an unaccounted effect, like imperfect shape of the tip of the pencil or a small wind; however, the spontaneous breaking in its essence is not an ignorance of an unaccounted effect, but simply the implication of the necessity of solution selection which would destroy the original symmetry. As explained earlier, it is not of interest how this solution selection takes place in the fundamental theory since it is beyond reach; instead, the solution of the breaking, that is our universe, is examined by introducing LV into the conventional model in various ways, among which modifications in transformation laws\cite{21.378,8.497} and field theoretical approaches\cite{gr-qc/9809038,89.231301,65.103509,90.211601,97.021601,78.125011,77.025022} have been pursued in the literature, albeit such different approaches can be shown to be contained in a systematic field theoretical framework\cite{1308.4973,0905.0031}.

Let us summarize the situation at hand. The last century has witnessed relentless yet unsuccessful attempts to marry, if we use the jargon, the QM and GR, where it is currently assumed that there is indeed an ultimate theory valid upto Planck level, that is the Quantum theory of Gravity, or simply quantum gravity, for which SM and GR are simply limiting cases. However, it is almost impossible to probe into that realm to take hints regarding the validity of quantum gravity candidates, at least in today's technology; hence, instead, the low energy effects of this ultimate theory is searched through the exotic phenomena, one of which is the violation of Lorentz symmetry, where we will be concerning ourselves with the spontaneous breaking of this symmetry in this thesis. As for the means of introducing this violation to the conventional SM, we will be dealing with a model called \emph{Standard Model Extension} (SME).

Standard Model Extension\cite{hep-th/0312310,hep-ph/9703464,hep-ph/9809521} is a systematic framework for the exploration of Lorentz and CPT violations. This framework, which was constructed over 15 years ago, is an action level EFT approach in which LV is inserted to the model via background fields named \emph{Lorentz Violating Terms} (LVT), and has been analyzed and investigated both in theoretical and experimental fronts \cite{0801.0287}. Basically, it is assumed that the effective low energy description of the high energy fundamental theory can be expanded in energy over a mass scale, which is possibly related to the Planck scale. In this expansion, the lowest order term becomes the SM. With the EFT approach, next terms in this expansion can be examined with the field theoretical machinery built within the SM since the process is inherently perturbative as the deviations from the Lorentz invariance should be very small due to the current experimental bounds.

The leading term after the lowest order term is called \emph{minimal Standard Model Extension} (\emph{m}SME), and considers all possible LVT acceptable in SME with the restriction of renormalizability\cite{hep-ph/9809521}. The examination of \emph{m}SME has been quite thorough in all sectors, including the photon sector for which the experimental bounds are quite impressive, almost enough to rule out any possible LV\cite{0801.0287}. In contrast, there are almost no bounds in some LVT in the photon sector of so-called \emph{non-minimal Standard Model Extension} (\emph{nm}SME) in which all nonrenormalizable LVT acceptable in SME are accounted. As gravity itself is nonrenormalizable, it is reasonable to assume that \emph{nm}SME, with the operators of arbitrarily high mass dimensions, constitute the next term after the \emph{m}SME in the perturbation expansion of the fundamental theory.

The general introduction of SME and its constructional philosophy are left to the \appref{CHAPTER:LORENTZ} along with the alternative LV theories in the literature, where more related concepts such as the justification of using nonrenormalizable terms in the model and how SME ascertains that physics remains independent of choice of reference frame in the face of LV are covered in \charef{CHAPTER:PRELIMINARIES}. The last item is actually a subtle issue about breaking the Lorentz invariance, and at the same time, remaining independent of the reference system of the observer, which is handled in SME by inducing the difference between so-called \emph{Observer Lorentz Transformations} (OLT) and so-called \emph{Particle Lorentz Transformations} (PLT). In SME, LVT are chosen to make sure that the Lorentz symmetry is never broken under OLT but only under PLT. The discrimination between these two transformations are elaborated there.

The outline of the thesis is as follows. In \charef{CHAPTER:PRELIMINARIES}, the CPT-odd photon sector of SME is briefly introduced. The necessary notations and definitions that will be used throughout the thesis are covered; the ongoing test methods and relevant current bounds are discussed, and possible special models, such as \emph{Vacuum Orthogonal Model} (VOM), are mentioned. Then, in \charef{CHAPTER:CPT ODD PHOTON}, the quantum field theoretical properties of the CPT-odd photon sector, namely the dispersion relations, polarization vectors, and the propagator,  are investigated. It is shown that only a particular subset of the LVT can produce physical results, and is also shown that the modified propagator can be brought to the diagonal form for this particular subspace. In the next chapter, \charef{vacuum orthogonal model}, coefficient space is further restricted to vacuum-orthogonal LVT only. It is demonstrated that the dispersion relations for this models split into two sets, non-conventional and conventional; and, non-conventional dispersion relations are shown to be spurious whereas conventional dispersion relations are shown to accept conventional polarization vectors. This means that vacuum orthogonal model remains vacuum orthogonal at all orders; that is, vacuum orthogonal LVT do not produce any effect on vacuum propagation whatsoever; hence, this proves that there are possible ways for the Lorentz symmetry to be broken for the photon although it seems to be intact in the vacuum experiments as there would be neither birefringence nor dispersive effects. In \charef{CHAPTER:SPECIAL MODEL}, the focus of the coefficient space is further restricted, with which some special cases are analyzed. It is demonstrated that there exists a nontrivial coefficient subspace satisfying the results found in \charef{vacuum orthogonal model}. Finally, the results are discussed and the thesis is concluded with a conclusion.
\chapter{Preliminaries}
\label{CHAPTER:PRELIMINARIES}
\section{Lorentz and CPT Symmetries}
\label{Section: lorentz and cpt}
In physics, the mathematical formulation of observation of a phenomena requires an observer dependent tool, like the coordinate system. Then, the findings of different observers observing the same phenomena can be translated to each other by a proper transformation.

Before Special Relativity, a concept called absolute time was considered to be valid; that is, each and every observer measures the time exactly the same. The proper transformation of space coordinates respecting this absolute time then was called \emph{Galilean Transformations}\footnotemark\footnotetext{More precisely, the transformation between two \emph{inertial frames} with the absolute time is called Galilean Transformation.}. With the Special Relativity, however, concept of absolute time had become inconsistent with the finite speed limit, which Einstein claimed to be the speed of light in vacuum. It was realized by \emph{Minkowski} in 1907 that the reference frames should actually be four dimensional inertial frames\cite{Rindler}.

Lorentz symmetry, named after \emph{Hendrik Lorentz}, is the associated symmetry of the mathematical group governing these four dimensional transformations. This group, also called Lorentz group, ensures that the relative orientation or the relative velocity (boost) of the laboratory in space do not affect the experimental results.

CPT invariance is the law that requires the physics to be unaffected under the combined operations of charge conjugation (C), parity inversion (P), and time reversal (T). In a nutshell, charge conjugation, parity inversion, and time reversal correspond to the interchange between the particle and antiparticle, the inversion of the direction of all true vectors in the coordinate space (like position and velocity), and the flip of flow of the time, respectively. In that sense, CPT invariance bluntly guarantees the physics to be exact for a particle in a spacetime and for the antiparticle in the inverted spacetime.

The above relation between a particle and antiparticle suggested by CPT invariance lead many natural expectations, regarding the symmetry between matter and antimatter properties, such as mass, charge, decay rate, gyromagnetic ratio of elementary particles, and spectra and particle-reaction processes of atoms\cite{hep-ph/0611177}. The test of whether these properties are indeed identical for matter and antimatter ascertains a probe into the test of CPT invariance.

As explained in the introduction, the CPT theorem is invalid once the Lorentz symmetry is broken; hence, one could see CPT violation if Lorentz symmetry is violated, and definitely expects Lorentz violation if CPT symmetry is broken, as long as locality is respected. In this thesis, we will deal with a model which breaks both CPT and Lorentz symmetries, but respects the locality nonetheless.

\section{Observer and Particle Lorentz Transformations}
\label{Section: olt and plt}

In the context of \emph{Special Relativity}, there are two types of transformations: \emph{passive transformations}, and \emph{active transformations}. The passive transformations are those in which the observer is boosted or rotated for an unchanged particle, and the active transformations are those in which the observer is left invariant for a boosted or rotated particle.

In the context of \emph{Quantum Field Theory}, the passive Lorentz transformations carry over as they are, with the name \emph{Observer Lorentz Transformations} (OLT); however, the active transformations are redefined as \emph{Particle Lorentz Transformations} (PLT) such that, crudely, the background fields as well as the observer are left invariant under PLT.

In the more precise terminology, constant background fields in the theory transform as Lorentz tensors under OLT, yet transform as scalars under PLT. If these constant background fields couple to usual currents that transform as Lorentz tensors under both transformations, the Lorentz symmetry is broken under PLT\cite{hep-ph/9809521}.

There is a subtle point about whether observer or particle Lorentz symmetry is broken. The observer Lorentz symmetry represents the fact that Physics is independent of the coordinate system that we use. This is vital as it would be nonsense if two observers observing the same phenomena obtain different results just because they labeled their origin or their axis differently. The SME guarantees that observer Lorentz symmetry is intact. This is ensured firstly because all background fields are proper Lorentz tensors under OLT, and secondly because the Lagrangian consists only of properly contracted terms.

The origin of these background fields are beyond the scope of SME, however they are expected to be \emph{vacuum expectation values} of some Lorentz tensors in the underlying theory. The reasoning for this conclusion is that the other option, that they arise from localized experimental conditions, is vetoed due to the fact that these background fields transform as scalars under PLT\cite{hep-ph/9809521}.

The breaking of Lorentz symmetry under PLT on the other hand, does not pose any conceptual problem. It merely means that the vacuum may have preferred directions. Since the breaking is also spontaneous, we can simply conclude that even though the underlying theory is Lorentz invariant, the vacuum solution is not on the symmetry point, hence the spontaneous symmetry breaking occurs, giving some background tensors expectation values\footnotemark\footnotetext{For more details, please refer to \appref{CHAPTER:SPONTANEOUS}.}. 

A direct example to distinguish between OLT and PLT might be hard to find; however, an analogy can be given for a charged particle in a cyclotron motion due to an homogeneous magnetic field. For the purpose of the illustration, let us accept the magnetic field as a background field, which would then transform as a scalar under particle boost, that is the change of the speed of the charged particle. Hence, under particle boost, the charged particle still accepts a cyclotron motion, only with a different radius. In contrast, the magnetic field transforms as a second rank tensor under OLT, which would then yields a nonzero electric field; therefore, the charged particle moves in a helix. The situation is depicted in \figref{OLT_vs_PLT}.
\begin{figure}
\sfigure{OLT_vs_PLT}{1.5}{Particle versus Observer Lorentz Transformations}{Particle versus Observer Lorentz Transformations. The effect of the difference between the behavior of the background field in OLT and PLT can be demonstrated for the cyclotron motion of an electron in a constant homogeneous magnetic field. If this magnetic field is taken as a background field, it would transform as scalar under particle boost, after which the electron is still in a cyclotron motion, albeit with a different radius. After an observer boost on the other hand, the magnetic field which behaves as a second rank Lorentz tensor under OLT transforms into a mixture of electric and magnetic fields in which the electron now follows a helical motion. That illustrates how PLT and OLT leave the system in different states.}
\end{figure}

\section{Effective Field Theoretical Approach and the Photon Sector}
\label{Section: eft and photon}

There are various approaches in literature that deals with Lorentz and CPT violations. A brief introduction of the model that will be used in this thesis, Standard Model Extension, and some other approaches are listed in \appref{CHAPTER:LORENTZ}. Here, we will briefly mention the \emph{effective field theoretical} (EFT) approach, and its relevance to nonrenormalizable LV.

Effective field theory is basically a field theory which is considered as an approximation to an underlying fundamental theory whose structure is unknown. In some systems, low energy behavior can be quite independent of high energy states. Therefore, a model of low energy states can be used as an effective theory which is valid upto a cutoff energy. For other systems, the high energy model can be unknown, which leads to construction of an EFT by a "bottom-up" approach where candidate Lagrangian's are built with symmetry and naturalness constraints. In either case, EFT turns out to be quite useful, and is used as a tool in particle physics and statistical mechanics\cite{EFT}.

Renormalization, in the context of Quantum Field Theory, is the technique to cure the infinities in the calculation of interested quantities, like energy or mass. Therefore, renormalizability of a theory is its ability to cope with the renormalization techniques. Thus, if a theory is nonrenormalizable, there arise infinities in some calculations that cannot be avoided. That is one of the main reasons why Gravity is still not properly quantized: It is a nonrenormalizable theory.

In the past, nonrenormalizable theories were considered to be ill-natured; however, this attitude has changed once nonrenormalizable theories are seen as effective field theories. Indeed, the appearance of the infinities in the calculations indicates this: Nonrenormalizable theories are not the fundamentally valid ones, they are merely approximations to their underlying theories.

Let us briefly show the usefulness of EFT approach in a nonrenormalizable theory by considering Euler-Heisenberg theory\cite{1312.4916}. In their theory, Hans Heinrich Euler and Werner Heisenberg introduce nonlinear photon behavior in vacuum. But we can see how this quantum effect can arise in the Lagrangian of the classical electromagnetic theory via the nonrenormalizable EFT approach. Clearly, we can include any higher order terms like $F^4$ or $\left(F\tilde{F}\right)^2\;$ to the classical Lagrangian $F^2$ once we abandon renormalizability\footnotemark\footnotetext{Here, $F$ denotes the second rank field strength tensor and $\tilde{F}$ denotes the dual field strength tensor.}. However, these terms represent light-by-light scattering, which are completely in quantum nature. Therefore, without knowing the underlying quantum theory, we can extract the light-by-light scattering information in the classical regime using an EFT approach. That could be seen as the power of nonrenormalizable theories.

\subsection{Standard Model Extension}
\label{Section: sme}

There are several methodologies to obtain a model with Lorentz and CPT violations. In this thesis, we work with an EFT model called \emph{Standard Model Extension}, which was constructed over almost two decades ago by the seminal paper \emph{Lorentz violating extension of the standard model}\cite{hep-ph/9809521}.

For the case of LV, EFT is quite a convenient framework. The power of EFT comes from the fact that it is expected to be applicable only in a validity range. Therefore, if we are dealing with small deviations from a known model, an effective model can easily be built upon it to explain these deviations. As current bounds on LV are quite stringent, we expect very small deviations in the interested low energy regimes, thus can easily use EFT.

Constructing an EFT extension of SM with the relaxation of Lorentz symmetry is quite straightforward: All combinations of contractions of usual SM field operators with some constant background tensors, respecting still intact symmetries like gauge invariance, are added to SM Lagrangian. As explained in \secref{Section: olt and plt}, these constant background tensors called LV coefficients are assumed to be the vacuum expectation values of some fields in the underlying theory, hence any contraction with these terms, called LVT, break the Lorentz symmetry under PLT. The details regarding to the construction of SME and the alternative theories in the literature are presented in \appref{CHAPTER:LORENTZ}; nonetheless, let us quote the Lagrangian of the renormalizable part of the photon sector in SME, that is the photon sector Lagrangian of \emph{m}SME:
\begin{equation*}
\mathcal{L}^{\text{\emph{m}SME}}_{\text{photon}}=-\frac{1}{4}F_{\mu\nu}F^{\mu\nu}-\frac{1}{4}(k_F)_{\kappa\lambda\mu\nu}F^{\kappa\lambda}F^{\mu\nu}+\frac{1}{2}(k_{AF})^\kappa\epsilon_{\kappa\lambda\mu\nu}A^\lambda F^{\mu\nu}\,,
\end{equation*}
from \equref{Eq: Minimal QED Lagrangian}.

The constant background tensors mentioned above correspond to the coefficients $(k_F)_{\kappa\lambda\mu\nu}$ and $(k_{AF})^\kappa$, which are simple constants\footnotemark\footnotetext{In the \emph{nm}SME, they will not be simple constants, but operators comprising of differentiations and contracted coefficients, as will be seen in the next section.} to be bounded by experiments. These coefficients are of mass dimensions $4$ and $3$ respectively; hence, they satisfy power counting renormalizability as expected, since they constitute the minimal portion of the SME.

In this thesis, however, we will be dealing with the nonrenormalizable part of the SME, called \emph{nonminimal Standard Model Extension} (\emph{nm}SME), despite that we will still be interested in the photon sector only. We will briefly discuss the extraction of the Lagrangian for \emph{nm}SME below, and analyze its CPTV part throughout the thesis. It should be remarked that although there are potential effects of other sectors, like fermion sector, on the photon sector LV, these effects are beyond the scope of this thesis.

\subsection{Examination of Photon Lagrangian}
\label{Section: examination of photon lagrangian}
The generic LV Lagrangian of the photon sector can be extracted from the most general LV action
\begin{equation*}
S_{(d)}=\int d^4x\; \mathcal{K}_{(d)}^{\alpha_1\alpha_2\alpha_3...\alpha_d}A_{\alpha_1}\partial_{\alpha_3}...\partial_{\alpha_d}A_{\alpha_2}\,.
\end{equation*}

Conservation of the energy and the momentum restrict $\mathcal{K}$ to be constant. Additionally, the inherit symmetries of $\mathcal{K}$ and the requirement of \emph{U(1) gauge invariance} reduce the possible representations of $\mathcal{K}$ into two sets only\cite{0905.0031}. Glossing over the details, we can make the following definitions
\begin{equation*}
\begin{aligned}
(k_{AF}^{(d)})_\kappa^{\;\alpha_1...\alpha_{d-3}}{}&{}:=\frac{1}{3!}\epsilon_{\kappa\mu\nu\rho}\mathcal{K}_{(d)}^{\mu\nu\rho\alpha_1\alpha_2...\alpha_{d-3}}\,,\\
(k_{F}^{(d)})^{\kappa\lambda\mu\nu\alpha_1...\alpha_{d-4}}{}&{}:=\mathcal{K}_{(d)}^{\kappa\mu\lambda\nu\alpha_1\alpha_2...\alpha_{d-4}}
\end{aligned}
\end{equation*}
for convenience, which in turn lead to the Lagrangian
\begin{equation}
\mathcal{L}=-\frac{1}{4}F_{\mu\nu}F^{\mu\nu}+\frac{1}{2}\epsilon^{\kappa\lambda\mu\nu}A_\lambda(\hat{k}_{AF})_\kappa F_{\mu\nu}-\frac{1}{4}F_{\kappa\lambda}(\hat{k}_{F})^{\kappa\lambda\mu\nu}F_{\mu\nu}\,,\label{eq:General Lagrangian}
\end{equation}
where the differential operators\footnotemark\footnotetext{The quantities with a hat are operators, constituting of infinitely many coefficients of all possible dimensions.} $(\hat{k}_{AF})_\kappa$ and $(\hat{k}_{F})^{\kappa\lambda\mu\nu}$ are defined as series expansion of the initial LV coefficients $(k_{F}^{(d)})$ and $(k_{AF}^{(d)})$:
\begin{subequations}
\begin{align}
(\hat{k}_{AF})_\kappa{}&{}=\sum_{d=\text{odd}\ge 3}(k_{AF}^{(d)})_\kappa^{\;\alpha_1...\alpha_{d-3}}\partial_{\alpha_1}...\partial_{\alpha_{(d-3)}}\,,\\
(\hat{k}_{F})^{\kappa\lambda\mu\nu}{}&{}=\sum_{d=\text{even}\ge 4}(k_{F}^{(d)})^{\kappa\lambda\mu\nu\alpha_1...\alpha_{d-4}}\partial_{\alpha_1}...\partial_{\alpha_{(d-4)}}\,.
\end{align}\label{Eq:k_AF and k_F}
\end{subequations}

In this notation, the subscript $AF$ refers to the fact that LV introduced by the respective coefficient comes from a contraction with a photon field $A$ and a field strength tensor $F$; hence, the corresponding LV brings also CPTV. Hence, the operator $\hat{k}_{AF}$ is CPT-odd. Contrarily, the operator $\hat{k}_{F}$, as its subscript denotes, contracts only with $F$'s, hence does not violate CPT invariance. Therefore, it is called CPT-even.

In operators, the quantities are denoted by a hat, and does not carry a $d$ index. The coefficients on the other hand carry this index, which denotes the dimension of the associated field operator in the corresponding LVT. Therefore, the \emph{m}SME is already contained within \emph{nm}SME, and can be obtained from it if we restrict $d$ to its minimum value, $d=3$ for $(k_{AF}^{(d)})_\kappa^{\;\alpha_1...\alpha_{d-3}}$ and $d=4$ for $(k_{F}^{(d)})^{\kappa\lambda\mu\nu\alpha_1...\alpha_{d-4}}$.

Due to the way they are constructed, there are various symmetry conditions on the coefficients. These conditions are as follows.
\begin{itemize}
\item $(k_{AF}^{(d)})_{\kappa}^{\;\alpha_1...\alpha_{d-3}}$
	\begin{itemize}
	\item is totally symmetric on its last d-3 indices.
	\item obeys the trace condition $(k_{AF}^{(d)})_{\alpha_1}^{\;\alpha_1...\alpha_{d-3}}=0$.
	\end{itemize}
\item $(k_{F}^{(d)})^{\kappa\lambda\mu\nu\alpha_1...\alpha_{d-4}}$
	\begin{itemize}
	\item has the symmetries of Riemann tensor in its first 4 indices.
	\item is totally symmetric on its last d-4 indices.
	\item is equal to zero if any 3 indices of it is antisymmetrized.
	\end{itemize}	
\end{itemize}

These conditions taken into account, the overall number of independent components for each coefficient becomes
\begin{equation*}
\begin{aligned}
N^{(d)}_{AF}{}&{}= \frac{1}{2}(d+1)(d-1)(d-2)\,,\\
N^{(d)}_{F}{}&{}=(d+1)d(d-3)\,,
\end{aligned}
\end{equation*}
where this counting includes the total trace term of $k_{F}^{(d)}$ which is Lorentz invariant.

Although it will not be explicitly utilized in this thesis, a new field tensor $G^{\mu\nu}$ can be defined such that the form of the conventional \emph{Equations of Motion} (EOM)
\begin{equation*}
\partial_\nu F^{\mu\nu}=0
\end{equation*}
can be preserved even in the presence of LV,
\begin{equation*}
\partial_\nu G^{\mu\nu}=0\,.
\end{equation*}
This can be done if $G^{\mu\nu}$ is defined as
\begin{equation*}
G^{\mu\nu}:=F^{\mu\nu}-2\epsilon^{\mu\nu\alpha\beta}(\hat{k}_{AF})_\alpha A_\beta+(\hat{k}_{F})^{\mu\nu\alpha\beta}F_{\alpha\beta}\,.
\end{equation*}
Here, the EOM is gauge invariant even though the definition of $G^{\mu\nu}$ is gauge dependent.

Lastly, it is possible to write this LV field tensor $G^{\mu\nu}$ in terms of the conventional $F^{\mu\nu}$ and $A^\mu$. It can be shown that
\begin{equation}
G^{\mu\nu}=\hat{\chi}^{\mu\nu\rho\sigma}F_{\rho\sigma}+2\hat{X}^{\mu\nu\rho}A_\rho\,,\label{Eq:constitutive relation}
\end{equation}
where
\begin{subequations}
\begin{align}
\hat{\chi}^{\mu\nu\rho\sigma}{}&{}=\frac{1}{2}(\eta^{\mu\rho}\eta^{\nu\sigma}-\eta^{\nu\rho}\eta^{\mu\sigma})+(\hat{k}_{F})^{\mu\nu\rho\sigma}\,,\\
\hat{X}^{\mu\nu\rho}{}&{}=\epsilon^{\mu\nu\rho\sigma}(\hat{k}_{AF})_\sigma\,. 
\end{align}\label{Most general constitutive relations}
\end{subequations}
The relation given by \equref{Eq:constitutive relation} is called \emph{constitutive relation}, where $\hat{\chi}^{\mu\nu\rho\sigma}$ and $\hat{X}^{\mu\nu\rho}$, defined by \equref{Most general constitutive relations}, are called \emph{constitutive 4 tensor} and \emph{constitutive 3 tensor} respectively.

\subsection{Spherical Decomposition}
\label{Section: spherical decomposition}

The LV coefficients, given by \equref{Eq:k_AF and k_F}, are in coordinate-free form as they are; however, what is measured in the experiments are the components of these coefficients with respect to a specific coordinate system. The Cartesian coordinate system is mostly the usual choice for a field theory; however, it is more convenient to work in so-called helicity basis\footnotemark\footnotetext{For more details, please refer to \appref{CHAPTER:HELICITY}.} for the photon.

In helicity basis, which is actually nothing more than a complex spherical coordinate system, the coefficients decompose into spin weighted spherical harmonics\footnotemark\footnotetext{This process is called \emph{spherical decomposition}. Since its details are not necessary, the process of that decomposition is not included in this thesis; however, an interested reader can consult to \cite{0905.0031}.}. This is the main reason to use, and main advantage of using, the helicity basis, as spherical decomposition is a natural classification with direct relevance to observations and experiments. Then, the LV operators $\hat{k}_{F}$ and $\hat{k}_{AF}$ decompose as
\begin{equation*}
\begin{aligned}
\hat{k}_{AF}{}&{}\longrightarrow\left\{
\begin{aligned}
(k_{AF}^{(d)})_{njm}^{(0B)}\\(k_{AF}^{(d)})_{njm}^{(1B)}\\(\vok_{AF}^{(d)})_{njm}^{(1E)}
\end{aligned}\,,\right. \\
\hat{k}_F{}&{}\longrightarrow\left\{
\begin{aligned}
(c_F^{(d)})_{njm}^{(0E)}\\(k_F^{(d)})_{njm}^{(0E)}\\(\vok_F^{(d)})_{njm}^{(1E)}\\(\vok_F^{(d)})_{njm}^{(2E)}\\(k_F^{(d)})_{njm}^{(1B)}\\(\vok_F^{(d)})_{njm}^{(2B)}\\
\end{aligned}\,.\right.
\end{aligned}
\end{equation*}

In its spherically decomposed version, the notation becomes a little bit nontrivial. How to read them is as follows.
\begin{itemize}
\item The symbol $k$ ($c$) means that the associated operators are birefringent (nonbirefringent).
\item As usual, the subscript $AF$ ($F$) means that the associated LVT is CPT-odd (CPT-even).
\item As usual, the superscript $d$ shows the dimension of associated field operator.
\item The presence of a negation diacritic $\neg$ indicates that the associated LVT is \emph{vacuum orthogonal}, a term to be precisely defined below.
\item In the subscript $njm$, $n$ determines the frequency dependence, $j$ determines the total angular momentum, and $m$ determines the $z$-component of the total angular momentum.
\item In the superscript $(iA)$, $i$ ($A$) denotes the spin-weight (parity) of the associated operator\footnotemark\footnotetext{The parity $(-1)^j$ is labeled as $E$ whereas the parity $(-1)^{j+1}$ is labeled as $B$.}.
\end{itemize}

\begin{table}
\caption[Spherically Decomposed Coefficients And Their Vacuum Properties]{Spherically Decomposed Coefficients And Their Vacuum Properties. The Lorentz violating coefficients in the photon Sector of \emph{nm}SME can be divided to their CPT and vacuum properties like below. Those in the left column are CPT-odd, whereas those in the right column are CPT-even. Likewise, top row corresponds to the coefficients with leading order vacuum effects, and bottom row corresponds to the coefficients with no leading order vacuum effect.}
\label{Table-sdc}
\begin{tabularx}{\textwidth}{lXX}
\hline\hline\\[-0.2in]
& $\mathbf{\hat{k}_{AF}}$ & $\mathbf{\hat{k}_{F}}$\\\hline\\[-0.15in]
Vacuum Coefficients &$k_{(V)jm}^{(d)}$&$c_{(I)jm}^{(d)}$, $k_{(E)jm}^{(d)}$, $k_{(B)jm}^{(d)}$\\\\[-0.15in]
Vacuum Orthogonal \mbox{Coefficients}&$(\vok_{AF}^{(d)})_{njm}^{(0B)}$, $(\vok_{AF}^{(d)})_{njm}^{(1B)}$, $(\vok_{AF}^{(d)})_{njm}^{(1E)}$&$(\voc_{F}^{(d)})_{njm}^{(0E)}$, $(\vok_{F}^{(d)})_{njm}^{(0E)}$, $(\vok_{F}^{(d)})_{njm}^{(1E)}$, $(\vok_{F}^{(d)})_{njm}^{(2E)}$, $(\vok_{F}^{(d)})_{njm}^{(1B)}$, $\,\,(\vok_{F}^{(d)})_{njm}^{(2B)}$ \\\hline\hline
\end{tabularx}
\end{table}

There is also a further decomposition of these coefficients according to their effects on the vacuum propagation: Those with leading order dispersive and birefringence effects, and those with neither leading order dispersive nor leading order birefringence effects. The first group is called vacuum coefficients, as their effects are detectable in the vacuum propagation of the photon. The second group is called vacuum-orthogonal coefficients, as these coefficients are constructed from the complement coefficient subspace of vacuum coefficients'  subspace. They are denoted by a negation diacritic as can be seen in \tabref{Table-sdc}.

The decomposition according to the vacuum properties is actually quite important for the \emph{nm}SME because it turns out that the vacuum orthogonal coefficients are nonzero only for $d>4$; hence, they represent LV effects not possible in the \emph{m}SME. That is another solid reason to consider a nonrenormalizable theory as there are indeed some kinds of LV that cannot be obtained in the renormalizable model.

\subsection{Tests and Bounds}
\label{Section: tests and bounds}

The most important aspect of a scientific theory in the view of Popper's falsifiability is just how much that theory can be tested with empiric experiments; in the same spirit, the most vital part of the LV theories is how much they can be tested as well. However, there is a subtle issue in the tests of LV:  As the effect is expected to be very suppressed, and since there is no actual lower bound for the violation, a test with a result of no LV is incapable of invalidating LV. The best that such a test can do is to eliminate the possibility of LV down to the order of the accuracy of the test, which is called the bound on the particular type of LV that the test was addressing.

There are quite different types of tests that have been actively engaged in the search of LV, which is actually a necessity as there are a variety of possible forms for LV in SME to be bounded. Yet, all of the tests search for any interaction between the particles and the background tensors explained in \appref{CHAPTER:SPONTANEOUS}. Such interactions can be found via effects depending on the couplings and the particle properties like velocity, spin, and flavor\cite{1309.3761}.

The bounds on the LV coefficients of SME are collected under data tables\cite{0801.0287}, which is first published in 2008 and annually updated ever since. According to the tables, current bounds on the CPTV photon sector of \emph{m}SME are extremely stringent, as can be seen in the \tabref{Table-bounds}. However, the bounds loosen as the dimension increases; in fact, there is no bounds at all on the vacuum orthogonal coefficients, which are absent in the minimal theory.

\begin{table}
\caption[Some of the Maximal Bounds on the CPT Violating Photon Sector]{Some of the Maximal Bounds on the CPT Violating Photon Sector. The table to the left demonstrates the bounds on the minimal photon sector while the table to the right demonstrates those on the nonminimal photon sector (Table S3. in\cite{0801.0287}). Clearly,  the possibility of LV in the renormalizable part is almost wiped out, whereas there is more and more room for LV with increasing dimensions in the nonrenormalizable part. Also, all of the bounds in the photon sector are for the vacuum coefficients: There is yet no bounds on the vacuum orthogonal ones.}
\label{Table-bounds}
\centering
\begin{minipage}{6.5cm}
\begin{tabular}{l l l}
\hline\hline
$d=3$ & \textbf{Coefficient} & \textbf{Sensitivity}\\\hline
&$k_{(V)00}^{(3)}$ & $10^{-43}$ GeV\\
&$k_{(V)10}^{(3)}$ & $10^{-42}$ GeV\\
& Re$k_{(V)11}^{(3)}$ & $10^{-42}$ GeV\\
& Im$k_{(V)11}^{(3)}$ & $10^{-42}$ GeV\\[-0.2in]
\\\hline\hline
\end{tabular}
\end{minipage}
\begin{minipage}{6.5cm}
\begin{tabular}{l l l}
\hline\hline
$d\ge5$ & \textbf{Coefficient} & \textbf{Sensitivity}\\\hline
&$k_{(V)00}^{(5)}$ & $10^{-34}$ GeV$^{-1}$\\
&$k_{(V)00}^{(7)}$ & $10^{-27}$ GeV$^{-3}$\\
&$k_{(V)00}^{(9)}$ & $10^{-21}$ GeV$^{-5}$\\
&&
\\\hline\hline
\end{tabular}
\end{minipage}
\end{table}
\chapter{CPT-odd Photon Sector}
\label{CHAPTER:CPT ODD PHOTON}

The construction of a general photon sector with spontaneous Lorentz violation is discussed and the motivation for considerations of nonrenormalizable terms are presented in \charef{CHAPTER:PRELIMINARIES}. In this chapter, we will elaborate on the photon sector with CPT violating terms only, by extracting its field theoretical properties; firstly the dispersion relation in a covariant and in explicitly helicity basis form, then the polarization vectors associated with the roots of the dispersion relations, and finally the propagator in finite order covariant form and in exact but explicitly helicity basis form.

The CPT-odd part of the general {\it nm}SME photon Lagrangian, given by \equref{eq:General Lagrangian}, can be extracted by imposing the condition $\hat{k}_{F}=0$, which in turn yields
\begin{equation}
\boxed{\mathcal{L}_{\text{modified}}={}-\frac{1}{4}F_{\mu\nu}F^{\mu\nu}+\frac{1}{2}\epsilon^{\kappa\lambda\mu\nu}A_\lambda(\hat{k}_{AF})_\kappa F_{\mu\nu}}\label{general cpt-odd lagrangian}
\end{equation}
Then, one can directly write the corresponding CPT-odd action as
\begin{equation*}
\mathcal{S}_{\text{modified}}={}-\frac{1}{4}\int d^4x\left(F_{\mu\nu}F^{\mu\nu}-2\epsilon^{\kappa\lambda\mu\nu}A_\lambda(\hat{k}_{AF})_\kappa F_{\mu\nu}+2(\partial_\mu A^\mu)^2\right)\,,\label{general action}
\end{equation*}
where $\zeta=1$  Feynman 't Hooft gauge fixing term is used. Then,
\begin{equation*}
\begin{aligned}
\mathcal{S}_{\text{modified}}=-\frac{1}{4}\int d^4x\Big({}&{}(\partial_\mu A_\nu-\partial_\nu A_\mu)(\partial^\mu A^\nu-\partial^\nu A^\mu)\\{}&{}-2\epsilon^{\kappa\lambda\mu\nu}A_\lambda(\hat{k}_{AF})_\kappa(\partial_\mu A_\nu-\partial_\nu A_\mu)+2(\partial_\mu A^\mu)(\partial_\nu A^\nu)\Big)\,,
\end{aligned}
\end{equation*}
\begin{equation*}
\begin{aligned}
\mathcal{S}_{\text{modified}}=-\frac{1}{4}\int d^4x\Big({}&{}\partial_\mu A_\nu\partial^\mu A^\nu-\partial_\nu A_\mu\partial^\mu A^\nu-\partial_\mu A_\nu\partial^\nu A^\mu+\partial_\nu A_\mu\partial^\nu A^\mu\\{}&{}+2\partial_\mu A^\mu\partial_\nu A^\nu-4A_\lambda(\epsilon^{\kappa\lambda\mu\nu}(\hat{k}_{AF})_\kappa\partial_\mu)A_\nu\Big)\,.
\end{aligned}
\end{equation*}
If the terms in the first column are rewritten and the surface terms are eliminated, the most general form of modified action arises:
\begin{equation}
\mathcal{S}_{\text{modified}}=\frac{1}{2}\int d^4x A_\mu\left(\eta^{\mu\nu}\partial^2-2\epsilon^{\mu\kappa\lambda\nu}(\hat{k}_{AF})_\kappa\partial_\lambda\right)A_\nu\label{Eqn:general_modified_action}\,.
\end{equation}

As the gauge field propagator is of the form
\begin{equation}
\mathcal{S}_{\text{modified}}=\frac{1}{2}\int d^4x A_\mu(\hat{G}^{-1})^{\mu\nu}A_\nu\,,\nonumber
\end{equation}
the modified action immediately gives the propagator in the configuration space, that is
\begin{equation}
(\hat{G}^{-1})^{\mu\nu}=\eta^{\mu\nu}\partial^2-2\epsilon^{\mu\kappa\lambda\nu}(\hat{k}_{AF})_\kappa\partial_\lambda\,.\nonumber
\end{equation}
But one can go to the momentum space with the prescription $\partial_\mu\rightarrow-ip_\mu$; hence, the most general form of the modified inverse propagator in the momentum space becomes
\begin{equation}
\boxed{(\hat{G}^{-1})^{\mu\nu}=-\eta^{\mu\nu}(p_\sigma p^\sigma)+2i\epsilon^{\mu\kappa\lambda\nu}(\hat{k}_{AF})_\kappa p_\lambda}\label{Momentum space inverse propagator}
\end{equation}

The next thing that should be sought is the form of the equations of motions. But, they are given by the Euler-Lagrange equations\footnotemark\footnotetext{It is actually not that straightforward to use these equations as there arises Ostrogradski instabilities. It is discussed in \cite{0905.0031} in detail, and the EOM that will be derived here is provided there as well. We will perform a simplistic approach disregarding the details, and obtain the correct result nonetheless.}
\begin{equation*}
\partial_\nu\frac{\partial\mathcal{L}}{\partial(\partial_\nu A_\mu)}=\frac{\partial \mathcal{L}}{\partial A_\mu}\,,
\end{equation*}

and from \equref{Eqn:general_modified_action}, the equivalent Lagrangian we have is
\begin{equation*}
\mathcal{L}=A_\mu\left(\eta^{\mu\nu}\partial^2-2\epsilon^{\mu\kappa\lambda\nu}(\hat{k}_{AF})_\kappa\partial_\lambda\right)A_\nu\,,
\end{equation*} however, this is the Lagrangian in which the gauge is fixed; hence, it would not yield the gauge solution. To be able to obtain all solutions, we restore back the term which the gauge fixing removed; therefore,
\begin{equation}
\mathcal{L}_{\text{equivalent}}=A_\mu\left(\eta^{\mu\nu}\partial^2-\partial^\mu\partial^\nu-2\epsilon^{\mu\kappa\lambda\nu}(\hat{k}_{AF})_\kappa\partial_\lambda\right)A_\nu\,.\nonumber
\end{equation}
Clearly,
\begin{equation}
\begin{aligned}
\frac{\partial\mathcal{L}}{\partial A_\mu}={}&{}\left(\eta^{\mu\nu}\partial^2-\partial^\mu\partial^\nu-2\epsilon^{\mu\kappa\lambda\nu}(\hat{k}_{AF})_\kappa\partial_\lambda\right)A_\nu\,,\\
\frac{\partial\mathcal{L}}{\partial (\partial_\nu A_\mu)}={}&{} 0\,.
\end{aligned}\nonumber
\end{equation}
Here, although equation \equref{Eq:k_AF and k_F} suggests that $(\hat{k}_{AF})_\kappa$ has a constant term in it which corresponds to $d=3$, as the main focus in this thesis is in nonrenormalizable part which starts with dimension $5$, that experimentally highly suppressed constant term can be ignored, meaning that $(\hat{k}_{AF})_\kappa$ is an operator which comprises of at least two derivatives; hence, there is no first derivative of the field $A_\mu$ in $\mathcal{L}_{\text{equivalent}}$\footnotemark\footnotetext{The assumption, or simply the shrink of the focus to nonrenormalizable part, in the derivation is not actually essential. In \cite{0905.0031}, the same EOM's are derived under no assumption by a different methodology unlike the usual variation of the action. Nonetheless, the important point is that the resultant EOM is established in the literature.}. Hence,
\begin{equation*}
\left(\eta^{\mu\nu}\partial^2-\partial^\mu\partial^\nu-2\epsilon^{\mu\kappa\lambda\nu}(\hat{k}_{AF})_\kappa\partial_\lambda\right)A_\nu=0\,.
\end{equation*}

But with the adoption of the plane wave ansatz
\begin{equation*}
A_\mu(x)=A_\mu(p)e^{-ix.p}\,,
\end{equation*}
equations of motion take the form 
\begin{equation}
M^{\mu\nu}A_\nu=0 \label{eq:EOM}
\end{equation}
for
\begin{equation}
\boxed{M^{\mu\nu}={}\eta^{\mu\nu}p_\alpha p^\alpha-p^\mu p^\nu-2i\epsilon^{\mu\nu\alpha\beta}(\hat{k}_{AF})_\alpha p_\beta} \label{general form of M}
\end{equation}

Finally, the constitutive relations\footnotemark\footnotetext{In general, a constitutive relation is roughly an equation which specifies the response of a material to a physical disturbance. In the context of standard electromagnetism, the constitutive relations are those which relate the usual \textbf{E} and \textbf{B} fields to the displacement field \textbf{D} and magnetic field \textbf{H}. In the four vector notation, the constitutive relation relates $G^{\mu\nu}$, which is the solution of the EOM in the material for the photon field, to the usual field strength tensor $F^{\mu\nu}$ where $G^{\mu\nu}=F^{\mu\nu}$ in the vacuum. In the presence of the Lorentz violation, $G^{\mu\nu}$ is different from $F^{\mu\nu}$ and their relation is given by the constitutive tensors via $G^{\mu\nu}=\hat{\chi}^{\mu\nu\rho\sigma}F_{\rho\sigma}+2\hat{X}^{\mu\nu\rho}A_\rho$ \cite{0905.0031}.} for the CPT-odd case are simply
\begin{subequations}
\begin{align}
\hat{\chi}^{\mu\nu\rho\sigma}{}&{}=\frac{1}{2}\left(\eta^{\mu\rho}\eta^{\nu\sigma}-\eta^{\nu\rho}\eta^{\mu\sigma}\right)\,,\\
\hat{X}^{\mu\nu\rho}{}&{}=\epsilon^{\mu\nu\rho\sigma}(\hat{k}_{AF})_\sigma\,,
\end{align}\label{eq:CPT odd constitute}
\end{subequations}
which can be obtained from the general relations, \equref{Most general constitutive relations}, by the substitution \mbox{$\hat{k}_F=0$}.

\section{The Dispersion Relation}
\label{Section: cpt odd dispersion}
In a classical theory, the dispersion relation is the functional form of the energy in terms of the momentum for the corresponding system; for example, $E=\frac{1}{2m}p^2$, \mbox{$E=\sqrt{p^2c^2+m^2c^4}$}, and $E=pc$ are the dispersion relations for a nonrelativistic particle, relativistic particle, and light respectively.

In the field theory, the dispersion relation becomes the constraint on the four momentum space for which there exists a nontrivial solution for the field living on that four-momentum space. For example, \equref{eq:EOM} indicates that there exists a nontrivial photon field $A_\mu$ satisfying the EOM only if the determinant of $M$ is zero; hence, $\det M=0$ will be the dispersion relation.

It can be easily checked that this dispersion relation is actually nothing but the classical dispersion relation which relates energy and momentum. For example, the EOM of the conventional photon is
\begin{equation*}
\partial^\mu F_{\mu\nu}=0\,.
\end{equation*}
With the usual Lorenz gauge choice $\partial_\rho A^\rho=0$, this equation becomes
\begin{equation*}
\partial_\mu\partial^\mu A_\nu=0\,,
\end{equation*}
which is
\begin{equation*}
p_\mu p^\mu A_\nu=0
\end{equation*}
in the momentum space, meaning that $M=p_\mu p^\mu$. But $\det M=0$ then simply means $p_\mu p^\mu=0$, which is nothing but $E=pc$ in the covariant form.

\subsection{Extraction of CPT-odd Sector}
\label{Section: extraction of cpt odd}

The dispersion relation for the Lagrangian given by \equref{general cpt-odd lagrangian} can be obtained by a similar, standard procedure: Firstly the gauge is fixed, and then the determinant of the reduced linear equations is calculated. However, this methodology breaks the covariance due to the explicit gauge choice; thus, one can alternatively use the rank-nullity of the EOM to find the covariant form of dispersion relations without sacrificing the gauge invariance. In \cite{0905.0031}, this procedure is employed, and the covariant dispersion relation
\begin{equation}
\begin{aligned}
0={}&{}-\frac{1}{3}\epsilon_{\mu_1\mu_2\mu_3\mu_4}\epsilon_{\nu_1\nu_2\nu_3\nu_4}p_{\rho_1}p_{\rho_2}p_{\rho_3}p_{\rho_4}\hat{\chi}^{\mu_1\mu_2\nu_1\rho_1}\hat{\chi}^{\nu_2\rho_2\rho_3\mu_3}\hat{\chi}^{\rho_4\mu_4\nu_3\nu_4}\\{}&{}+8p_\alpha p_\beta(\hat{k}_{AF})_\mu(\hat{k}_{AF})_\nu\hat{\chi}^{\alpha\mu\beta\nu}
\end{aligned}\label{eq:covariant dispersion relation}
\end{equation}
is obtained. But, from here, the dispersion relation for the CPT-odd model can be deduced by imposing the constitutive relations, that are \equref{eq:CPT odd constitute}:
\begin{align*}
0={}&{}-\frac{1}{24}\epsilon_{\mu_1\mu_2\mu_3\mu_4}\epsilon_{\nu_1\nu_2\nu_3\nu_4}p_{\rho_1}p_{\rho_2}p_{\rho_3}p_{\rho_4}\left(\eta^{\mu_1\nu_1}\eta^{\mu_2\rho_1}-\eta^{\mu_2\nu_1}\eta^{\mu_1\rho_1}\right)\\{}&{}\times\left(\eta^{\nu_2\rho_3}\eta^{\rho_2\mu_3}-\eta^{\rho_2\rho_3}\eta^{\nu_2\mu_3}\right)\left(\eta^{\rho_4\nu_3}\eta^{\mu_4\nu_4}-\eta^{\mu_4\nu_3}\eta^{\rho_4\nu_4}\right)\\{}&{}+4p_\alpha p_\beta(\hat{k}_{AF})_\mu(\hat{k}_{AF})_\nu\left(\eta^{\alpha\beta}\eta^{\mu\nu}-\eta^{\mu\beta}\eta^{\alpha\nu}\right)\,,\\
0={}&{}-\frac{1}{24}\epsilon_{\mu_1\mu_2\mu_3\mu_4}\epsilon_{\nu_1\nu_2\nu_3\nu_4}p_{\rho_1}p_{\rho_2}p_{\rho_3}p_{\rho_4}\Big(\eta^{\mu_1\nu_1}\eta^{\mu_2\rho_1}\eta^{\nu_2\rho_3}\eta^{\rho_2\mu_3}\eta^{\rho_4\nu_3}\eta^{\mu_4\nu_4}\\{}&{}-\eta^{\mu_1\nu_1}\eta^{\mu_2\rho_1}\eta^{\nu_2\rho_3}\eta^{\rho_2\mu_3}\eta^{\mu_4\nu_3}\eta^{\rho_4\nu_4}-\eta^{\mu_1\nu_1}\eta^{\mu_2\rho_1}\eta^{\rho_2\rho_3}\eta^{\nu_2\mu_3}\eta^{\rho_4\nu_3}\eta^{\mu_4\nu_4}\\{}&{}+\eta^{\mu_1\nu_1}\eta^{\mu_2\rho_1}\eta^{\rho_2\rho_3}\eta^{\nu_2\mu_3}\eta^{\mu_4\nu_3}\eta^{\rho_4\nu_4}-\eta^{\mu_2\nu_1}\eta^{\mu_1\rho_1}\eta^{\nu_2\rho_3}\eta^{\rho_2\mu_3}\eta^{\rho_4\nu_3}\eta^{\mu_4\nu_4}\\{}&{}+\eta^{\mu_2\nu_1}\eta^{\mu_1\rho_1}\eta^{\nu_2\rho_3}\eta^{\rho_2\mu_3}\eta^{\mu_4\nu_3}\eta^{\rho_4\nu_4}+\eta^{\mu_2\nu_1}\eta^{\mu_1\rho_1}\eta^{\rho_2\rho_3}\eta^{\nu_2\mu_3}\eta^{\rho_4\nu_3}\eta^{\mu_4\nu_4}\\{}&{}-\eta^{\mu_2\nu_1}\eta^{\mu_1\rho_1}\eta^{\rho_2\rho_3}\eta^{\nu_2\mu_3}\eta^{\mu_4\nu_3}\eta^{\rho_4\nu_4}\Big)+4p_\alpha p^\alpha(\hat{k}_{AF})_\mu(\hat{k}_{AF})^\mu-4\left( p_\alpha(\hat{k}_{AF})^\alpha\right)^2\,.
\end{align*}

Before the explicit calculation, it is worthwhile to note the relation between the covariant and contravariant Levi-Civita tensor densities. As the underlying space is of Minkowski metric,
\begin{equation}
\epsilon^{\mu_1\nu_1\rho_1\sigma_1}=-\eta^{\mu_1\mu_2}\eta^{\nu_1\nu_2}\eta^{\rho_1\rho_2}\eta^{\sigma_1\sigma_2}\epsilon_{\mu_2\nu_2\rho_2\sigma_2}\label{Equ:levi civita}\,.
\end{equation}
With that in mind, one can calculate the terms in the dispersion relation one by one.
\begin{enumerate}
\item $-\frac{1}{24}\epsilon_{\mu_1\mu_2\mu_3\mu_4}\epsilon_{\nu_1\nu_2\nu_3\nu_4}p_{\rho_1}p_{\rho_2}p_{\rho_3}p_{\rho_4}\eta^{\mu_1\nu_1}\eta^{\mu_2\rho_1}\eta^{\nu_2\rho_3}\eta^{\rho_2\mu_3}\eta^{\rho_4\nu_3}\eta^{\mu_4\nu_4}$
\begin{equation}
\begin{aligned}
={}&{}\frac{1}{24}\epsilon_{\mu_1\mu_2\mu_3\mu_4}\epsilon^{\mu_1\nu_2\nu_3\mu_4}p^{\mu_2}p^{\mu_3}p_{\nu_2}p_{\nu_3}\nonumber\\
={}&{}\frac{1}{12}\left(\delta_{\mu_2}^{\nu_2}\delta_{\mu_3}^{\nu_3}-\delta_{\mu_2}^{\nu_3}\delta_{\mu_3}^{\nu_2}\right)p^{\mu_2}p^{\mu_3}p_{\nu_2}p_{\nu_3}\nonumber\\={}&{} 0
\end{aligned}
\end{equation}

\item
$\frac{1}{24}\epsilon_{\mu_1\mu_2\mu_3\mu_4}\epsilon_{\nu_1\nu_2\nu_3\nu_4}p_{\rho_1}p_{\rho_2}p_{\rho_3}p_{\rho_4}\eta^{\mu_1\nu_1}\eta^{\mu_2\rho_1}\eta^{\nu_2\rho_3}\eta^{\rho_2\mu_3}\eta^{\mu_4\nu_3}\eta^{\rho_4\nu_4}$
\begin{equation}
\begin{aligned}
={}&{}-\frac{1}{24}\epsilon_{\mu_1\mu_2\mu_3\mu_4}\epsilon^{\mu_1\nu_2\mu_4\nu_4}p^{\mu_2}p^{\mu_3}p_{\nu_2}p_{\nu_4}\nonumber\\
={}&{}\frac{1}{12}\left(\delta_{\mu_2}^{\nu_2}\delta_{\mu_3}^{\nu_4}-\delta_{\mu_2}^{\nu_4}\delta_{\mu_3}^{\nu_2}\right)p^{\mu_2}p^{\mu_3}p_{\nu_2}p_{\nu_4}\nonumber\\={}&{} 0
\end{aligned}
\end{equation}

\item
$\frac{1}{24}\epsilon_{\mu_1\mu_2\mu_3\mu_4}\epsilon_{\nu_1\nu_2\nu_3\nu_4}p_{\rho_1}p_{\rho_2}p_{\rho_3}p_{\rho_4}\eta^{\mu_1\nu_1}\eta^{\mu_2\rho_1}\eta^{\rho_2\rho_3}\eta^{\nu_2\mu_3}\eta^{\rho_4\nu_3}\eta^{\mu_4\nu_4}$
\begin{equation}
\begin{aligned}
={}&{}-\frac{1}{24}\epsilon_{\mu_1\mu_2\mu_3\mu_4}\epsilon^{\mu_1\mu_3\nu_3\mu_4}p^{\mu_2}p_{\rho_2}p^{\rho_2}p_{\nu_3}\nonumber\\
={}&{}\frac{1}{4}\delta_{\mu_2}^{\nu_3}p^{\mu_2}p_{\rho_2}p^{\rho_2}p_{\nu_3}\nonumber\\
={}&{}\frac{1}{4}\big(p_\alpha p^\alpha\big)^2
\end{aligned}
\end{equation}

\item
$-\frac{1}{24}\epsilon_{\mu_1\mu_2\mu_3\mu_4}\epsilon_{\nu_1\nu_2\nu_3\nu_4}p_{\rho_1}p_{\rho_2}p_{\rho_3}p_{\rho_4}\eta^{\mu_1\nu_1}\eta^{\mu_2\rho_1}\eta^{\rho_2\rho_3}\eta^{\nu_2\mu_3}\eta^{\mu_4\nu_3}\eta^{\rho_4\nu_4}$
\begin{equation}
\begin{aligned}
={}&{}\frac{1}{24}\epsilon_{\mu_1\mu_2\mu_3\mu_4}\epsilon^{\mu_1\mu_3\mu_4\nu_4}p^{\mu_2}p_{\rho_2}p^{\rho_2}p_{\nu_4}\nonumber\\
={}&{}\frac{1}{4}\delta_{\mu_2}^{\nu_4}p^{\mu_2}p_{\rho_2}p^{\rho_2}p_{\nu_4}\nonumber\\
={}&{}\frac{1}{4}\big(p_\alpha p^\alpha\big)^2
\end{aligned}
\end{equation}

\item
$\frac{1}{24}\epsilon_{\mu_1\mu_2\mu_3\mu_4}\epsilon_{\nu_1\nu_2\nu_3\nu_4}p_{\rho_1}p_{\rho_2}p_{\rho_3}p_{\rho_4}\eta^{\mu_2\nu_1}\eta^{\mu_1\rho_1}\eta^{\nu_2\rho_3}\eta^{\rho_2\mu_3}\eta^{\rho_4\nu_3}\eta^{\mu_4\nu_4}$
\begin{equation}
\begin{aligned}
={}&{}-\frac{1}{24}\epsilon_{\mu_1\mu_2\mu_3\mu_4}\epsilon^{\mu_2\nu_2\nu_3\mu_4}p^{\mu_1}p^{\mu_3}p_{\nu_2}p_{\nu_3}\nonumber\\
={}&{}\frac{1}{12}\left(\delta_{\mu_1}^{\nu_2}\delta_{\mu_3}^{\nu_3}-\delta_{\mu_1}^{\nu_3}\delta_{\mu_3}^{\nu_2}\right)p^{\mu_1}p^{\mu_3}p_{\nu_2}p_{\nu_3}\nonumber\\
={}&{} 0\nonumber
\end{aligned}
\end{equation}

\item
$-\frac{1}{24}\epsilon_{\mu_1\mu_2\mu_3\mu_4}\epsilon_{\nu_1\nu_2\nu_3\nu_4}p_{\rho_1}p_{\rho_2}p_{\rho_3}p_{\rho_4}\eta^{\mu_2\nu_1}\eta^{\mu_1\rho_1}\eta^{\nu_2\rho_3}\eta^{\rho_2\mu_3}\eta^{\mu_4\nu_3}\eta^{\rho_4\nu_4}$
\begin{equation}
\begin{aligned}
={}&{}\frac{1}{24}\epsilon_{\mu_1\mu_2\mu_3\mu_4}\epsilon^{\mu_2\nu_2\mu_4\nu_4}p^{\mu_1}p^{\mu_3}p_{\nu_2}p_{\nu_4}\nonumber\\
={}&{}\frac{1}{12}\left(\delta_{\mu_1}^{\nu_2}\delta_{\mu_3}^{\nu_4}-\delta_{\mu_3}^{\nu_2}\delta_{\mu_1}^{\nu_4}\right)p^{\mu_1}p^{\mu_3}p_{\nu_2}p_{\nu_4}\nonumber\\
={}&{} 0\nonumber
\end{aligned}
\end{equation}

\item
$-\frac{1}{24}\epsilon_{\mu_1\mu_2\mu_3\mu_4}\epsilon_{\nu_1\nu_2\nu_3\nu_4}p_{\rho_1}p_{\rho_2}p_{\rho_3}p_{\rho_4}\eta^{\mu_2\nu_1}\eta^{\mu_1\rho_1}\eta^{\rho_2\rho_3}\eta^{\nu_2\mu_3}\eta^{\rho_4\nu_3}\eta^{\mu_4\nu_4}$
\begin{equation}
\begin{aligned}
={}&{}\frac{1}{24}\epsilon_{\mu_1\mu_2\mu_3\mu_4}\epsilon^{\mu_2\mu_3\nu_3\mu_4}p^{\mu_1}p_{\rho_2}p^{\rho_2}p_{\nu_3}\nonumber\\
={}&{}\frac{1}{4}\delta_{\mu_1}^{\nu_3}p^{\mu_1}p_{\rho_2}p^{\rho_2}p_{\nu_3}\nonumber\\
={}&{}\frac{1}{4}\big(p_\alpha p^\alpha\big)^2\nonumber
\end{aligned}
\end{equation}

\item
$\frac{1}{24}\epsilon_{\mu_1\mu_2\mu_3\mu_4}\epsilon_{\nu_1\nu_2\nu_3\nu_4}p_{\rho_1}p_{\rho_2}p_{\rho_3}p_{\rho_4}\eta^{\mu_2\nu_1}\eta^{\mu_1\rho_1}\eta^{\rho_2\rho_3}\eta^{\nu_2\mu_3}\eta^{\mu_4\nu_3}\eta^{\rho_4\nu_4}$
\begin{equation}
\begin{aligned}
={}&{}-\frac{1}{24}\epsilon_{\mu_1\mu_2\mu_3\mu_4}\epsilon^{\mu_2\mu_3\mu_4\nu_4}p^{\mu_1}p_{\rho_2}p^{\rho_2}p_{\nu_4}\nonumber\\
={}&{}\frac{1}{4}\delta_{\mu_1}^{\nu_4}p^{\mu_1}p_{\rho_2}p^{\rho_2}p_{\nu_4}\nonumber\\
={}&{}\frac{1}{4}\big(p_\alpha p^\alpha\big)^2\nonumber
\end{aligned}
\end{equation}
\end{enumerate}

Hence, the dispersion relation \equref{eq:covariant dispersion relation} becomes
\begin{equation}
\boxed{0=\left(p_\mu p^\mu\right)^2+4p_\alpha p^\alpha(\hat{k}_{AF})_\mu(\hat{k}_{AF})^\mu-4\left( p_\mu(\hat{k}_{AF})^\mu\right)^2} \label{General CPT-odd Dispersion Relation}
\end{equation}

This dispersion relation of CPT-odd nonrenormalizable photon sector is not available in the literature, as there is no study of general CPT-odd photon sector; however, its leading order form can be checked with the general leading order photon dispersion relation
\begin{equation*}
\left(p_\mu p^\mu-(\hat{c}_F)^{\mu\nu}p_\mu p_\nu\right)^2-2(\hat{\chi}_\omega)^{\alpha\beta\gamma\delta}(\hat{\chi}_\omega)_{\alpha\mu\gamma\nu}p_\beta p_\delta p^\mu p^\nu -4\left(p^\mu(\hat{k}_{AF})_\mu\right)^2\simeq 0\,.
\end{equation*}

But the CPT-odd part can be extracted simply by imposing $\hat{k}_F=0$, which in turn yields\footnotemark\footnotetext{Here, $\hat{\chi}_\omega$ is trace-free Weyl component of constitutive tensor $\hat{\chi}$. As the focus of this thesis is the CPT-odd part only, any further information such as what Weyl component means is beyond the scope at hand.}
\begin{equation*}
\begin{aligned}
(\hat{\chi}_\omega)_{\alpha\mu\gamma\nu}=0\,,\\\hat{c}_F=0\,,
\end{aligned}
\end{equation*}
hence
\begin{equation*}
0\simeq\left(p_\mu p^\mu\right)^2-4\left( p_\mu(\hat{k}_{AF})^\mu\right)^2\,,
\end{equation*}
which is exactly our dispersion relation above in the leading order limit\footnotemark\footnotetext{In the leading order limit, the condition $\omega\simeq p$ is enforced only on the LVT; in other words, $\omega$ and $p$ are still free variables in the dispersion relation. Nonetheless, the deviation is expected to be small because we are in an EFT regime. Then, we can take $\omega=p(1+\epsilon)$ for $\epsilon<1$ in the dispersion relation and check the order of each term in \equref{General CPT-odd Dispersion Relation}. Clearly, $p_\mu p^\mu=\omega^2-p^2$ is of order $\mathcal{O}(\epsilon)$, hence the first term is of $\mathcal{O}(\epsilon)$, and the second term is of $\mathcal{O}(\epsilon\lvert\hat{k}\rvert^2)$ where $\lvert\hat{k}\rvert\ll1$ is the regime of the EFT at hand. The last term on the other hand is of $\mathcal{O}(\lvert\hat{k}\rvert^2)$; hence, with the leading order approximation, the lowest term that is the second one vanishes.} Further details regarding these formulae can be obtained in \cite{0905.0031}.

\subsection{Spherical Decomposition}
\label{Section: cpt odd spherical decomposition}

Special models such as vacuum, general vacuum-orthogonal and camouflage models can be most transparently applied if the spherical decomposition method is employed. To do that, we first set the helicity basis as the space part of the coordinate system. 

The details of helicity basis is available in the \appref{CHAPTER:HELICITY}. As it can be seen, in this basis with the order $\{0,+,r,-\}$, the four momentum takes the form
\begin{equation}
p^\mu\doteq\begin{pmatrix}
\omega\\0\\p\\0
\end{pmatrix}\,,\label{eq:p in helicity basis}
\end{equation}
where $\omega$ is the usual frequency and $p$ is the magnitude of the space part of four momentum: $p:=\lvert\mathbf{p}\rvert$. Hence, we have
\begin{equation*}
\begin{aligned}
(\hat{k}_{AF})_\mu(\hat{k}_{AF})^\mu{}&{}=\left((\hat{k}_{AF})_0\right)^2-(\hat{k}_{AF})_+(\hat{k}_{AF})_--(\hat{k}_{AF})_r(\hat{k}_{AF})_r-(\hat{k}_{AF})_-(\hat{k}_{AF})_+\,,\\
p_\mu(\hat{k}_{AF})^\mu{}&{}=\omega(\hat{k}_{AF})_0-p(\hat{k}_{AF})_r\,.
\end{aligned}
\end{equation*}

Here, $(\hat{k}_{AF})_i$ are just summations of components of $p^\mu$, spin-weighted spherical harmonics and LV coefficients as can be seen from \equref{Equ: k_i}, hence they are \emph{not} operators in the momentum space, meaning that they commute with each other. Therefore, the dispersion relation \equref{General CPT-odd Dispersion Relation} becomes
\begin{align*}
0={}&{}\left(p_\mu p^\mu\right)^2+4\left(\omega^2-p^2\right)\left(\left((\hat{k}_{AF})_0\right)^2-2(\hat{k}_{AF})_+(\hat{k}_{AF})_--\left((\hat{k}_{AF})_r\right)^2\right)\\{}&{}-4\left(\omega(\hat{k}_{AF})_0-p(\hat{k}_{AF})_r\right)^2\,,\\
0={}&{}\left(p_\mu p^\mu\right)^2+4\Big(\omega^2\left((\hat{k}_{AF})_0\right)^2-2\omega^2(\hat{k}_{AF})_+(\hat{k}_{AF})_--\omega^2\left((\hat{k}_{AF})_r\right)^2\\{}&{}-p^2\left((\hat{k}_{AF})_0\right)^2+2p^2(\hat{k}_{AF})_+(\hat{k}_{AF})_-+p^2\left((\hat{k}_{AF})_r\right)^2\\{}&{}-\left(\omega(\hat{k}_{AF})_0-p(\hat{k}_{AF})_r\right)^2\Big)\,,\\
0={}&{}\left(p_\mu p^\mu\right)^2+4\Big(-2\omega^2(\hat{k}_{AF})_+(\hat{k}_{AF})_--\omega^2\left((\hat{k}_{AF})_r\right)^2-p^2\left((\hat{k}_{AF})_0\right)^2\\{}&{}+2p^2(\hat{k}_{AF})_+(\hat{k}_{AF})_-+2\omega p(\hat{k}_{AF})_0(\hat{k}_{AF})_r\Big)\,,\\
0={}&{}\left(p_\mu p^\mu\right)^2-4\left(\omega^2\left((\hat{k}_{AF})_r\right)^2+p^2\left((\hat{k}_{AF})_0\right)^2-2\omega p(\hat{k}_{AF})_0(\hat{k}_{AF})_r\right)\\{}&{}-8\left(\omega^2-p^2\right)(\hat{k}_{AF})_+(\hat{k}_{AF})_-\,,
\end{align*}
which gives
\begin{equation}
\boxed{0={}\left(p_\mu p^\mu\right)^2-4\left(p(\hat{k}_{AF})_0-\omega(\hat{k}_{AF})_r\right)^2-8p_\mu p^\mu(\hat{k}_{AF})_+(\hat{k}_{AF})_-}\label{General CPT-odd Spherical Dispersion Relation}
\end{equation}
that is the General CPT-odd Dispersion Relation in the helicity basis.

Here, $(\hat{k}_{AF})_i$ can be expanded over spin-weighted spherical harmonics. The prescription for this expansion is 
\begin{subequations}
\begin{align}
(\hat{k}_{AF})_0 {}&{}=\sum\limits_{dnjm}\omega^{d-3-n}p^n\prescript{}{0}{Y}_{jm}(\mathbf{\hat{p}})(k_{AF}^{(d)})^{(0B)}_{njm}\,,\\
(\hat{k}_{AF})_r {}&{}=\sum\limits_{dnjm}\omega^{d-3-n}p^n\prescript{}{0}{Y}_{jm}(\mathbf{\hat{p}})\frac{-1}{n+2}\left((k_{AF}^{(d)})^{(1B)}_{njm}+(d-2-n)(k_{AF}^{(d)})^{(0B)}_{(n-1)jm}\right)\,,\\
(\hat{k}_{AF})_\pm {}&{}=\sum\limits_{dnjm}\omega^{d-3-n}p^n\prescript{}{\pm 1}{Y}_{jm}(\mathbf{\hat{p}})\frac{1}{\sqrt{2j(j+1)}}\left(\pm(k_{AF}^{(d)})^{(1B)}_{njm}+i(\vok_{AF}^{(d)})^{(1E)}_{njm}\right)\,,
\end{align}\label{Equ: k_i}
\end{subequations}
whose details can be checked in Ref. \cite{0905.0031}. Then, \equref{General CPT-odd Spherical Dispersion Relation} becomes
\begin{equation*}
\begin{aligned}
0 ={}&{}\left(p_\mu p^\mu\right)^2-4\Bigg(\sum\limits_{dnjm}\omega^{d-3-n}p^n\prescript{}{0}{Y}_{jm}(\mathbf{\hat{p}})\bigg(p(k_{AF}^{(d)})^{(0B)}_{njm}+\frac{\omega}{n+2}\Big((k_{AF}^{(d)})^{(1B)}_{njm}\\{}&{}+(d-2-n)(k_{AF}^{(d)})^{(0B)}_{(n-1)jm}\Big)\bigg)\Bigg)^2-8p_\mu p^\mu\sum\limits_{d_1n_1j_1m_1}\omega^{d_1-3-n_1}p^{n_1}\prescript{}{+1}{Y}_{j_1m_1}(\mathbf{\hat{p}})\\{}&{}\times\frac{1}{\sqrt{2j_1(j_1+1)}}\left((k_{AF}^{(d_1)})^{(1B)}_{n_1j_1m_1}+i(\vok_{AF}^{(d_1)})^{(1E)}_{n_1j_1m_1}\right)
\sum\limits_{d_2n_2j_2m_2}\omega^{d_2-3-n_2}p^{n_2}\\{}&{}\times\prescript{}{-1}{Y}_{j_2m_2}(\mathbf{\hat{p}})\frac{1}{\sqrt{2j_2(j_2+1)}}\left(-(k_{AF}^{(d_2)})^{(1B)}_{n_2j_2m_2}+i(\vok_{AF}^{(d_2)})^{(1E)}_{n_2j_2m_2}\right)\,.
\end{aligned}
\end{equation*}

Hence, finally,
\begin{equation}
\begin{aligned}
0 ={}&{}\left(p_\mu p^\mu\right)^2-4\left(\sum\limits_{dnjm}\omega^{d-3-n}p^n\prescript{}{0}{Y}_{jm}(\mathbf{\hat{p}})\left(\frac{dp}{n+3}(k_{AF}^{(d)})^{(0B)}_{njm}+\frac{\omega}{n+2}(k_{AF}^{(d)})^{(1B)}_{njm}\right)\right)^2\\{}&{}-8p_\mu p^\mu\sum\limits_{d_1d_2n_1n_2j_1j_2m_1m_2}\omega^{d_1+d_2-6-n_1-n_2}p^{n_1+n_2}\prescript{}{+1}{Y}_{j_1m_1}(\mathbf{\hat{p}})\prescript{}{-1}{Y}_{j_2m_2}(\mathbf{\hat{p}})\\
{}&{}\times\frac{1}{\sqrt{4j_1j_2(j_1+1)(j_2+1)}}\left((k_{AF}^{(d_1)})^{(1B)}_{n_1j_1m_1}+i(\vok_{AF}^{(d_1)})^{(1E)}_{n_1j_1m_1}\right)\\
{}&{}\times\left(-(k_{AF}^{(d_2)})^{(1B)}_{n_2j_2m_2}+i(\vok_{AF}^{(d_2)})^{(1E)}_{n_2j_2m_2}\right)\,. 
\end{aligned}\label{General CPT-odd Dispersion Relation2}
\end{equation}

This is the most general dispersion relation for CPT-odd photon sector of {\it nm}SME. As it stands, it is quite complicated; however, we will show in the next section that the last term will drop so as to have a corresponding physical polarization vector.

\section{Polarization Vectors}
\label{Section: cpt odd polarization vectors}

In order to determine the photon field $A_\mu$, one needs to solve the EOM, \mbox{$M^{\mu\nu}A_\nu=0$}. The necessary condition for non-trivial solution is $det(M)=0$, through which one finds the dispersion relations. The standard method is to apply these conditions on $M$ and find the corresponding polarization vectors. As extracting the generic explicit forms of the dispersion relation out of the implicit formula given by \equref{General CPT-odd Dispersion Relation2} is quite formidable, we will pursue an alternative way here. We will calculate the rank of $M$ using a generic frequency $\omega$, and obtain the constraints from the requirement $M$ having at most rank 2.\footnotemark\footnotetext{A rank-3 $M$ gives the gauge solution only, and a rank-4 $M$ gives the trivial solution, that is $A^{\mu}=0$. That one solution should be the gauge solution can be understood as there is still the gauge freedom on $M$\cite{0905.0031}.} Then, these constraints will be applied to the dispersion relations, which we already worked out, in order to determine whether there exists a nontrivial coefficient subspace with a physical polarization vector obeying the general dispersion relation \equref{General CPT-odd Dispersion Relation2}. 

In the beginning of the chapter, the general form of the tensor $M$ is derived in the covariant form, which is readily given by \equref{general form of M}. This covariant form can be expanded with the explicit choice of a basis, which is chosen to be the helicity basis conveniently. Then,
\begin{align*}
M^{\mu\nu}={}&{}\eta^{\mu\nu}p_\alpha p^\alpha-p^\mu p^\nu-2i\left(\epsilon^{\mu\nu\alpha\beta}\right)(\hat{k}_{AF})_\alpha p_\beta\\
={}&{}\eta^{\mu\nu}p_\alpha p^\alpha-p^\mu p^\nu-2i\left(\delta^\mu_0\delta^\nu_i\delta^\alpha_j\delta^\beta_k-\delta^\mu_i\delta^\nu_0\delta^\alpha_j\delta^\beta_k+\delta^\mu_i\delta^\nu_j\delta^\alpha_0\delta^\beta_k-\delta^\mu_i\delta^\nu_j\delta^\alpha_k\delta^\beta_0\right)\\{}&{}\times\epsilon^{ijk}(\hat{k}_{AF})_\alpha p_\beta \\
={}&{}\eta^{\mu\nu}p_\alpha p^\alpha-p^\mu p^\nu-2i\epsilon^{ijk}\Big(\delta^\mu_0\delta^\nu_i(\hat{k}_{AF})_j p_k-\delta^\mu_i\delta^\nu_0(\hat{k}_{AF})_j p_k\\{}&{}+\delta^\mu_i\delta^\nu_j(\hat{k}_{AF})_0 p_k-\delta^\mu_i\delta^\nu_j(\hat{k}_{AF})_k p_0\Big) \\
={}&{}\eta^{\mu\nu}p_\alpha p^\alpha-p^\mu p^\nu-2i\epsilon^{ijr}p\left(\delta^\mu_0\delta^\nu_i(\hat{k}_{AF})_j -\delta^\mu_i\delta^\nu_0(\hat{k}_{AF})_j+\delta^\mu_i\delta^\nu_j(\hat{k}_{AF})_0\right)\\{}&{}+2i\epsilon^{ijk}\delta^\mu_i\delta^\nu_j(\hat{k}_{AF})_k \omega \\
={}&{}\eta^{\mu\nu}p_\alpha p^\alpha-p^\mu p^\nu+2\left(\delta^i_+\delta^j_--\delta^i_-\delta^j_+\right)p\Big(\delta^\mu_0\delta^\nu_i(\hat{k}_{AF})_j -\delta^\mu_i\delta^\nu_0(\hat{k}_{AF})_j\\
{}&{}+\delta^\mu_i\delta^\nu_j(\hat{k}_{AF})_0\Big)+2\Big(\delta^i_+\delta^j_r\delta^k_--\delta^i_+\delta^j_-\delta^k_r+\delta^i_-\delta^j_+\delta^k_r-\delta^i_-\delta^j_r\delta^k_++\delta^i_r\delta^j_-\delta^k_+\\
{}&{}-\delta^i_r\delta^j_+\delta^k_-\Big)\delta^\mu_i\delta^\nu_j(\hat{k}_{AF})_k \omega \\
={}&{} \eta^{\mu\nu}p_\alpha p^\alpha-p^\mu p^\nu+2\delta^\mu_0\delta^\nu_+(\hat{k}_{AF})_-p -2\delta^\mu_+\delta^\nu_0(\hat{k}_{AF})_-p+2\delta^\mu_+\delta^\nu_-(\hat{k}_{AF})_0p\\{}&{}-2\delta^\mu_0\delta^\nu_-(\hat{k}_{AF})_+p +2\delta^\mu_-\delta^\nu_0(\hat{k}_{AF})_+p-2\delta^\mu_-\delta^\nu_+(\hat{k}_{AF})_0p
+2\delta^\mu_+\delta^\nu_r(\hat{k}_{AF})_- \omega\\{}&{}-2\delta^\mu_+\delta^\nu_-(\hat{k}_{AF})_r \omega+2\delta^\mu_-\delta^\nu_+(\hat{k}_{AF})_r \omega-2\delta^\mu_-\delta^\nu_r(\hat{k}_{AF})_+ \omega+2\delta^\mu_r\delta^\nu_-(\hat{k}_{AF})_+ \omega\\{}&{}-2\delta^\mu_r\delta^\nu_+(\hat{k}_{AF})_- \omega\\
={}&{}\eta^{\mu\nu}p_\alpha p^\alpha-p^\mu p^\nu+2\left(\delta^\mu_0\delta^\nu_+-\delta^\mu_+\delta^\nu_0\right)(\hat{k}_{AF})_-p+2\left(\delta^\mu_-\delta^\nu_0-\delta^\mu_0\delta^\nu_-\right)(\hat{k}_{AF})_+p\\
{}&{}+2\left(\delta^\mu_+\delta^\nu_--\delta^\mu_-\delta^\nu_+\right)\left((\hat{k}_{AF})_0p-(\hat{k}_{AF})_r \omega\right)+2\left(\delta^\mu_+\delta^\nu_r-\delta^\mu_r\delta^\nu_+\right)(\hat{k}_{AF})_-\omega\\{}&{}+2\left(\delta^\mu_r\delta^\nu_--\delta^\mu_-\delta^\nu_r\right)(\hat{k}_{AF})_+\omega
\end{align*}

As $p^\mu$ has only temporal and radial components in helicity basis
\begin{equation*}
p^\mu p^\nu=\delta^\mu_0\delta^\nu_0\omega^2+\left(\delta^\mu_0\delta^\nu_r+\delta^\mu_r\delta^\nu_0\right)p\omega+\delta^\mu_r\delta^\nu_rp^2\,.
\end{equation*}

Once this is inserted and the first index is lowered, $M$ becomes
\begin{equation*}
\begin{aligned}
M_\mu^{\;\nu}={}&{}\delta_\mu^{\;\nu}(\omega^2-p^2)-\eta_{	\mu 0}\delta^\nu_0\omega^2-\left(\eta_{\mu 0}\delta^\nu_r+\eta_{\mu r}\delta^\nu_0\right)p\omega-\eta_{\mu r}\delta^\nu_rp^2\\{}&{}+2\left(\eta_{\mu 0}\delta^\nu_+-\eta_{\mu +}\delta^\nu_0\right)(\hat{k}_{AF})_-p+2\left(\eta_{\mu -}\delta^\nu_0-\eta_{\mu 0}\delta^\nu_-\right)(\hat{k}_{AF})_+p\\{}&{}+2\left(\eta_{\mu +}\delta^\nu_--\eta_{\mu -}\delta^\nu_+\right)\left((\hat{k}_{AF})_0p-(\hat{k}_{AF})_r \omega\right)+2\left(\eta_{\mu +}\delta^\nu_r-\eta_{\mu r}\delta^\nu_+\right)(\hat{k}_{AF})_-\omega\\{}&{}+2\left(\eta_{\mu r}\delta^\nu_--\eta_{\mu -}\delta^\nu_r\right)(\hat{k}_{AF})_+\omega
\,.\end{aligned}
\end{equation*}

Although index form is useful on its own, it is somehow more insightful to go to the matrix representation. Hence, we employ matrix representation convention $M_\rho^{\;\;\nu}$ in $(0,+,r,-)$ basis:
\begin{equation}
M\doteq
\begin{pmatrix}
-p^2 {}&{} 2(\hat{k}_{AF})_-p {}&{} -p\omega {}&{} -2(\hat{k}_{AF})_+p\\
-2(\hat{k}_{AF})_+p {}&{} \omega^2-p^2+2p(\hat{k}_{AF})_s {}&{} 2(\hat{k}_{AF})_+\omega {}&{} 0\\
p\omega {}&{} 2(\hat{k}_{AF})_-\omega {}&{} \omega^2 {}&{} -2(\hat{k}_{AF})_+\omega \\
2(\hat{k}_{AF})_-p {}&{} 0 {}&{} -2(\hat{k}_{AF})_-\omega {}&{} \omega^2-p^2-2p(\hat{k}_{AF})_s
\end{pmatrix}\label{conventional rooted M in helicity basis}\,.
\end{equation}
where
\begin{equation}
(\hat{k}_{AF})_s:={}(\hat{k}_{AF})_0-\frac{\omega}{p}(\hat{k}_{AF})_r \label{definition of k_s}
\end{equation}
is defined for brevity.
\subsection*{No LV Limit}
One can test the method explained above in the conventional case. The standard procedure for the derivation of dispersion relation for the conventional photon is to set an explicit gauge choice, and to extract the required condition for nontrivial photon field $A^\mu$, which was shown for the Lorenz gauge $\partial_\mu A^\mu=0$ in the beginning of \secref{Section: cpt odd dispersion}.

In our alternative method, we check the rank of the matrix $M$ for the dispersion relation without an explicit gauge choice. It is obvious that the matrix $M$ of the conventional photon will be \equref{conventional rooted M in helicity basis} under the condition $\hat{k}_{AF}=0$, as the transition from the SME to SM is a smooth one; hence,
\begin{equation*}
M\doteq
\begin{pmatrix}
-p^2 {}&{} 0 {}&{} -p\omega {}&{} 0\\
0 {}&{} \omega^2-p^2 {}&{} 0 {}&{} 0\\
p\omega {}&{} 0 {}&{} \omega^2 {}&{} 0 \\
0 {}&{} 0 {}&{} 0 {}&{} \omega^2-p^2
\end{pmatrix}\,. \label{No LV M without omega=p}
\end{equation*}

The condition for a physical solution to emerge is that $M$ should have rank not greater than two. However, this matrix is of rank 3 unless $\omega=p$; hence, the required dispersion relation for the conventional photon obtained by rank-nullity is $\omega=p$.

Under the dispersion relation condition, $M$ takes the form
\begin{equation*}
M\doteq
\begin{pmatrix}
-p^2 {}&{} 0 {}&{} -p^2 {}&{} 0\\
0 {}&{} 0 {}&{} 0 {}&{} 0\\
p^2 {}&{} 0 {}&{} p^2 {}&{} 0 \\
0 {}&{} 0 {}&{} 0 {}&{} 0
\end{pmatrix}\,, \label{No LV M}
\end{equation*}
for which $MA=0$ yields the following solutions
\begin{equation*}
A_\mu\in\left\{\begin{pmatrix}1\\0\\-1\\0
\end{pmatrix}, \begin{pmatrix}0\\1\\0\\0
\end{pmatrix}, \begin{pmatrix}0\\0\\0\\1
\end{pmatrix}\right\}\,.
\end{equation*}

This is the desired result for the conventional photon. As can be seen, the dispersion relation associates the scalar and longitudinal polarizations; in fact, the dispersion relation and the first solution can be written as
\begin{equation*}
\omega A_0+pA_r=0 \longrightarrow p^\mu A_\mu=0 \longrightarrow p_\mu A^\mu=0\,.
\end{equation*}
But this is also true for other two equations as well, since $p_\pm=0$. But $p_\mu A^\mu=0$ under the prescription $p_\mu\rightarrow i\partial_\mu$ is simply the Lorenz gauge $\partial_\mu A^\mu=0$. That means, even though we did not specify any gauge condition, the rank requirement of $M$ with the resultant dispersion relation yielded the corresponding gauge. Clearly, when one further applies the Coulomb gauge, $A^0=0$, first solution will die out, and the theory will remain with two transverse polarization vectors with a common dispersion relation $\omega=p$.

\subsection*{General Case}
The rank nullity approach described and demonstrated in the conventional case above can be directly applied for the general case given by \equref{conventional rooted M in helicity basis}. Firstly, we write the equations of motion
\begin{equation*}\small
\begin{pmatrix}
-p^2 {}&{} 2(\hat{k}_{AF})_-p {}&{} -p\omega {}&{} -2(\hat{k}_{AF})_+p\\
-2(\hat{k}_{AF})_+p {}&{} p_\mu p^\mu+2p(\hat{k}_{AF})_s {}&{} 2(\hat{k}_{AF})_+\omega {}&{} 0\\
p\omega {}&{} 2(\hat{k}_{AF})_-\omega {}&{} \omega^2 {}&{} -2(\hat{k}_{AF})_+\omega \\
2(\hat{k}_{AF})_-p {}&{} 0 {}&{} -2(\hat{k}_{AF})_-\omega {}&{} p_\mu p^\mu-2p(\hat{k}_{AF})_s
\end{pmatrix}\begin{pmatrix}
A_0 \\ -A_+ \\ -A_r \\ -A_-
\end{pmatrix}=\begin{pmatrix}
0\\0\\0\\0
\end{pmatrix}\,,
\end{equation*}
where the minus signs in the space components of the field vector is a result of the convention used, where $MA=0$ reads $M_\mu^{\;\nu}A_\nu=0$ hence $A$ should be in its covariant form.

From now on, one can apply the rank-nullity approach by calculating the determinant of $M$ and finding the conditions for its rank to be less than three. However, the mathematical burden of the approach can be greatly reduced by invoking the fact that the rank of a matrix is invariant under row operations. By playing with the first and third row of \equref{conventional rooted M in helicity basis}, one can obtain the equivalent $M$ as follows
\begin{equation}
M_{\text{eq}}=\begin{pmatrix}
-p {}&{} 0 {}&{} -\omega {}&{} 0\\
-2(\hat{k}_{AF})_+p {}&{} \omega^2-p^2+2p(\hat{k}_{AF})_s {}&{} 2(\hat{k}_{AF})_+\omega {}&{} 0\\
0 {}&{} (\hat{k}_{AF})_- {}&{} 0 {}&{} -(\hat{k}_{AF})_+ \\
2(\hat{k}_{AF})_-p {}&{} 0 {}&{} -2(\hat{k}_{AF})_-\omega {}&{} \omega^2-p^2-2p(\hat{k}_{AF})_s
\end{pmatrix}\,.\label{simplified Equations of Motion for photon field}
\end{equation}

Further simplifications with the row operations are possible; however, they vary regarding whether $(\hat{k}_{AF})_\pm=0$. Therefore, it is preferable to examine each case separately.
\subsubsection*{The case for which $(\hat{k}_{AF})_+\ne0$ \& $(\hat{k}_{AF})_-\ne0$ :}
For this case, $(\hat{k}_{AF})_\pm$ multiples of the first row of \equref{simplified Equations of Motion for photon field} can be added to the second and forth rows, resulting in the following form
\begin{equation}
M_{\text{eq}}=\begin{pmatrix}
-p {}&{} 0 {}&{} -\omega {}&{} 0\\
0 {}&{} \omega^2-p^2+2p(\hat{k}_{AF})_s {}&{} 4(\hat{k}_{AF})_+\omega {}&{} 0\\
0 {}&{} (\hat{k}_{AF})_- {}&{} 0 {}&{} -(\hat{k}_{AF})_+ \\
0 {}&{} 0 {}&{} -4(\hat{k}_{AF})_-\omega {}&{} \omega^2-p^2-2p(\hat{k}_{AF})_s
\end{pmatrix}\,.
\end{equation}
It is clear that this matrix is of rank greater than two; hence, there cannot be any physical solutions for this case.
\subsubsection*{The case for which $(\hat{k}_{AF})_+\ne0$ \& $(\hat{k}_{AF})_-=0$ :}
For this case, $-2(\hat{k}_{AF})_+$ multiple of the first row of \equref{simplified Equations of Motion for photon field} can be added to the second row, and $(\hat{k}_{AF})_-=0$ is imposed, which results in the upper triangular form
\begin{equation}
M_{\text{eq}}=\begin{pmatrix}
-p {}&{} 0 {}&{} -\omega {}&{} 0\\
0 {}&{} \omega^2-p^2+2p(\hat{k}_{AF})_s {}&{} 4(\hat{k}_{AF})_+\omega {}&{} 0\\
0 {}&{} 0 {}&{} 0 {}&{} -(\hat{k}_{AF})_+ \\
0 {}&{} 0 {}&{} 0 {}&{} \omega^2-p^2-2p(\hat{k}_{AF})_s
\end{pmatrix}\,.
\end{equation}
Not unlike the earlier one, this matrix is of rank greater than two as well; hence, there cannot be any physical solutions for this case either.
\subsubsection*{The case for which $(\hat{k}_{AF})_+=0$ \& $(\hat{k}_{AF})_-\ne0$ :}
For this case, $2(\hat{k}_{AF})_-$ multiple of the first row of \equref{simplified Equations of Motion for photon field} can be added to the fourth row, and $(\hat{k}_{AF})_+=0$ is imposed, which results in the form
\begin{equation}
M_{\text{eq}}=\begin{pmatrix}
-p & 0 & -\omega & 0\\
0 & \omega^2-p^2+2p(\hat{k}_{AF})_s & 0 {}&{} 0\\
0 {}&{} (\hat{k}_{AF})_- {}&{} 0 {}&{} 0 \\
0 {}&{} 0 {}&{} -4(\hat{k}_{AF})_-\omega  {}&{} \omega^2-p^2-2p(\hat{k}_{AF})_s
\end{pmatrix}\,.
\end{equation}
Again, this matrix is of rank greater than two; hence, there cannot be any physical solutions for this case as well.
\subsubsection*{The case for which $(\hat{k}_{AF})_+=0$ \& $(\hat{k}_{AF})_-=0$ :}
For this case, we simply impose $2(\hat{k}_{AF})_\pm=0$ into \equref{simplified Equations of Motion for photon field}, hence
\begin{equation}
M_{\text{eq}}=\begin{pmatrix}
-p {}&{} 0 {}&{} -\omega {}&{} 0\\
0 {}&{} \omega^2-p^2+2p(\hat{k}_{AF})_s {}&{} 0 {}&{} 0\\
0 {}&{} 0 {}&{} 0 {}&{} 0 \\
0 {}&{} 0 {}&{} 0 {}&{} \omega^2-p^2-2p(\hat{k}_{AF})_s
\end{pmatrix}\,. \label{Equ: M under k_pm=0}
\end{equation}
This matrix is indeed of rank 2 if one of the equations $\omega^2-p^2\pm2p(\hat{k}_{AF})_s=0$ hold, and of rank 1 if both equations hold at the same time. 

The first case in which there are two different rank 2 matrices with two different conditions physically means that there are two physical solutions with two different dispersion relations. This is clear since each rank two matrix is associated with one gauge and one physical solution, and since for each physical solution the required condition, that is the dispersion relation, is different.

Let us find the associated physical solutions for each dispersion relation. Under the condition $(\hat{k}_{AF})_\pm=0$ with the dispersion relation $\omega^2-p^2-2p(\hat{k}_{AF})_s=0$, \equref{Equ: M under k_pm=0} gives the EOM

\begin{equation*}
M_{\text{eq}}=\begin{pmatrix}
-p {}&{} 0 {}&{} -\omega {}&{} 0\\
0 {}&{} \omega^2-p^2+2p(\hat{k}_{AF})_s {}&{} 0 {}&{} 0\\
0 {}&{} 0 {}&{} 0 {}&{} 0 \\
0 {}&{} 0 {}&{} 0 {}&{} 0
\end{pmatrix}\begin{pmatrix}
A_0 \\ -A_+ \\ -A_r \\ -A_-
\end{pmatrix}=\begin{pmatrix}
0\\0\\0\\0
\end{pmatrix}\,.
\end{equation*}
This equation gives the solutions 
\begin{equation*}
\left\{\begin{pmatrix}
\omega\\0\\p\\0
\end{pmatrix} , \begin{pmatrix}
0\\0\\0\\1
\end{pmatrix}\right\}\,,
\end{equation*}
where the first one is the gauge solution and the second one is the left-circularly polarized photon.

Similarly, under the condition $(\hat{k}_{AF})_\pm=0$,  \equref{Equ: M under k_pm=0} with the dispersion relation\\ \mbox{$\omega^2-p^2+2p(\hat{k}_{AF})_s=0$} results in
\begin{equation*}
M_{\text{eq}}=\begin{pmatrix}
-p {}&{} 0 {}&{} -\omega {}&{} 0\\
0 {}&{} 0 {}&{} 0 {}&{} 0\\
0 {}&{} 0 {}&{} 0 {}&{} 0 \\
0 {}&{} 0 {}&{} 0 {}&{} \omega^2-p^2-2p(\hat{k}_{AF})_s
\end{pmatrix}\begin{pmatrix}
A_0 \\ -A_+ \\ -A_r \\ -A_-
\end{pmatrix}=\begin{pmatrix}
0\\0\\0\\0
\end{pmatrix}\,.
\end{equation*}
This equation gives the solutions 
\begin{equation*}
\left\{\begin{pmatrix}
\omega\\0\\p\\0
\end{pmatrix} , \begin{pmatrix}
0\\1\\0\\0
\end{pmatrix}\right\}\,,
\end{equation*}
where the first one is the gauge solution and the second one is the right-circularly polarized photon.

Therefore, the model have two physical polarization vectors, which are both transverse solutions, however they obey different dispersion relations. That means, the photon field is birefringent under Lorenz violation. 

The two dispersion relations coincide, at which the rank of $M$ reduces to one as stated above, if these constraints are equal to each other; that is,
\begin{equation*}
\begin{aligned}
\omega^2-p^2+2p(\hat{k}_{AF})_s=0\,,\\
\omega^2-p^2-2p(\hat{k}_{AF})_s=0\,.
\end{aligned}
\end{equation*}
However, this is possible only if
\begin{equation*}
\begin{aligned}
\omega{}&{}=p\,,\\
(\hat{k}_{AF})_r{}&{}=(\hat{k}_{AF})_0\,.
\end{aligned}
\end{equation*}
which means that the physical solutions arising with the LV under the restrictions $(\hat{k}_{AF})_\pm=0$ and $(\hat{k}_{AF})_r=(\hat{k}_{AF})_0$ are conventional transverse solutions obeying the same conventional dispersion relation $\omega=p$.

Concisely, once the $\hat{k}_{AF}$ is chosen, whether there arise any physical solutions and whether these solutions are birefringent or not can immediately be determined by examining the components of $\hat{k}_{AF}$ in the helicity basis. The whole possible coefficient space for CPT-odd photon sector of {\it nm}SME then can be decomposed into three subspaces: one with conventional solutions, one with birefringent solutions and one with no physical solutions. We will denote these coefficient spaces as $\hat{k}_{AF}^{(cn)}$, $\hat{k}_{AF}^{(bf)}$, and $\hat{k}_{AF}^{(np)}$ where $cn$, $bf$,  and $np$ refer to nature of resultant polarization vectors: conventional, birefringent, and nonphysical, respectively.

That physical solutions of photon field do exist vetoes the possibility $\hat{k}_{AF}\in\hat{k}_{AF}^{(np)}$; hence, it is merely considered for completeness. The results are summarized in \tabref{Table-cscops}.

\section{The Propagator}
\label{Section: cpt odd propagator}
In the beginning of the chapter, the form of the inverse propagator is calculated in the covariant form as \equref{Momentum space inverse propagator}. Therefore, in index notation, the propagator is given by the equation
\begin{equation}
\left(-\eta^{\mu\nu}(p_\sigma p^\sigma)+2i\epsilon^{\mu\kappa\lambda\nu}(\hat{k}_{AF})_\kappa p_\lambda\right)\hat{G}_{\nu\rho}=\delta^{\mu}_{\;\rho}\,.\label{eq: definifion of the propagator}
\end{equation}

\begin{table}
\caption[Coefficient Space of CPT-odd Photon Sector]{Coefficient Space of CPT-odd Photon Sector. In \secref{Section: cpt odd polarization vectors}, it is shown that any vector $(\hat{k}_{AF})_\mu$ can be categorized according to the relations between its components in the helicity basis. These relations are sufficient to determine whether the model with that LVT will have physical solutions, whether these solutions will be birefringent, and what dispersion relations they will obey, as listed below.}
\label{Table-cscops}
\begin{tabularx}{\textwidth}{l@{\hskip -0.2cm}ccc}
\hline\hline \begin{tabular}{c}
{\hskip -0.15in}Coefficient\\{\hskip -0.15in}Subspace 
\end{tabular}& Conditions & Dispersion Relation & Polarization Vectors \\\hline\\[-0.15in]
$\hat{k}_{AF}^{(bf)}$
{}&{}$\begin{aligned}
(\hat{k}_{AF})_+={}&{} 0\\(\hat{k}_{AF})_-={}&{} 0\\(\hat{k}_{AF})_0\ne{}&{}(\hat{k}_{AF})_r
\end{aligned}${}&{}{\hskip -0.1in}$\begin{array}{l}
p_\mu p^\mu-2p(\hat{k}_{AF})_s={}0\\p_\mu p^\mu+2p(\hat{k}_{AF})_s={}0
\end{array}${}&{} $\footnotesize \begin{array}{c}
\left\{\begin{pmatrix}
\omega\\0\\p\\0
\end{pmatrix} , \begin{pmatrix}
0\\0\\0\\1
\end{pmatrix}\right\}\\\left\{\begin{pmatrix}
\omega\\0\\p\\0
\end{pmatrix} , \begin{pmatrix}
0\\1\\0\\0
\end{pmatrix}\right\}
\end{array}$\\\\[-0.15in]
$\hat{k}_{AF}^{(cn)}$
{}&{}$\begin{aligned}
(\hat{k}_{AF})_+={}&{} 0\\(\hat{k}_{AF})_-={}&{} 0\\(\hat{k}_{AF})_0={}&{}(\hat{k}_{AF})_r
\end{aligned}${}&{}$\omega=p${}&{}$\footnotesize \left\{\begin{pmatrix}
1\\0\\1\\0
\end{pmatrix}, \begin{pmatrix}
0\\1\\0\\0
\end{pmatrix}, \begin{pmatrix}
0\\0\\0\\1
\end{pmatrix}\right\}$\\\\[-0.15in]
$\hat{k}_{AF}^{(np)}$
{}&{} $\begin{array}{l}
\{(\hat{k}_{AF})_+\ne0\}\,\lor \\\{(\hat{k}_{AF})_-\ne0\}
\end{array}$ & \equref{General CPT-odd Dispersion Relation2} & \text{Gauge Solution Only}\\\hline\hline
\end{tabularx}
\end{table}

In the conventional field theory, corresponding equations for propagators are somehow manageable, and one simply uses some index tricks to obtain the analytic and covariant form of the propagator. In this case however, the equation is quite formidable and may even not have a both analytic and covariant solution. Therefore, we should either give up the analytic form by using some approximations, or sacrifice the covariance by choosing an explicit basis. We will try the first approach in following two sections, and the second approach in the last section.

\subsection{Propagator Ansatz}
\label{Section: cpt odd ansatz}

In theory, one can construct an ansatz in a judiciously judged form for the propagator by using the available tensors in the momentum space. In our case, these are $p^\mu$, $(\hat{k}_{AF})_\mu$ and the metric $\eta_{\rho\mu}$. Then, following ansatz can be proposed for the propagator
\begin{equation}
\hat{G}_{\rho\mu}={}\tilde{a}\,\eta_{\rho\mu}+\tilde{b}\,p_\rho p_\mu+\tilde{c}\,(\hat{k}_{AF})_{\{\rho} p_{\mu\}}+\tilde{d}\,(\hat{k}_{AF})_{[\rho} p_{\mu]}+\tilde{e}\,(\hat{k}_{AF})_\rho(\hat{k}_{AF})_\mu\,. \label{Propagator ansatz}
\end{equation}
where $(\hat{k}_{AF})_{\{\rho} p_{\mu\}}$ and $(\hat{k}_{AF})_{[\rho} p_{\mu]}$ are symmetric and antisymmetric combinations of $(\hat{k}_{AF})_\rho$ and $p_\mu$ respectively. In our notation, we define these contributions such that
\begin{equation*}
(\hat{k}_{AF})_{\{\rho} p_{\mu\}}+(\hat{k}_{AF})_{[\rho} p_{\mu]} = (\hat{k}_{AF})_{\rho} p_{\mu}\,,
\end{equation*}
which is ensured if
\begin{equation*}
\begin{aligned}
(\hat{k}_{AF})_{\{\rho} p_{\mu\}} {}&{} =\frac{1}{2}\left((\hat{k}_{AF})_{\rho} p_{\mu}+(\hat{k}_{AF})_{\mu} p_{\rho}\right)\,,\\
(\hat{k}_{AF})_{[\rho} p_{\mu]} {}&{} =\frac{1}{2}\left((\hat{k}_{AF})_{\rho} p_{\mu}-(\hat{k}_{AF})_{\mu} p_{\rho}\right)\,.
\end{aligned}
\end{equation*}
From \equref{eq: definifion of the propagator}:
\begin{align*}
\delta_\rho^\nu={}&{}\left(\tilde{a}\eta_{\rho\mu}+\tilde{b}p_\rho p_\mu+\tilde{c}(\hat{k}_{AF})_{\{\rho} p_{\mu\}}+\tilde{d}(\hat{k}_{AF})_{[\rho} p_{\mu]}+\tilde{e}(\hat{k}_{AF})_\rho(\hat{k}_{AF})_\mu\right)\\{}&{}\times\left(-\eta^{\mu\nu}(p_\sigma p^\sigma)+2i\epsilon^{\mu\kappa\lambda\nu}(\hat{k}_{AF})_\kappa p_\lambda\right)\,,\\
\delta_\rho^\nu={}&{}-\tilde{a}(p_\sigma p^\sigma)\delta_\rho^\nu-\tilde{b}(p_\sigma p^\sigma)p_\rho p^\nu-\tilde{c}(p_\sigma p^\sigma)(\hat{k}_{AF})_{\{\rho} p_{\mu\}}\eta^{\mu\nu}-\tilde{d}(p_\sigma p^\sigma)(\hat{k}_{AF})_{[\rho} p_{\mu]}\eta^{\mu\nu}\\{}&{}-\tilde{e}(p_\sigma p^\sigma)(\hat{k}_{AF})_\rho (\hat{k}_{AF})^\nu+2i\tilde{a}\eta_{\rho\mu}\epsilon^{\mu\kappa\lambda\nu}(\hat{k}_{AF})_\kappa p_\lambda+2i\tilde{b}\epsilon^{\mu\kappa\lambda\nu}(\hat{k}_{AF})_\kappa p_\lambda p_\rho p_\mu\\
{}&{}+2i\tilde{c}\epsilon^{\mu\kappa\lambda\nu}(\hat{k}_{AF})_\kappa (\hat{k}_{AF})_{\{\rho} p_{\mu\}} p_\lambda+2i\tilde{d}\epsilon^{\mu\kappa\lambda\nu}(\hat{k}_{AF})_\kappa (\hat{k}_{AF})_{[\rho} p_{\mu]} p_\lambda \\
{}&{}+2i\tilde{e}\epsilon^{\mu\kappa\lambda\nu}(\hat{k}_{AF})_\kappa (\hat{k}_{AF})_\rho (\hat{k}_{AF})_\mu p_\lambda\,.
\end{align*}

As $(\hat{k}_{AF})_\mu$ are just functions of $\omega$, $p$ and LV coefficients, they commute with one another, meaning that $(\hat{k}_{AF})_{[\mu}(\hat{k}_{AF})_{\nu]}=0$. Similarly, $p_{[\mu}p_{\nu]}=0$, hence
\begin{equation*}
\begin{aligned}
\delta_\rho^\nu={}&{}-\tilde{a}(p_\sigma p^\sigma)\delta_\rho^\nu-\tilde{b}(p_\sigma p^\sigma)p_\rho p^\nu-\tilde{c}(p_\sigma p^\sigma)(\hat{k}_{AF})_{\{\rho} p_{\mu\}}\eta^{\mu\nu}-\tilde{d}(p_\sigma p^\sigma)(\hat{k}_{AF})_{[\rho} p_{\mu]}\eta^{\mu\nu}\\
{}&{}-\tilde{e}(p_\sigma p^\sigma)(\hat{k}_{AF})_\rho (\hat{k}_{AF})^\nu+2i\tilde{a}\eta_{\rho\mu}\epsilon^{\mu\kappa\lambda\nu}(\hat{k}_{AF})_\kappa p_\lambda\,.
\end{aligned}
\end{equation*}

If the attention is restricted to the leading order LV in the propagator, the $5^\text{th}$ term can be discarded; in other words,
\begin{equation*}
\begin{aligned}
\delta_\rho^\nu={}&{}-\tilde{a}(p_\sigma p^\sigma)\delta_\rho^\nu-\tilde{b}(p_\sigma p^\sigma)p_\rho p^\nu-\tilde{c}(p_\sigma p^\sigma)(\hat{k}_{AF})_{\{\rho} p_{\mu\}}\eta^{\mu\nu}-\tilde{d}(p_\sigma p^\sigma)(\hat{k}_{AF})_{[\rho} p_{\mu]}\eta^{\mu\nu}\\{}&{}+2i\tilde{a}\eta_{\rho\mu}\epsilon^{\mu\kappa\lambda\nu}(\hat{k}_{AF})_\kappa p_\lambda\,.
\end{aligned}
\end{equation*}
Let us raise $\rho$:
\begin{equation}
\begin{aligned}
\eta^{\rho\nu}={}&{}-\tilde{a}(p_\sigma p^\sigma)\eta^{\rho\nu}-\tilde{b}(p_\sigma p^\sigma)p^\rho p^\nu-\tilde{c}(p_\sigma p^\sigma)(\hat{k}_{AF})^{\{\rho} p^{\nu\}}\\{}&{}-\tilde{d}(p_\sigma p^\sigma)(\hat{k}_{AF})^{[\rho} p^{\nu]}+2i\tilde{a}\delta^\rho_\mu\epsilon^{\mu\kappa\lambda\nu}(\hat{k}_{AF})_\kappa p_\lambda\,.
\end{aligned}
\label{propagator coefficient equation}
\end{equation}

Since the left hand side is symmetric over \{$\rho,\nu$\}, so must the right hand side. As the last two terms are antisymmetric, they should cancel each other, hence
\begin{equation}
\tilde{d}(p_\sigma p^\sigma)(\hat{k}_{AF})^{[\rho} p^{\nu]}=2i\tilde{a}\delta^\rho_\mu\epsilon^{\mu\kappa\lambda\nu}(\hat{k}_{AF})_\kappa p_\lambda\,. \label{Eq: propagator ansatz antisymmetric}
\end{equation}

Then, \equref{propagator coefficient equation} becomes
\begin{equation*}
\eta^{\rho\nu}=-\tilde{a}(p_\sigma p^\sigma)\eta^{\rho\nu}-\tilde{b}(p_\sigma p^\sigma)p^\rho p^\nu-\tilde{c}(p_\sigma p^\sigma)(\hat{k}_{AF})^{\{\rho} p^{\nu\}}\,.
\end{equation*}

This equality is satisfied for all $p^\mu$ and $(\hat{k}_{AF})^\mu$ only if
\begin{equation*}
\begin{aligned}
\tilde{a}={}&{}-\frac{1}{(p_\sigma p^\sigma)}\,,\\
\tilde{b}={}&{} 0\,,\\
\tilde{c}={}&{} 0\,.
\end{aligned}
\end{equation*}

Thus, above equations with \equref{Eq: propagator ansatz antisymmetric} turns \equref{Propagator ansatz} into
\begin{equation*}
G_{\rho\mu}=-\frac{1}{p_\alpha p^\alpha}\eta_{\rho\mu}-2i\frac{1}{(p_\alpha p^\alpha)^2}\eta_{\rho\sigma}\eta_{\alpha\mu}\epsilon^{\sigma\kappa\lambda\alpha}(\hat{k}_{AF})_\kappa p_\lambda\,.
\end{equation*}

With \equref{Equ:levi civita}, the general leading order covariant form of the propagator for the CPT-odd modified photon in {\it nm}SME becomes
\begin{equation}
\boxed{\hat{G}_{\rho\mu}\simeq{}-\frac{1}{p_\alpha p^\alpha}\eta_{\rho\mu}+2i\frac{1}{(p_\alpha p^\alpha)^2}\epsilon_{\rho\mu\kappa\lambda}(\hat{k}_{AF})^\kappa p^\lambda} \label{LO Propagator}
\end{equation}
We see that the propagator smoothly reduces to the conventional one for no LV case.

\subsection{Perturbation Expansion}
\label{Section: cpt odd perturbation}
Another covariant extraction method of the propagator out of \equref{eq: definifion of the propagator} would be a perturbation expansion in the powers of $(\hat{k}_{AF})_\mu$. In this method, the series formulation of $\hat{G}_{\rho\mu}$ can be written as
\begin{equation}
\hat{G}_{\rho\mu}={}\sum\limits_{n=0}^{\infty}G^{(n)}_{\rho\mu}\,, \label{propagator expansion}
\end{equation}
where $G^{(n)}_{\rho\nu}$ is the term in the propagator with $n^\text{th}$ order $(\hat{k}_{AF})_\mu$ only. Then, with \equref{Momentum space inverse propagator}, \equref{eq: definifion of the propagator} gives
\begin{align*}
\delta^\nu_\rho={}&{}\sum\limits_{n=0}^{\infty}G^{(n)}_{\rho\mu}\left(-\eta^{\mu\nu}(p_\sigma p^\sigma)+2i\epsilon^{\mu\kappa\lambda\nu}(\hat{k}_{AF})_\kappa p_\lambda\right)\,, \\
\delta^\nu_\rho={}&{}-(p_\sigma p^\sigma)\sum\limits_{n=0}^{\infty}(G^{(n)})_{\rho}^\nu+2i\sum\limits_{n=0}^{\infty}G^{(n)}_{\rho\mu}\epsilon^{\mu\kappa\lambda\nu}(\hat{k}_{AF})_\kappa p_\lambda\,,\\
\delta^\nu_\rho={}&{}-(p_\sigma p^\sigma)(G^{(0)})_{\rho}^\nu-(p_\sigma p^\sigma)\sum\limits_{n=1}^{\infty}(G^{(n)})_{\rho}^\nu+2i\sum\limits_{n=0}^{\infty}G^{(n)}_{\rho\mu}\epsilon^{\mu\kappa\lambda\nu}(\hat{k}_{AF})_\kappa p_\lambda\,.
\end{align*}

The perturbation expansion inherently assumes a smooth transition to the conventional case, which can be invoked by turning off the Lorentz violation. Then, both summations vanish; hence,
\begin{equation*}
(G^{(0)})_{\rho}^\nu={}-\frac{1}{(p_\sigma p^\sigma)}\delta^\nu_\rho\,,
\end{equation*}
which dictates
\begin{equation*}
\begin{aligned}
0={}&{}-(p_\sigma p^\sigma)\sum\limits_{n=1}^{\infty}(G^{(n)})_{\rho}^\nu+2i\sum\limits_{n=0}^{\infty}G^{(n)}_{\rho\mu}\epsilon^{\mu\kappa\lambda\nu}(\hat{k}_{AF})_\kappa p_\lambda\,,\\
0={}&{}-(p_\sigma p^\sigma)\sum\limits_{n=0}^{\infty}(G^{(n+1)})_{\rho}^\nu+2i\sum\limits_{n=0}^{\infty}G^{(n)}_{\rho\mu}\epsilon^{\mu\kappa\lambda\nu}(\hat{k}_{AF})_\kappa p_\lambda\,,\\
0={}&{}\sum\limits_{n=0}^{\infty}\left((G^{(n+1)})_{\rho}^\nu-\frac{2i}{(p_\sigma p^\sigma)}G^{(n)}_{\rho\mu}\epsilon^{\mu\kappa\lambda\nu}(\hat{k}_{AF})_\kappa p_\lambda\right)\,.
\end{aligned}
\end{equation*}

Due to the nature of perturbation expansion, the last summation is satisfied only if each term is itself zero, which results in the recursion formula
\begin{equation*}
G^{(n+1)}_{\rho\nu}=\frac{2i}{(p_\sigma p^\sigma)}\epsilon^{\mu\kappa\lambda\sigma}(\hat{k}_{AF})_\kappa p_\lambda\eta_{\nu\sigma}G^{(n)}_{\rho\mu}\,.
\end{equation*}
Clearly, the propagator can be written upto any order desired, where first few terms are listed in \tabref{Table-fftipeop}. Particularly, the leading order propagator is same with that obtained by the ansatz method in \secref{Section: cpt odd ansatz}, that is \equref{LO Propagator}, which is a good consistency check.
\begin{table}
\caption[First Few Terms in the Expansion of the CPT-odd Propagator]{First Few Terms in the Expansion of the CPT-odd Propagator. The covariant form of the propagator can be extracted from \equref{eq: definifion of the propagator} with a series expansion of the form \equref{propagator expansion}. In such an expansion, first few terms can be shown to be those below.}
\label{Table-fftipeop}
\begin{tabularx}{\textwidth}{XX}
\hline\hline N$^\text{th}$ Term & Expression\\\hline\\[-0.15in]
$G^{(0)} _{\mu\nu}$ {}&{} $\displaystyle-\frac{1}{(p_\alpha p^\alpha)}\eta_{\mu\nu}$\\\\[-0.15in]
$G^{(1)} _{\mu\nu}$ {}&{} $\displaystyle\frac{2i}{(p_\alpha p^\alpha)^2}\epsilon_{\mu\nu\kappa\lambda}(\hat{k}_{AF})^\kappa p^\lambda$\\\\[-0.15in]
$G^{(2)} _{\mu\nu}$ {}&{} $\displaystyle\frac{2}{(p_\alpha p^\alpha)^3}\eta_{\nu\sigma}\epsilon_{\rho\mu\kappa\lambda}\epsilon^{\rho\sigma\gamma\delta}(\hat{k}_{AF})^\kappa(\hat{k}_{AF})_\gamma p^\lambda p_\delta$\\\hline\hline
\end{tabularx}
\end{table}

\subsection{Helicity Basis Propagator}
\label{Section: cpt odd helicity propagator}

In sections \secref{Section: cpt odd ansatz} and \secref{Section: cpt odd perturbation}, the main focus was on the covariance of the propagator, that is, whether the form of the propagator makes any explicit reference to an explicit basis. As discussed at the beginning of \secref{Section: cpt odd propagator}, the complexity of the inverse propagator \equref{Momentum space inverse propagator} makes it formidable, if possible, to preserve both covariance and the analyticity; hence, the propagators extracted have been either in leading order or in an expansion form. In this section, we will instead sacrifice the covariance and find the analytic form of the propagator in a definite coordinate system.

It is sufficient, though not generally necessary, to choose any particular basis so as to find the analytic form of the propagator as one can always invoke the matrix representation in any chosen basis, and that non-singular matrices are always analytically invertible. For relevance to the case at hand, the explicit basis will be chosen as the helicity basis.

Let's start by decomposing the inverse propagator \equref{Momentum space inverse propagator} to its temporal and spatial components by utilizing
\begin{equation}
\epsilon^{\mu\kappa\lambda\nu}=\left(\delta^\mu_0\delta^\kappa_i\delta^\lambda_j\delta^\nu_k-\delta^\mu_i\delta^\kappa_0\delta^\lambda_j\delta^\nu_k+\delta^\mu_i\delta^\kappa_j\delta^\lambda_0\delta^\nu_k-\delta^\mu_i\delta^\kappa_j\delta^\lambda_k\delta^\nu_0\right)\epsilon^{ijk} \label{4 Levi Civita}\,,
\end{equation}
which in turn gives
\begin{align*}
\epsilon^{\mu\kappa\lambda\nu}(\hat{k}_{AF})_\kappa p_\lambda={}&{}\left(\delta^\mu_0\delta^\kappa_i\delta^\lambda_j\delta^\nu_k-\delta^\mu_i\delta^\kappa_0\delta^\lambda_j\delta^\nu_k+\delta^\mu_i\delta^\kappa_j\delta^\lambda_0\delta^\nu_k-\delta^\mu_i\delta^\kappa_j\delta^\lambda_k\delta^\nu_0\right)\epsilon^{ijk}(\hat{k}_{AF})_\kappa p_\lambda\,,\\
\epsilon^{\mu\kappa\lambda\nu}(\hat{k}_{AF})_\kappa p_\lambda={}&{}\delta^\mu_0\delta^\nu_k(\mathbf{(\hat{k}_{AF})}\times\mathbf{p})^k-\delta^\mu_i\delta^\nu_k\epsilon^{ijk}p_j(\hat{k}_{AF})_0+\delta^\mu_i\delta^\nu_k\epsilon^{ijk}\mathbf{(\hat{k}_{AF})}_j\omega\\{}&{}-\delta^\mu_i\delta^\nu_0(\mathbf{(\hat{k}_{AF})}\times\mathbf{p})^i\,,
\end{align*}
hence
\begin{equation*}
\begin{aligned}
(\hat{G}^{-1})^{\mu\nu}={}&{}-\eta^{\mu\nu}(p_\sigma p^\sigma)+2i\delta^\mu_0\delta^\nu_k(\mathbf{(\hat{k}_{AF})}\times\mathbf{p})^k-2i\delta^\mu_i\delta^\nu_k\epsilon^{ijk}p_j(\hat{k}_{AF})_0\\{}&{}+2i\delta^\mu_i\delta^\nu_k\epsilon^{ijk}\mathbf{(\hat{k}_{AF})}_j\omega-2i\delta^\mu_i\delta^\nu_0(\mathbf{(\hat{k}_{AF})}\times\mathbf{p})^i\,,
\end{aligned}
\end{equation*}
that is
\begin{equation}
(\hat{G}^{-1})^{\mu\nu}=-\eta^{\mu\nu}(p_\sigma p^\sigma)+4i\delta^\mu_{[0}\delta^\nu_{i]}(\mathbf{(\hat{k}_{AF})}\times\mathbf{p})^i+2i\delta^\mu_i\delta^\nu_j\epsilon^{ijk}(p_k(\hat{k}_{AF})_0-\mathbf{(\hat{k}_{AF})}_k\omega)\label{decomposed Inverse propagator}
\end{equation}
in a more compact manner.

Above equation, though it is in component form, does not yet refer to a specific basis for its space part. Helicity basis can be chosen for the space part by inserting the corresponding metric and the Levi-Civita tensor of the helicity basis, which are explicitly discussed in \appref{CHAPTER:HELICITY}.

To keep track of the terms, let us calculate $\left(\mathbf{(\hat{k}_{AF})}\times\mathbf{p}\right)^i$ and\\ $\epsilon^{ijk}\left(p_k(\hat{k}_{AF})_0-\mathbf{(\hat{k}_{AF})}_kp_0\right)$ one by one:
\begin{itemize}
\item $(\mathbf{(\hat{k}_{AF})}\times\mathbf{p})^i$:
\begin{align*}
={}&{}\epsilon^{ijk}(\hat{k}_{AF})_jp_k\nonumber\\
={}&{}\epsilon^{ijr}(\hat{k}_{AF})_jp_r\footnotemark=\delta^i_+\epsilon^{+-r}(\hat{k}_{AF})_-p_r+\delta^i_-\epsilon^{-+r}(\hat{k}_{AF})_+p_r \nonumber\\
={}&{} ip(\delta^i_+(\hat{k}_{AF})_--\delta^i_-(\hat{k}_{AF})_+)
\end{align*}
\footnotetext{In helicity basis, the three momentum has only radial component by construction, which is why $p_r$ is the only nonzero component of three vector $p_k$, which can also be seen from \equref{eq:p in helicity basis}.}
\item $\epsilon^{ijk}(p_k(\hat{k}_{AF})_0-\mathbf{(\hat{k}_{AF})}_kp_0)$:
\begin{align*}
={}&{}\epsilon^{ijk}p_k(\hat{k}_{AF})_0-\epsilon^{ijk}\mathbf{(\hat{k}_{AF})}_kp_0\\
={}&{}-\epsilon^{ijr}(\omega(\hat{k}_{AF})_r-p(\hat{k}_{AF})_0)-\epsilon^{ij+}\omega(\hat{k}_{AF})_+-\epsilon^{ij-}\omega(\hat{k}_{AF})_-\\
={}&{} i(\delta^i_-\delta^j_+-\delta^i_+\delta^j_-)(\omega(\hat{k}_{AF})_r-p(\hat{k}_{AF})_0)+i(\delta^i_r\delta^j_--\delta^i_-\delta^j_r)\omega(\hat{k}_{AF})_+\\{}&{}+i(\delta^i_+\delta^j_r-\delta^i_r\delta^j_+)\omega(\hat{k}_{AF})_-
\end{align*}
\end{itemize}
where $p_0=\omega$ and $p_r=p$ are simply the usual frequency and the magnitude of the momentum respectively. Once these are inserted into \equref{decomposed Inverse propagator}),
\begin{equation}
\begin{aligned}
(\hat{G}^{-1})^{\mu\nu}={}&{}-\eta^{\mu\nu}(p_\sigma p^\sigma)-4p\delta^\mu_{[0}\delta^\nu_{i]}(\delta^i_+(\hat{k}_{AF})_--\delta^i_-(\hat{k}_{AF})_+)\\{}&{}-2\delta^\mu_i\delta^\nu_j\Big((\delta^i_-\delta^j_+-\delta^i_+\delta^j_-)(\omega(\hat{k}_{AF})_r-p(\hat{k}_{AF})_0)\\{}&{}+(\delta^i_r\delta^j_--\delta^i_-\delta^j_r)\omega(\hat{k}_{AF})_++(\delta^i_+\delta^j_r-\delta^i_r\delta^j_+)\omega(\hat{k}_{AF})_-\Big)\,,
\end{aligned}
\end{equation}
which can be expanded component by component as follows:
\begin{align*}
(\hat{G}^{-1})^{\mu\nu}={}&{}-\eta^{\mu\nu}(p_\sigma p^\sigma)-2\delta^\mu_0\delta^\nu_+p(\hat{k}_{AF})_-+2\delta^\mu_+\delta^\nu_0p(\hat{k}_{AF})_-+2\delta^\mu_0\delta^\nu_-p(\hat{k}_{AF})_+\\
{}&{}-2\delta^\mu_-\delta^\nu_0p(\hat{k}_{AF})_+-2\delta^\mu_-\delta^\nu_+(\omega(\hat{k}_{AF})_r-p(\hat{k}_{AF})_0)\\
{}&{}+2\delta^\mu_+\delta^\nu_-(\omega(\hat{k}_{AF})_r-p(\hat{k}_{AF})_0)-2\delta^\mu_r\delta^\nu_-\omega(\hat{k}_{AF})_+\\
{}&{}+2\delta^\mu_-\delta^\nu_r\omega(\hat{k}_{AF})_+-2\delta^\mu_+\delta^\nu_r\omega(\hat{k}_{AF})_-+2\delta^\mu_r\delta^\nu_+\omega(\hat{k}_{AF})_-\,.
\end{align*}

As $(\hat{G}^{-1})_\rho^{\;\;\nu}=(\hat{G}^{-1})^{\mu\nu}\eta_{\mu\rho}$, above equation becomes
\begin{align*}
(\hat{G}^{-1})_\rho^{\;\;\nu}={}&{}-\delta_\rho^{\;\nu}(p_\sigma p^\sigma)-2\eta_{\rho 0}\delta^\nu_+p(\hat{k}_{AF})_-+2\eta_{\rho +}\delta^\nu_0p(\hat{k}_{AF})_-+2\eta_{\rho 0}\delta^\nu_-p(\hat{k}_{AF})_+\\
{}&{}-2\eta_{\rho -}\delta^\nu_0p(\hat{k}_{AF})_+-2\eta_{\rho -}\delta^\nu_+(\omega(\hat{k}_{AF})_r-p(\hat{k}_{AF})_0)\\
{}&{}+2\eta_{\rho +}\delta^\nu_-(\omega(\hat{k}_{AF})_r-p(\hat{k}_{AF})_0)-2\eta_{\rho r}\delta^\nu_-\omega(\hat{k}_{AF})_+\\
{}&{}+2\eta_{\rho -}\delta^\nu_r\omega(\hat{k}_{AF})_+-2\eta_{\rho +}\delta^\nu_r\omega(\hat{k}_{AF})_-+2\eta_{\rho r}\delta^\nu_+\omega(\hat{k}_{AF})_-\,,
\end{align*}
which can be represented with the matrix notation in the basis $\{\hat{e}_0, \hat{e}_+, \hat{e}_r, \hat{e}_-\}$ as
\begin{equation}\small
\begin{aligned}
(\hat{G}^{-1})\doteq
\begin{pmatrix}
-(p_\sigma p^\sigma){}&{}-2p(\hat{k}_{AF})_-{}&{} 0{}&{} 2p(\hat{k}_{AF})_+\\
2p(\hat{k}_{AF})_+{}&{}-(p_\sigma p^\sigma)+2p(\hat{k}_{AF})_s{}&{} -2\omega(\hat{k}_{AF})_+{}&{} 0\\
0{}&{} -2\omega(\hat{k}_{AF})_-{}&{}-(p_\sigma p^\sigma){}&{} 2\omega(\hat{k}_{AF})_+\\
-2p(\hat{k}_{AF})_-{}&{} 0{}&{} 2\omega(\hat{k}_{AF})_-{}&{}-(p_\sigma p^\sigma)-2p(\hat{k}_{AF})_s
\end{pmatrix}\,,
\end{aligned}\normalsize\label{inverse propagator in helicity}
\end{equation}
where the matrix representation convention implied and used is
\begin{equation*}
\begin{aligned}
(\hat{G})_\rho^{\;\;\nu}\doteq
\begin{pmatrix}
\hat{G}_{0}^{\;0}{}&{}\hat{G}_{0}^{\;+}{}&{}\hat{G}_{0}^{\;r}{}&{}\hat{G}_{0}^{\;-}\\
\hat{G}_{+}^{\;0}{}&{}\hat{G}_{+}^{\;+}{}&{}\hat{G}_{+}^{\;r}{}&{}\hat{G}_{+}^{\;-}\\
\hat{G}_{r}^{\;0}{}&{}\hat{G}_{r}^{\;+}{}&{}\hat{G}_{r}^{\;r}{}&{}\hat{G}_{r}^{\;-}\\
\hat{G}_{-}^{\;0}{}&{}\hat{G}_{-}^{\;+}{}&{}\hat{G}_{-}^{\;r}{}&{}\hat{G}_{-}^{\;-}\\
\end{pmatrix}
\end{aligned}
\end{equation*}
in component form.

The propagator is simply the inverse of this matrix and can easily be calculated analytically since each term in this matrix is simply a scalar function of $\omega$ and $p$. Nonetheless, we will not provide it here for two reasons: Firstly, it does not give any particular insight; and secondly, \equref{inverse propagator in helicity} can be further simplified once the attention is restricted to physical solutions.

What is meant with the last remark is related to the fact that not all possible component combinations of $(\hat{k}_{AF})_\mu$ yield a model with physical solutions, that is the polarization vectors, for the photon field. As a matter of fact, what restrictions on the combinations are required to limit the focus on the physical cases are already found and discussed in \secref{Section: cpt odd polarization vectors}. There, the splitting of the coefficient space into so-defined $\hat{k}_{AF}^{(bf)}$, $\hat{k}_{AF}^{(cn)}$ and $\hat{k}_{AF}^{(np)}$ is introduced; hence, all that is required is to discard the nonphysical coefficient subspace $\hat{k}_{AF}^{(np)}$, which is achieved by simply setting  $(\hat{k}_{AF})_\pm=0$. Then, from \equref{inverse propagator in helicity} we have
\begin{equation}
\begin{aligned}
(\hat{G}^{-1})\doteq{}
\begin{pmatrix}
-(p_\sigma p^\sigma){}&{} 0{}&{} 0{}&{} 0\\
0{}&{}-(p_\sigma p^\sigma)+2p(\hat{k}_{AF})_s{}&{} 0{}&{} 0\\
0{}&{} 0{}&{}-(p_\sigma p^\sigma){}&{} 0\\
0{}&{} 0{}&{} 0{}&{}-(p_\sigma p^\sigma)-2p(\hat{k}_{AF})_s
\end{pmatrix}\,.
\end{aligned}\label{inverse propagator in helicity2}
\end{equation}

This is the main superiority of explicit helicity basis over covariant approaches, and other possible basis choices for this section's analytic approach, as the nonphysical possibility $\hat{k}_{AF}^{(np)}$ can not be trivially eliminated in them. 

With the attention restricted to the coefficient subspace of physical solutions then, the diagonal inverse propagator, \equref{inverse propagator in helicity2}, can straightforwardly inverted, hence
\begin{equation}
\hat{G}\doteq{}\text{Diagonal}\left(-\frac{1}{p_\mu p^\mu}, -\frac{1}{(p_\mu p^\mu)+2p(\hat{k}_{AF})_s}, -\frac{1}{p_\mu p^\mu}, -\frac{1}{(p_\mu p^\mu)-2p(\hat{k}_{AF})_s}\right)\label{eq: Propagator}\,.
\end{equation}

\section{Photon Sector Special Models}
\label{Section: photon sector special models}
The general {\it nm}SME framework as it stands is quite complicated due to the vast number of LVT. Thus, it is generally appropriate to work with a subset of all possible coefficients, selection of which can be categorized regarding the main focus in each of these special models. Although a number of different such special models can be constructed, we will list only the most common ones, which can be found in Ref. \cite{0905.0031} as well.
\begin{enumerate}
\item \emph{Minimal SME:} This special model is actually the original version of the SME as it was introduced in 1998 \cite{hep-ph/9809521}. One can restrict the attention to minimal SME, or simply {\it m}SME, by allowing only the power counting renormalizable LV operators in the Lagrangian. Here, it should be stressed that although the general SME, or simply {\it nm}SME, is nonrenormalizable by construction, \textit{m}SME is \emph{not} necessarily renormalizable simply because all operators are of power counting renormalizable dimensions. Whether the model is indeed renormalizable should be verified explicitly for each sector, about which several work including that of QED have been conducted at least for one-loop renormalizability \cite{Kostelecky:2001jc, Colladay:2006rk, Colladay:2007aj, PhysRevD.79.125019}.

\item \emph{Isotropic Models:} While examining possible Lorentz violations, one can restrict the attention to LVT which would preserve the rotational symmetry of the system nonetheless. In nonminimal photon sector, this restriction translates into the condition that all spherical coefficients with nonzero $j$ are set to be zero in the preferred frame. This ensures that the LV, whatever it is, is rotationally invariant in the preferred frame. The subtle point about this special model is that this isotropy is valid only in one frame as the LV coefficients will mix once one boosts to another frame; hence, the preferred frame should be chosen wisely. The theoretically natural choice is that of \emph{Cosmic Microwave Background}, where the canonical sun-centered frame can be chosen for practical purposes as well.

\item \emph{Vacuum Models:} The general dispersion relation for photon in \textit{nm}SME is nontrivial, as can be seen from \equref{General CPT-odd Dispersion Relation2} for the CPT-odd case. However, one can replace the dispersion relation with the conventional $\omega=p$ in the leading order, as the electromagnetic fields in the vacuum can be approximated by vacuum plane waves. The imposition $\omega=p$ on $\hat{k}_{AF}$ and $\hat{k}_{F}$ then identifies which parts of these coefficients contribute in the leading order. These coefficients, which still contribute in the leading order under the conventional dispersion relation, are then named \emph{vacuum coefficients}.

In the CPT-odd case, the vacuum coefficients are obtained as the totally symmetric and traceless part of $(k_{AF}^{(d)})^{\kappa\alpha_1...\alpha_{(d-3)}}$. In terms of spherical decomposition, this reads as
\begin{equation*}
k^{(d)}_{(V)jm}=\sum_{n}(-1)^{j+1}\left(\frac{d}{n+3}(k^{(d)}_{AF})^{(0B)}_{njm}+\frac{1}{n+2}(k^{(d)}_{AF})^{(1B)}_{njm}\right)\,.
\end{equation*}
Further details regarding the decomposition of the coefficients with respect to the vacuum propagation properties are presented in \tabref{Table-sdc}.

\item \emph{Vacuum Orthogonal Models:} The so-called vacuum coefficients are defined above. The remaining coefficients which do not have a leading order birefringent or dispersive effect are called vacuum-orthogonal coefficients, and are denoted by a negation diacritic $\neg$ as can be seen in \tabref{Table-sdc}. Hence, any model whose focus is on the complementary part of the coefficient space to the subspace of vacuum coefficients is called \emph{Vacuum Orthogonal Model}.
\end{enumerate}		
\chapter{Vacuum Orthogonal Model for CPT-odd Photon}
\label{vacuum orthogonal model}

In \secref{Section: tests and bounds}, the data table \cite{0801.0287} is introduced in which the most up-to-date bounds on the various coefficients are listed. One curious thing about the data table is that there is no bound on any of the so-called vacuum orthogonal coefficients which are mentioned in the end of last chapter and whose complete list can be seen in \tabref{Table-sdc}. That is actually no coincidence, but a result of two important points: There is no vacuum orthogonal coefficient in the renormalizable dimensions, but only in the nonrenormalizable ones which are barely examined; and most of the bounds come from the astrophysical sources which are immune to the effects of vacuum orthogonal coefficients in the leading order.

That their effects are not bounded at all makes the vacuum orthogonal coefficients theoretically attractive. Moreover, that they are not present in \textit{m}SME which has been thoroughly studied raises the possibility of accompanying new effects. Despite these benefits, a model whose only LVT are vacuum orthogonal coefficient, or simply \emph{Vacuum Orthogonal Model} (VOM) has not been properly analyzed in the literature though, except Ref. \cite{Albayrak}. For the rest of the thesis then, that reference will be used as the main source.

The VOM can be analyzed in any basis; yet, the advantages of the helicity basis which have been stressed so far\footnotemark\footnotetext{One of the main advantages is that helicity basis enables spherical decomposition, as studied in \secref{Section: spherical decomposition}, which is a natural classification with direct relevance to observations and experiments. Another main advantage would be how it naturally divides the coefficient space into distinct subspaces, as derived in \secref{Section: cpt odd polarization vectors}. Finally, working explicitly in the helicity basis guarantees the removal of redundant components in the propagator, which was shown in \secref{Section: cpt odd propagator}.} apply here as well; hence, the helicity basis will be employed in this Chapter too. However, the VOM is simply a special case of the general model, which means that one should be able to extract the field theoretical quantities like dispersion relations or propagators of the VOM out of the same quantities of the general model under suitable restrictions. It turns out that there indeed exist some simple prescriptions for that, and the rest of the chapter is devoted to the application of these prescriptions and analysis of the results.

\section{Dispersion Relation and The Polarization Vectors}
\label{vom dispersion}

The restriction of the general model to the vacuum orthogonal coefficients only is achieved by imposing
\begin{subequations}
\begin{align}
(k_{AF}^{(d)})^{(0B)}_{njm}={}&{}\frac{(d-2-n)(n+3)}{d(d-2-n+j)}\left((\vok_{AF}^{(d)})^{(0B)}_{njm}-(\vok_{AF}^{(d)})^{(0B)}_{(n-2)jm}\right)\nonumber\\{}&{}-\frac{1}{n+1}(\vok_{AF}^{(d)})^{(1B)}_{(n-1)jm}\,,\\
(k_{AF}^{(d)})^{(1B)}_{njm}={}&{}\frac{j(n+2)}{d-3-n+j}\left((\vok_{AF}^{(d)})^{(0B)}_{(n+1)jm}-(\vok_{AF}^{(d)})^{(0B)}_{(n-1)jm}\right)\nonumber\\{}&{}+\frac{d}{n+4}(\vok_{AF}^{(d)})^{(1B)}_{njm}
\end{align} \label{Eq:general to vacuum orthogonal}
\end{subequations}
on the dispersion relation \equref{General CPT-odd Dispersion Relation2}. Then, it becomes
\begin{align*}
0={}&{}\left(p_\mu p^\mu\right)^2-4\Bigg\{\sum\limits_{dnjm}\omega^{d-3-n}p^n\prescript{}{0}{Y}_{jm}(\mathbf{\hat{p}})\Bigg(\frac{dp}{n+3}\bigg(\frac{(d-2-n)(n+3)}{d(d-2-n+j)}\\{}&{}\times\left((\vok_{AF}^{(d)})^{(0B)}_{njm}-(\vok_{AF}^{(d)})^{(0B)}_{(n-2)jm}\right)-\frac{1}{n+1}(\vok_{AF}^{(d)})^{(1B)}_{(n-1)jm}\bigg)+\frac{\omega}{n+2}\\{}&{}\times\left(\frac{j(n+2)}{d-3-n+j}\left((\vok_{AF}^{(d)})^{(0B)}_{(n+1)jm}-(\vok_{AF}^{(d)})^{(0B)}_{(n-1)jm}\right)+\frac{d}{n+4}(\vok_{AF}^{(d)})^{(1B)}_{njm}\right)\Bigg)\Bigg\}^2\\{}&{}-8p_\mu p^\mu\sum\limits_{d_1d_2n_1n_2j_1j_2m_1m_2}\omega^{d_1+d_2-6-n_1-n_2}p^{n_1+n_2}\prescript{}{+1}{Y}_{j_1m_1}(\mathbf{\hat{p}})\prescript{}{-1}{Y}_{j_2m_2}(\mathbf{\hat{p}})\\{}&{}\times\frac{1}{\sqrt{4j_1j_2(j_1+1)(j_2+1)}}\Bigg(\bigg(\frac{j_1(n_1+2)}{d_1-3-n_1+j_1}\Big((\vok_{AF}^{(d_1)})^{(0B)}_{(n_1+1)j_1m_1}\\
{}&{}-(\vok_{AF}^{(d_1)})^{(0B)}_{(n_1-1)j_1m_1}\Big)+\frac{d_1}{n_1+4}(\vok_{AF}^{(d_1)})^{(1B)}_{n_1j_1m_1}\bigg)+i(\vok_{AF}^{(d_1)})^{(1E)}_{n_1j_1m_1}\Bigg)\\{}&{}\times\Bigg(-\bigg(\frac{j_2(n_2+2)}{d_2-3-n_2+j_2}\left((\vok_{AF}^{(d_2)})^{(0B)}_{(n_2+1)j_2m_2}-(\vok_{AF}^{(d_2)})^{(0B)}_{(n_2-1)j_2m_2}\right)\\{}&{}+\frac{d_2}{n_2+4}(\vok_{AF}^{(d_2)})^{(1B)}_{n_2j_2m_2}\bigg)+i(\vok_{AF}^{(d_2)})^{(1E)}_{n_2j_2m_2}\Bigg)\,,
\end{align*}
hence
\begin{align*}
0={}&{}\left(p_\mu p^\mu\right)^2-4\Bigg\{\sum\limits_{dnjm}\omega^{d-3-n}p^n\prescript{}{0}{Y}_{jm}(\mathbf{\hat{p}})\Bigg((\vok_{AF}^{(d)})^{(0B)}_{njm}\frac{p}{\omega^2}\bigg(\frac{(d-2-n)\omega^2}{d-2-n+j}\\{}&{}-\frac{(d-4-n)p^2}{d-4-n+j}\bigg)
-\frac{dp^2}{(n+4)(n+2)\omega}(\vok_{AF}^{(d)})^{(1B)}_{njm}+(\vok_{AF}^{(d)})^{(0B)}_{njm}\frac{j}{p}\\{}&{}\times\left(\frac{\omega^2}{d-2-n+j}-\frac{p^2}{d-4-n+j}\right)+\frac{d\omega}{(n+2)(n+4)}(\vok_{AF}^{(d)})^{(1B)}_{njm}\Bigg)\Bigg\}^2\\{}&{}-8p_\mu p^\mu\sum\limits_{d_1d_2n_1n_2j_1j_2m_1m_2}\omega^{d_1+d_2-6-n_1-n_2}p^{n_1+n_2}\prescript{}{+1}{Y}_{j_1m_1}(\mathbf{\hat{p}})\prescript{}{-1}{Y}_{j_2m_2}(\mathbf{\hat{p}})\\
{}&{}\times\frac{1}{\sqrt{4j_1j_2(j_1+1)(j_2+1)}}\Bigg(\left(\frac{\omega j_1(n_1+1)}{p(d_1-2-n_1+j_1)}-\frac{pj_1(n_1+3)}{\omega(d_1-4-n_1+j_1)}\right)\\{}&{}\times(\vok_{AF}^{(d_1)})_{n_1j_1m_1}^{(0B)}+\frac{d_1}{n_1+4}(\vok_{AF}^{(d_1)})^{(1B)}_{n_1j_1m_1}+i(\vok_{AF}^{(d_1)})^{(1E)}_{n_1j_1m_1}\Bigg)\\{}&{}\times\Bigg(-\left(\frac{\omega j_2(n_2+1)}{p(d_2-2-n_2+j_2)}-\frac{pj_2(n_2+3)}{\omega(d_2-4-n_2+j_2)}\right)(\vok_{AF}^{(d_2)})_{n_2j_2m_2}^{(0B)}\\{}&{}-\frac{d_2}{n_2+4}(\vok_{AF}^{(d_2)})^{(1B)}_{n_2j_2m_2}+i(\vok_{AF}^{(d_2)})^{(1E)}_{n_2j_2m_2}\Bigg)\,.
\end{align*}

The last summation can be further simplified by invoking the symmetries. Clearly, last two rows are of the form $\left(A(t_1)+iB(t_1)\right)\left(-A(t_2)+iB(t_2)\right)$ where $t_i$ is the collective index for $\{d_i,n_i,j_i,m_i\}$. But this is equal to $-\left(A(t_1)A(t_2)+B(t_1)B(t_2)\right)+i\left(A(t_1)B(t_2)-A(t_2)B(t_1)\right)$, and since the fourth row is symmetric over $\{t_1,t_2\}$, its contraction with the second term, which is antisymmetric over $\{t_1,t_2\}$, gives zero; hence, the equation finally reduces to 
\begin{align}
0={}&{}\left(p_\mu p^\mu\right)^2-4\Bigg\{\sum\limits_{dnjm}\omega^{d-3-n}p^n\prescript{}{0}{Y}_{jm}(\mathbf{\hat{p}})\Bigg((\vok_{AF}^{(d)})^{(0B)}_{njm}\frac{p}{\omega^2}\bigg(\frac{(d-2-n)\omega^2}{d-2-n+j}\nonumber\\{}&{}-\frac{(d-4-n)p^2}{d-4-n+j}\bigg)-\frac{dp^2}{(n+4)(n+2)\omega}(\vok_{AF}^{(d)})^{(1B)}_{njm}+(\vok_{AF}^{(d)})^{(0B)}_{njm}\frac{j}{p}\bigg(\frac{\omega^2}{d-2-n+j}\nonumber\\{}&{}-\frac{p^2}{d-4-n+j}\bigg)+\frac{d\omega}{(n+2)(n+4)}(\vok_{AF}^{(d)})^{(1B)}_{njm}\Bigg)\Bigg\}^2+8p_\mu p^\mu\nonumber\\
{}&{}\times\sum\limits_{d_1d_2n_1n_2j_1j_2m_1m_2}\omega^{d_1+d_2-6-n_1-n_2}p^{n_1+n_2}\prescript{}{+1}{Y}_{j_1m_1}(\mathbf{\hat{p}})\prescript{}{-1}{Y}_{j_2m_2}(\mathbf{\hat{p}})\nonumber\\{}&{}\frac{1}{\sqrt{4j_1j_2(j_1+1)(j_2+1)}}\Bigg\{\Bigg(\bigg(\frac{\omega j_1(n_1+1)}{p(d_1-2-n_1+j_1)}-\frac{pj_1(n_1+3)}{\omega(d_1-4-n_1+j_1)}\bigg)\nonumber\\{}&{}\times(\vok_{AF}^{(d_1)})_{n_1j_1m_1}^{(0B)}+\frac{d_1}{n_1+4}(\vok_{AF}^{(d_1)})^{(1B)}_{n_1j_1m_1}\Bigg)\Bigg(\bigg(\frac{\omega j_2(n_2+1)}{p(d_2-2-n_2+j_2)}\nonumber\\{}&{}-\frac{pj_2(n_2+3)}{\omega(d_2-4-n_2+j_2)}\bigg)(\vok_{AF}^{(d_2)})_{n_2j_2m_2}^{(0B)}+\frac{d_2}{n_2+4}(\vok_{AF}^{(d_2)})^{(1B)}_{n_2j_2m_2}\Bigg)\nonumber\\{}&{}+(\vok_{AF}^{(d_1)})^{(1E)}_{n_1j_1m_1}(\vok_{AF}^{(d_2)})^{(1E)}_{n_2j_2m_2}\Bigg\}\,,\label{General Vacuum-orthogonal dispersion relation}
\end{align}
where the range for each term is summarized in \tabref{Table-vocir}.
\begin{table}
\caption[Index Ranges for Vacuum Orthogonal Coefficients]{Index Ranges for Vacuum Orthogonal Coefficients. Due to their construction, the symmetries they obey and the other constraints restrict the frequency dependence $n$ and total angular momentum $j$ of the coefficients as listed below, which can be extracted from the more comprehensive table, \emph{Table XVIII} of \cite{0905.0031}.}
\label{Table-vocir}
\begin{tabularx}{\textwidth}{XXXl}
\hline\hline Coefficient & d & n & j \\\hline\\[-0.15in]
$(\vok_{AF}^{(d)})^{(0B)}_{njm}$& odd$\ge5$&$0,1,...,d-4$&$n,n-2,n-4,...,\ge0$\\\\[-0.15in]
$(\vok_{AF}^{(d)})^{(1B)}_{njm}$& odd$\ge5$&$0,1,...,d-4$&$n+1,n-1,n-3,...,\ge1$\\\\[-0.15in]
$(\vok_{AF}^{(d)})^{(1E)}_{njm}$& odd$\ge5$&$1,2,...,d-3$&$n,n-2,n-4,...,\ge1$
\\\hline\hline
\end{tabularx}
\end{table}

This equation does not give much insight as it stands. However, we will show that it can be further simplified and be written in a more compact form. To do that, one needs to reorganize the second term in \equref{General Vacuum-orthogonal dispersion relation}.
\begin{align*}
2^{\text{nd}} \text{ Term}={}&{} 4\Bigg\{\sum\limits_{dnjm}\omega^{d-3-n}p^n\prescript{}{0}{Y}_{jm}(\mathbf{\hat{p}})\Bigg((\vok_{AF}^{(d)})^{(0B)}_{njm}\frac{p}{\omega^2}\bigg(\frac{(d-2-n)\omega^2}{d-2-n+j}\\{}&{}-\frac{(d-4-n)p^2}{d-4-n+j}\bigg)-\frac{dp^2}{(n+4)(n+2)\omega}(\vok_{AF}^{(d)})^{(1B)}_{njm}+(\vok_{AF}^{(d)})^{(0B)}_{njm}\frac{j}{p}\\{}&{}\times\left(\frac{\omega^2}{d-2-n+j}-\frac{p^2}{d-4-n+j}\right)+\frac{d\omega}{(n+2)(n+4)}(\vok_{AF}^{(d)})^{(1B)}_{njm}\Bigg)\Bigg\}^2\\
={}&{} 4\Bigg\{\sum\limits_{dnjm}\omega^{d-3-n}p^n\prescript{}{0}{Y}_{jm}(\mathbf{\hat{p}})\Bigg[ (\vok_{AF}^{(d)})^{(1B)}_{njm}\bigg(-\frac{dp^2}{(n+4)(n+2)\omega}\\{}&{}+\frac{d\omega}{(n+2)(n+4)}\bigg)
(\vok_{AF}^{(d)})^{(0B)}_{njm}\Bigg(\frac{p}{\omega^2}\left(\frac{(d-2-n)\omega^2}{d-2-n+j}-\frac{(d-4-n)p^2}{d-4-n+j}\right)\\{}&{}+\frac{j}{p}\left(\frac{\omega^2}{d-2-n+j}-\frac{p^2}{d-4-n+j}\right)\Bigg)\Bigg]\Bigg\}^2\,,
\end{align*}
thus,
\begin{equation}
\begin{aligned}
2^{\text{nd}} \text{ Term}={}&{} 4\bigg(\sum\limits_{dnjm}\omega^{d-4-n}p^n\prescript{}{0}{Y}_{jm}(\mathbf{\hat{p}})(\vok_{AF}^{(d)})^{(1B)}_{njm}\frac{d(p_\mu p^\mu)}{(n+2)(n+4)}\\{}&{}+\sum\limits_{dnjm}\omega^{d-5-n}p^{n-1}\prescript{}{0}{Y}_{jm}(\mathbf{\hat{p}})(\vok_{AF}^{(d)})^{(0B)}_{njm}\mathbf{K}_{dnjm}(\omega,p)\bigg)^2\,,
\end{aligned}\label{Eqn: Second term}
\end{equation}
where
\begin{equation*}
\begin{aligned}
\mathbf{K}_{dnjm}(\omega,p):={}&{}p^2\left(\frac{(d-2-n)\omega^2}{d-2-n+j}-\frac{(d-4-n)p^2}{d-4-n+j}\right)\\{}&{}+j\omega^2\left(\frac{\omega^2}{d-2-n+j}-\frac{p^2}{d-4-n+j}\right)
\end{aligned}
\end{equation*}
is defined for convenience.

Simplification of the second term is now reduced to the simplification of the function $\mathbf{K}_{dnjm}(\omega,p)$. But this is a simple algebraic calculation, which can be carried out as follows.

\begin{align*}
&\begin{aligned}
\mathbf{K}_{dnjm}(\omega,p)={}&{} \frac{1}{(d-2-n+j)(d-4-n+j)}\Big(j\omega^4(d-4-n+j)+\omega^2 p^2\\
{}&{}\times\big((d-2-n)(d-4-n+j)-j(d-2-n+j)\big)\\
{}&{}-p^4(d-4-n)(d-2-n+j)\Big)
\end{aligned}
\\
&=\frac{1}{(d-2-n+j)(d-4-n+j)}\left(\begin{array}{r l}
\omega^4 {}&{} \times(d-4-n)j\\
+\omega^4 {}&{} \times(-1)\\
+\omega^2p^2{}&{}\times(d-2-n)(d-4-n)\\
+\omega^2p^2{}&{}\times(+1)\\
+p^4{}&{}\times-(d-2-n)(d-4-n)\\
+p^4{}&{}\times-j(d-4-n)
\end{array}\right)\,,
\end{align*}
which can be put in the form
\begin{equation}
\mathbf{K}_{dnjm}(\omega,p)={}(p_\mu p^\mu)\left(\omega^2\frac{j}{d-2-n+j}+p^2\frac{d-4-n}{d-4-n+j}\right)\,. \label{Eqn: function k}
\end{equation}

Once \equref{Eqn: function k} is inserted into \equref{Eqn: Second term}, the second term becomes
\begin{align}
2^{\text{nd}} \text{ Term}={}&{}
4(p_\mu p^\mu)^2\Bigg(\sum\limits_{dnjm}\omega^{d-4-n}p^n\prescript{}{0}{Y}_{jm}(\mathbf{\hat{p}})\left((\vok_{AF}^{(d)})^{(1B)}_{njm}\frac{d}{(n+2)(n+4)}\right)\nonumber\\{}&{}+\omega^{d-5-n}p^{n-1}\prescript{}{0}{Y}_{jm}(\mathbf{\hat{p}})(\vok_{AF}^{(d)})^{(0B)}_{njm}\nonumber\\{}&{}\times\left(\omega^2\frac{j}{d-2-n+j}+p^2\frac{d-4-n}{d-4-n+j}\right)\Bigg)^2\,.\label{eq: Second term}
\end{align}

This was the compact form which was sought in the first place. With that, the most general dispersion relation of the CPT-odd vacuum orthogonal nonrenormalizable photon, \equref{General Vacuum-orthogonal dispersion relation}, can be cast into form 
\begin{equation}
\boxed{0={}(p_\mu p^\mu)\times\left((p_\mu p^\mu)\mathcal{P}(\omega,p)+\mathcal{Q}(\omega,p)\right)}\label{General Vacuum-orthogonal dispersion relation2}
\end{equation}
where $\mathcal{P}$ and $\mathcal{Q}$ are defined as
\begin{subequations}
\begin{align}
\mathcal{P}(\omega,p):={}&{} 1-4\Bigg(\sum\limits_{dnjm}\omega^{d-4-n}p^n\prescript{}{0}{Y}_{jm}(\mathbf{\hat{p}})\left((\vok_{AF}^{(d)})^{(1B)}_{njm}\frac{d}{(n+2)(n+4)}\right)\nonumber\\{}&{}+\omega^{d-5-n}p^{n-1}\prescript{}{0}{Y}_{jm}(\mathbf{\hat{p}})(\vok_{AF}^{(d)})^{(0B)}_{njm}\nonumber\\{}&{}\times\left(\omega^2\frac{j}{d-2-n+j}+p^2\frac{d-4-n}{d-4-n+j}\right)\Bigg)^2\,,\\
\mathcal{Q}(\omega,p):={}&{} 8\sum\limits_{d_1d_2n_1n_2j_1j_2m_1m_2}\omega^{d_1+d_2-6-n_1-n_2}p^{n_1+n_2}\prescript{}{+1}{Y}_{j_1m_1}(\mathbf{\hat{p}})\prescript{}{-1}{Y}_{j_2m_2}(\mathbf{\hat{p}})\nonumber\\
{}&{}\times\frac{1}{\sqrt{4j_1j_2(j_1+1)(j_2+1)}}\Bigg\{\Bigg(\bigg(\frac{\omega j_1(n_1+1)}{p(d_1-2-n_1+j_1)}\nonumber\\{}&{}-\frac{pj_1(n_1+3)}{\omega(d_1-4-n_1+j_1)}\bigg)(\vok_{AF}^{(d_1)})_{n_1j_1m_1}^{(0B)}+\frac{d_1}{n_1+4}(\vok_{AF}^{(d_1)})^{(1B)}_{n_1j_1m_1}\Bigg)\nonumber\\{}&{}\times\Bigg(\bigg(\frac{\omega j_2(n_2+1)}{p(d_2-2-n_2+j_2)}-\frac{pj_2(n_2+3)}{\omega(d_2-4-n_2+j_2)}\bigg)(\vok_{AF}^{(d_2)})_{n_2j_2m_2}^{(0B)}\nonumber\\{}&{}+\frac{d_2}{n_2+4}(\vok_{AF}^{(d_2)})^{(1B)}_{n_2j_2m_2}\Bigg)+(\vok_{AF}^{(d_1)})^{(1E)}_{n_1j_1m_1}(\vok_{AF}^{(d_2)})^{(1E)}_{n_2j_2m_2}\Bigg\}\,.
\end{align}\label{P and Q of dispersion relation}
\end{subequations}

The form of dispersion relation \equref{General Vacuum-orthogonal dispersion relation2} is quite suggestive: For the VOM, the LV takes a multiplicative form instead of an additive form in the dispersion relation; that is, the conventional root $p_\mu p^\mu=0$ remains as a valid root despite the violation of the Lorentz symmetry.

In the general model, we showed that conventional dispersion relation indeed raises for certain coefficients, whose combination comprises the so-defined coefficient subspace $\hat{k}_{AF}^{(cn)}$. However, there arises other cases, as shown in the \tabref{Table-cscops}, in which the conventional dispersion relation does not hold. This seems to be in contrast to what we find here; after all, VOM is only the special case of the general CPT-odd photon, hence that conventional solution exists for all possible coefficients in the VOM is possible if the only physically relevant coefficient subspace is $\hat{k}_{AF}^{(cn)}$; in other words, if $\hat{k}_{AF}^{(bf)}$ is no longer a physical subspace for the VOM\footnotemark\footnotetext{$\hat{k}_{AF}^{(np)}$ is already a physically irrelevant case in the general model, and therefore is so in the vacuum orthogonal one as well.}. But this is precisely the case as we will demonstrate now.

Let the focus be restricted to the physical solutions, for which that $\hat{k}_{AF}\in\hat{k}_{AF}^{(np)}$ can be discarded. This translates into the restriction that $(\hat{k}_{AF})_\pm=0$, which turns into the constraint $\mathcal{Q}(\omega,p)=0$ for the vacuum orthogonal subspace as can be seen by comparing \equref{General CPT-odd Spherical Dispersion Relation}, \equref{General Vacuum-orthogonal dispersion relation} and \equref{P and Q of dispersion relation}. Then \equref{General Vacuum-orthogonal dispersion relation2} becomes
\begin{equation}
0={}\left(p_\mu p^\mu\right)^2\left(1+\mathcal{R}(\omega,p)\right)\left(1-\mathcal{R}(\omega,p)\right)\,,\label{General Vacuum-orthogonal dispersion relation3}
\end{equation}
where $\mathcal{R}$ is defined as
\begin{equation}
\begin{aligned}
\mathcal{R}(\omega,p):={}&{} 2\sum\limits_{dnjm}\Bigg(\omega^{d-4-n}p^n\prescript{}{0}{Y}_{jm}(\mathbf{\hat{p}})\left((\vok_{AF}^{(d)})^{(1B)}_{njm}\frac{d}{(n+2)(n+4)}\right)\\{}&{}+\omega^{d-5-n}p^{n-1}\prescript{}{0}{Y}_{jm}(\mathbf{\hat{p}})(\vok_{AF}^{(d)})^{(0B)}_{njm}\\{}&{}\times\left(\omega^2\frac{j}{d-2-n+j}+p^2\frac{d-4-n}{d-4-n+j}\right)\Bigg)\,.
\end{aligned}\label{definition of R}
\end{equation}

The dispersion relation in \equref{General Vacuum-orthogonal dispersion relation3} has three roots: $\omega=p$ and $\mathcal{R}(\omega,p)\pm1=0$. Now, it is claimed that $\omega=p$ is the dispersion relation which is associated with $\hat{k}_{AF}^{(cn)}$, and $\mathcal{R}(\omega,p)\pm1=0$ are dispersion relations which are associated with $\hat{k}_{AF}^{(bf)}$. Then, the coefficient subspace $\hat{k}_{AF}^{(bf)}$ indeed becomes nonphysical as those dispersion relations are not acceptable dispersion relations. But before we go into that, let us first prove that they are indeed the dispersion relations associated with $\hat{k}_{AF}^{(bf)}$.

Actually, without any calculation, one can straightforwardly relate $\mathcal{R}(\omega,p)\pm1=0$ to $\hat{k}_{AF}^{(bf)}$ from \tabref{Table-cscops} by the fact that $\mathcal{R}(\omega,p)\pm1=0$ are birefringent solutions which can be raised only in $\hat{k}_{AF}^{(bf)}$. Yet, let us carry out the details. From \tabref{Table-cscops} and \equref{definition of k_s}, it is clear that one needs to calculate $p(\hat{k}_{AF})_0-\omega(\hat{k}_{AF})_r$ in the VOM. Judging from \equref{eq: Second term} then
\begin{align*}
p(\hat{k}_{AF})_0-\omega(\hat{k}_{AF})_r=(p_\mu p^\mu)\Bigg(\sum\limits_{dnjm}\omega^{d-4-n}p^n\prescript{}{0}{Y}_{jm}(\mathbf{\hat{p}})\left((\vok_{AF}^{(d)})^{(1B)}_{njm}\frac{d}{(n+2)(n+4)}\right)\\+\omega^{d-5-n}p^{n-1}\prescript{}{0}{Y}_{jm}(\mathbf{\hat{p}})(\vok_{AF}^{(d)})^{(0B)}_{njm}\left(\omega^2\frac{j}{d-2-n+j}+p^2\frac{d-4-n}{d-4-n+j}\right)\Bigg)\,,
\end{align*}

hence,
\begin{equation}
p(\hat{k}_{AF})_s={}\frac{1}{2}(p_\mu p^\mu)\mathcal{R}(\omega,p)\label{eq: k_s in vom}
\end{equation}
from \equref{definition of R}. But this simply means that the dispersion relation of $\hat{k}_{AF}^{(bf)}$ are
\begin{equation*}
\omega^2-p^2\pm 2p(\hat{k}_{AF})_s=0\quad \longrightarrow\quad (p_\mu p^\mu)\left(1\pm\mathcal{R}(\omega,p)\right)=0\,.
\end{equation*}

From this equation, $p_\mu p^\mu=0$ looks like a root but is actually not. This is due to the fact that $\omega=p$ forces $(\hat{k}_{AF})_s=0$, which is $(\hat{k}_{AF})_r-(\hat{k}_{AF})_0=0$ under $\omega=p$. But this contradicts with the defining constraint of $\hat{k}_{AF}^{(bf)}$; thus $p_\mu p^\mu\ne0$, meaning that \mbox{$\mathcal{R}(\omega,p)\pm1=0$} are the dispersion relations of $\hat{k}_{AF}^{(bf)}$, which is exactly the earlier claim.

Now, as promised, let us go into the details of these dispersion relations. Without any calculation, it is clear that \mbox{$\mathcal{R}(\omega,p)\pm1=0$} cannot be satisfied for finite $\omega$ and $p$ if we turn off the LV. That is, for finite energy and momentum, \equref{definition of R} dictates that $\mathcal{R}(\omega,p)\rightarrow0$ as  $\{(\vok_{AF}^{(d)})^{(0B)}_{njm},(\vok_{AF}^{(d)})^{(1B)}_{njm}\}\rightarrow\{0,0\}$, which contradicts with the dispersion relation. Actually, as it will be explicitly demonstrated in specific models in \charef{CHAPTER:SPECIAL MODEL}, this implicit dispersion relation can be converted into an explicit dispersion relation only if $\omega$ has an expansion in which one of the terms has a LV coefficient in the denominator. These kind of solutions are called \emph{spurious solutions} because they blow up in the no LV limit. They are shown to be the artifacts of the fundamental theory in the low energy regimes that are simply to be ignored. The physically interested solutions on the other hand are supposed smoothly to reduce to the conventional solution in the no LV limit, and hence are named \emph{perturbative solutions}.

That birefringent coefficient subspace is no longer physical in VOM has an interesting consequence aside the fact that it explains how LV can take a multiplicative form in the dispersion relation without raising contradictions. By the very construction of VOM, the solutions do not have leading order birefringence effects; however, whether they have higher order birefringence effects or not is not generally explored in the literature. For CPT-odd case on the other hand, as it is demonstrated above, which is based on \cite{Albayrak}, the VOM does not produce any physical birefringent solution whatsoever.

Let us wrap it up. In \charef{CHAPTER:CPT ODD PHOTON}, it was demonstrated that the coefficient space of {\it nm}SME CPT-odd photon sector can be divided into three subspaces: $\hat{k}_{AF}^{(bf)}$, $\hat{k}_{AF}^{(cn)}$, and $\hat{k}_{AF}^{(np)}$. The last one is physically irrelevant and is listed only for completeness, whereas the first two ones produce birefringent and conventional solutions respectively. In the general VOM on the other hand, $\hat{k}_{AF}^{(bf)}$ produces spurious solutions only, hence becomes physically irrelevant as well. Therefore, the only physical solutions that ever arise in the general VOM are the conventional solutions; but this also means that solutions are nonbirefringent at all orders. The situations is summarized as \emph{the vacuum orthogonal model is vacuum orthogonal at all orders}, and \emph{all polarization vectors and their dispersion relations remain conventional in vacuum orthogonal model}.

\section{The Coefficient Subspace}
\label{vom coefficient}
According to \tabref{Table-cscops}, the coefficient space of the VOM, that is $\hat{k}_{AF}^{(cn)}$, satisfies the restrictions $(\hat{k}_{AF})_\pm=0$ and $(\hat{k}_{AF})_r=(\hat{k}_{AF})_0$. In this section, we will convert these restrictions into the language of vacuum orthogonal coefficients. In \secref{Section: cpt odd spherical decomposition}, the spherical decomposition of these components were given as
\begin{equation*}
\begin{aligned}
(\hat{k}_{AF})_0 {}&{}=\sum\limits_{dnjm}\omega^{d-3-n}p^n\prescript{}{0}{Y}_{jm}(\mathbf{\hat{p}})(k_{AF}^{(d)})^{(0B)}_{njm}\,,\\
(\hat{k}_{AF})_r {}&{}=\sum\limits_{dnjm}\omega^{d-3-n}p^n\prescript{}{0}{Y}_{jm}(\mathbf{\hat{p}})\frac{-1}{n+2}\left((k_{AF}^{(d)})^{(1B)}_{njm}+(d-2-n)(k_{AF}^{(d)})^{(0B)}_{(n-1)jm}\right)\,,\\
(\hat{k}_{AF})_\pm {}&{}=\sum\limits_{dnjm}\omega^{d-3-n}p^n\prescript{}{\pm 1}{Y}_{jm}(\mathbf{\hat{p}})\frac{1}{\sqrt{2j(j+1)}}\left(\pm(k_{AF}^{(d)})^{(1B)}_{njm}+i(\vok_{AF}^{(d)})^{(1E)}_{njm}\right)\,,
\end{aligned}
\end{equation*}
in \equref{Equ: k_i}. Additionally, in the beginning of this chapter, the prescription to limit the focus to the vacuum orthogonal coefficients is given as well:
\begin{align*}
(k_{AF}^{(d)})^{(0B)}_{njm}={}&{}\frac{(d-2-n)(n+3)}{d(d-2-n+j)}\left((\vok_{AF}^{(d)})^{(0B)}_{njm}-(\vok_{AF}^{(d)})^{(0B)}_{(n-2)jm}\right)\nonumber\\{}&{}-\frac{1}{n+1}(\vok_{AF}^{(d)})^{(1B)}_{(n-1)jm}\,,\nonumber\\\nonumber
(k_{AF}^{(d)})^{(1B)}_{njm}={}&{}\frac{j(n+2)}{d-3-n+j}\left((\vok_{AF}^{(d)})^{(0B)}_{(n+1)jm}-(\vok_{AF}^{(d)})^{(0B)}_{(n-1)jm}\right)\\{}&{}+\frac{d}{n+4}(\vok_{AF}^{(d)})^{(1B)}_{njm}\,,
\end{align*} 
which can be seen in \equref{Eq:general to vacuum orthogonal}. Then, the helicity components of $\hat{k}_{AF}$ become
\begin{subequations}
\begin{align}
(\hat{k}_{AF})_0={}&{}\sum\limits_{dnjm}\omega^{d-3-n}p^n\prescript{}{0}{Y}_{jm}(\hat{\mathbf{p}})\frac{(d-2-n)(n+3)}{d(d-2-n+j)}\nonumber\\{}&{}\times\left(\left((\vok_{AF}^{(d)})^{(0B)}_{njm}-(\vok_{AF}^{(d)})^{(0B)}_{(n-2)jm}\right)-\frac{1}{n+1}(\vok_{AF}^{(d)})^{(1B)}_{(n-1)jm}\right)\,,\\
(\hat{k}_{AF})_r={}&{}\sum\limits_{dnjm}\omega^{d-3-n}p^n\prescript{}{0}{Y}_{jm}(\hat{\mathbf{p}})\bigg(\frac{j}{d-3-n+j}\Big((\vok_{AF}^{(d)})^{(0B)}_{(n+1)jm}\nonumber\\{}&{}-(\vok_{AF}^{(d)})^{(0B)}_{(n-1)jm}\Big)+\frac{d}{(n+4)(n+2)}(\vok_{AF}^{(d)})^{(1B)}_{njm}\nonumber\\{}&{}+\frac{(d-1-n)(d-2-n)}{d(d-1-n+j)}\Big((\vok_{AF}^{(d)})^{(0B)}_{(n-1)jm}-(\vok_{AF}^{(d)})^{(0B)}_{(n-3)jm}\Big)\nonumber\\{}&{}-\frac{1}{n}(\vok_{AF}^{(d)})^{(1B)}_{(n-2)jm}\bigg)\,,\\
(\hat{k}_{AF})_\pm={}&{}\sum\limits_{dnjm}\omega^{d-3-n}p^n\prescript{}{\pm 1}{Y}_{jm}(\hat{\mathbf{p}})\frac{1}{\sqrt{2j(j+1)}}\bigg(\pm\frac{j(n+2)}{d-3-n+j}\nonumber\\{}&{}\times\left((\vok_{AF}^{(d)})^{(0B)}_{(n+1)jm}-(\vok_{AF}^{(d)})^{(0B)}_{(n-1)jm}\right)\pm\frac{d}{n+4}(\vok_{AF}^{(d)})^{(1B)}_{njm}+i(\vok_{AF}^{(d)})^{(1E)}_{njm}\bigg)\,.
\end{align}\label{Vacuum orthogonal k's}
\end{subequations}

\subsection*{Analysis of $\mathbf{(\hat{k}_{AF})_0-(\hat{k}_{AF})_r=0}$ under the dispersion relation $\mathbf{\omega=p}$}
The analysis of $(\hat{k}_{AF})_0-(\hat{k}_{AF})_r=0$ is actually a straightforward calculation.\\\equref{Vacuum orthogonal k's} indicates that
\begin{align*}
(\hat{k}_{AF})_0-(\hat{k}_{AF})_r={}&{}\sum\limits_{dnjm}p^{d-3}\prescript{}{0}{Y}_{jm}(\hat{\mathbf{p}})\Bigg(\frac{(d-2-n)(n+3)}{d(d-2-n+j)}\\{}&{}\times\left(\left((\vok_{AF}^{(d)})^{(0B)}_{njm}-(\vok_{AF}^{(d)})^{(0B)}_{(n-2)jm}\right)-\frac{1}{n+1}(\vok_{AF}^{(d)})^{(1B)}_{(n-1)jm}\right)\\
{}&{}-\bigg(\frac{j}{d-3-n+j}\left((\vok_{AF}^{(d)})^{(0B)}_{(n+1)jm}-(\vok_{AF}^{(d)})^{(0B)}_{(n-1)jm}\right)\\
{}&{}+\frac{d}{(n+4)(n+2)}(\vok_{AF}^{(d)})^{(1B)}_{njm}+\frac{(d-1-n)(d-2-n)}{d(d-1-n+j)}\\
{}&{}\times\left((\vok_{AF}^{(d)})^{(0B)}_{(n-1)jm}-(\vok_{AF}^{(d)})^{(0B)}_{(n-3)jm}\right)-\frac{1}{n}(\vok_{AF}^{(d)})^{(1B)}_{(n-2)jm}\bigg)\Bigg)\\={}&{}0\,,
\end{align*}
where $\omega=p$ dispersion relation is imposed as there is no other dispersion relation valid in the VOM, which was discussed at the end of \secref{vom dispersion}. This equation can be rewritten as
\begin{align*}
0={}&{}\sum\limits_{d}p^{d-3}\sum\limits_{njm}\prescript{}{0}{Y}_{jm}(\hat{\mathbf{p}})\Bigg(\frac{(d-2-n)(n+3)}{d(d-2-n+j)}\bigg(\left((\vok_{AF}^{(d)})^{(0B)}_{njm}-(\vok_{AF}^{(d)})^{(0B)}_{(n-2)jm}\right)\\{}&{}-\frac{1}{n+1}(\vok_{AF}^{(d)})^{(1B)}_{(n-1)jm}\bigg)
-\bigg(\frac{j}{d-3-n+j}\left((\vok_{AF}^{(d)})^{(0B)}_{(n+1)jm}-(\vok_{AF}^{(d)})^{(0B)}_{(n-1)jm}\right)\\{}&{}+\frac{d}{(n+4)(n+2)}(\vok_{AF}^{(d)})^{(1B)}_{njm}
+\frac{(d-1-n)(d-2-n)}{d(d-1-n+j)}\\{}&{}\times\left((\vok_{AF}^{(d)})^{(0B)}_{(n-1)jm}-(\vok_{AF}^{(d)})^{(0B)}_{(n-3)jm}\right)-\frac{1}{n}(\vok_{AF}^{(d)})^{(1B)}_{(n-2)jm}\bigg)\Bigg)\,.
\end{align*}

As magnitude of the photon momentum $p$ is a free variable, above equation holds if
\begin{align*}
0={}&{}\sum\limits_{njm}\prescript{}{0}{Y}_{jm}(\hat{\mathbf{p}})\Bigg(\frac{(d-2-n)(n+3)}{d(d-2-n+j)}\bigg(\left((\vok_{AF}^{(d)})^{(0B)}_{njm}-(\vok_{AF}^{(d)})^{(0B)}_{(n-2)jm}\right)\\{}&{}-\frac{1}{n+1}(\vok_{AF}^{(d)})^{(1B)}_{(n-1)jm}\bigg)
-\bigg(\frac{j}{d-3-n+j}\left((\vok_{AF}^{(d)})^{(0B)}_{(n+1)jm}-(\vok_{AF}^{(d)})^{(0B)}_{(n-1)jm}\right)\\{}&{}+\frac{d}{(n+4)(n+2)}(\vok_{AF}^{(d)})^{(1B)}_{njm}
+\frac{(d-1-n)(d-2-n)}{d(d-1-n+j)}\\{}&{}\times\left((\vok_{AF}^{(d)})^{(0B)}_{(n-1)jm}-(\vok_{AF}^{(d)})^{(0B)}_{(n-3)jm}\right)-\frac{1}{n}(\vok_{AF}^{(d)})^{(1B)}_{(n-2)jm}\bigg)\Bigg)
\end{align*}
for each dimension $d$. In addition, that $\prescript{}{0}{Y}_{jm}(\hat{\mathbf{p}})$ are orthogonal functions for different $j,m$ values dictates that the multiplier of $\prescript{}{0}{Y}_{jm}(\hat{\mathbf{p}})$ should be itself zero for each $j,m$ values; that is,
\begin{align*}
0={}&{}\sum\limits_{n}\Bigg(\frac{(d-2-n)(n+3)}{d(d-2-n+j)}\bigg(\left((\vok_{AF}^{(d)})^{(0B)}_{njm}-(\vok_{AF}^{(d)})^{(0B)}_{(n-2)jm}\right)\\
{}&{}-\frac{1}{n+1}(\vok_{AF}^{(d)})^{(1B)}_{(n-1)jm}\bigg)-\bigg(\frac{j}{d-3-n+j}\left((\vok_{AF}^{(d)})^{(0B)}_{(n+1)jm}-(\vok_{AF}^{(d)})^{(0B)}_{(n-1)jm}\right)\\
{}&{}+\frac{d}{(n+4)(n+2)}(\vok_{AF}^{(d)})^{(1B)}_{njm}+\frac{(d-1-n)(d-2-n)}{d(d-1-n+j)}\\
{}&{}\times\left((\vok_{AF}^{(d)})^{(0B)}_{(n-1)jm}-(\vok_{AF}^{(d)})^{(0B)}_{(n-3)jm}\right)-\frac{1}{n}(\vok_{AF}^{(d)})^{(1B)}_{(n-2)jm}\bigg)\Bigg)\,.
\end{align*}

Let us try to regroup terms:
\begin{align*}
0={}&{}\sum\limits_{n}\frac{(d-2-n)(n+3)}{d(d-2-n+j)}(\vok_{AF}^{(d)})^{(0B)}_{njm}-\sum\limits_{n}\frac{(d-2-n)(n+3)}{d(d-2-n+j)}(\vok_{AF}^{(d)})^{(0B)}_{(n-2)jm}\\
{}&{}-\sum\limits_{n}\frac{(d-2-n)(n+3)}{d(d-2-n+j)(n+1)}(\vok_{AF}^{(d)})^{(1B)}_{(n-1)jm}-\sum\limits_{n}\frac{j}{d-3-n+j}(\vok_{AF}^{(d)})^{(0B)}_{(n+1)jm}\\
{}&{}+\sum\limits_{n}\frac{j}{d-3-n+j}(\vok_{AF}^{(d)})^{(0B)}_{(n-1)jm}-\sum\limits_{n}\frac{d}{(n+4)(n+2)}(\vok_{AF}^{(d)})^{(1B)}_{njm}\\
{}&{}-\sum\limits_{n}\frac{(d-1-n)(d-2-n)}{d(d-1-n+j)}(\vok_{AF}^{(d)})^{(0B)}_{(n-1)jm}\\
{}&{}+\sum\limits_{n}\frac{(d-1-n)(d-2-n)}{d(d-1-n+j)}(\vok_{AF}^{(d)})^{(0B)}_{(n-3)jm}+\sum\limits_{n}\frac{1}{n}(\vok_{AF}^{(d)})^{(1B)}_{(n-2)jm}\,.
\end{align*}
Since the summations are over all possible values, we can shift the parameter n as we want. Then,
\begin{align*}
0={}&{}\sum\limits_{n}\frac{(d-2-n)(n+3)}{d(d-2-n+j)}(\vok_{AF}^{(d)})^{(0B)}_{njm}-\sum\limits_{n}\frac{(d-4-n)(n+5)}{d(d-4-n+j)}(\vok_{AF}^{(d)})^{(0B)}_{njm}\\
{}&{}-\sum\limits_{n}\frac{(d-3-n)(n+4)}{d(d-3-n+j)(n+2)}(\vok_{AF}^{(d)})^{(1B)}_{njm}-\sum\limits_{n}\frac{j}{d-2-n+j}(\vok_{AF}^{(d)})^{(0B)}_{njm}\\
{}&{}+\sum\limits_{n}\frac{j}{d-4-n+j}(\vok_{AF}^{(d)})^{(0B)}_{njm}-\sum\limits_{n}\frac{d}{(n+4)(n+2)}(\vok_{AF}^{(d)})^{(1B)}_{njm}\\
{}&{}-\sum\limits_{n}\frac{(d-2-n)(d-3-n)}{d(d-2-n+j)}(\vok_{AF}^{(d)})^{(0B)}_{njm}\\
{}&{}+\sum\limits_{n}\frac{(d-4-n)(d-5-n)}{d(d-4-n+j)}(\vok_{AF}^{(d)})^{(0B)}_{njm}+\sum\limits_{n}\frac{1}{n+2}(\vok_{AF}^{(d)})^{(1B)}_{njm}\,,
\end{align*}
hence:
\begin{align*}
0={}&{}\sum\limits_{n}(\vok_{AF}^{(d)})^{(0B)}_{njm}\bigg(\frac{(d-2-n)(n+3)}{d(d-2-n+j)}-\frac{(d-4-n)(n+5)}{d(d-4-n+j)}\\
{}&{}-\frac{j}{d-2-n+j}+\frac{j}{d-4-n+j}-\frac{(d-2-n)(d-3-n)}{d(d-2-n+j)}\\
{}&{}+\frac{(d-4-n)(d-5-n)}{d(d-4-n+j)}\bigg)+\sum\limits_{n}(\vok_{AF}^{(d)})^{(1B)}_{njm}\bigg(\frac{1}{n+2}-\frac{d}{(n+4)(n+2)}\\
{}&{}-\frac{(d-3-n)(n+4)}{d(d-3-n+j)(n+2)}\bigg)\,.
\end{align*}

Let us simplify this a little bit using the equality
\begin{align*}
{}&{}\frac{(d-2-n)(n+3)}{d(d-2-n+j)}-\frac{(d-4-n)(n+5)}{d(d-4-n+j)}-\frac{j}{d-2-n+j}+\frac{j}{d-4-n+j}\\
{}&{}-\frac{(d-2-n)(d-3-n)}{d(d-2-n+j)}+\frac{(d-4-n)(d-5-n)}{d(d-4-n+j)}\\{}&{}=-\frac{4}{d}+\frac{4j(d+1+j)}{d(d-2-n+j)(d-4-n+j)}\,.
\end{align*}
Then, the first condition becomes as follows.
\begin{equation}
\boxed{\begin{aligned}
0={}&{}\sum\limits_{n}(\vok_{AF}^{(d)})^{(0B)}_{njm}\left(-\frac{4}{d}+\frac{4j(d+1+j)}{d(d-2-n+j)(d-4-n+j)}\right)\\
{}&{}+\sum\limits_{n}(\vok_{AF}^{(d)})^{(1B)}_{njm}\left(\frac{1}{n+2}-\frac{d}{(n+4)(n+2)}-\frac{(d-3-n)(n+4)}{d(d-3-n+j)(n+2)}\right)
\end{aligned}}\label{first condition for vacuum polarization}
\end{equation}
\subsection*{Analysis of $\mathbf{(\hat{k}_{AF})_\pm=0}$ under the dispersion relation $\mathbf{\omega=p}$}
It is straightforward that the conditions $(\hat{k}_{AF})_\pm=0$ are
\begin{align*}
(\hat{k}_{AF})_\pm={}&{}\sum\limits_{dnjm}p^{d-3}\prescript{}{\pm 1}{Y}_{jm}(\hat{\mathbf{p}})\frac{1}{\sqrt{2j(j+1)}}\bigg(\pm\frac{j(n+2)}{d-3-n+j}\\{}&{}\times\left((\vok_{AF}^{(d)})^{(0B)}_{(n+1)jm}-(\vok_{AF}^{(d)})^{(0B)}_{(n-1)jm}\right)\pm\frac{d}{n+4}(\vok_{AF}^{(d)})^{(1B)}_{njm}+i(\vok_{AF}^{(d)})^{(1E)}_{njm}\bigg)
\end{align*}
under the dispersion relation $\omega=p$ from \equref{Vacuum orthogonal k's}. Not unlike the earlier case, that the magnitude of momentum $p$ is a free variable and that spin weighted spherical harmonics $\prescript{}{\pm1}{Y}_{jm}(\hat{\mathbf{p}})$ are orthogonal to each other for different values of $j,m$ can be invoked, resulting in the simplified conditions
\begin{align*}
0={}&{}\sum\limits_{n}\frac{1}{\sqrt{2j(j+1)}}\bigg(\pm\frac{j(n+2)}{d-3-n+j}\left((\vok_{AF}^{(d)})^{(0B)}_{(n+1)jm}-(\vok_{AF}^{(d)})^{(0B)}_{(n-1)jm}\right)\\{}&{}\pm\frac{d}{n+4}(\vok_{AF}^{(d)})^{(1B)}_{njm}+i(\vok_{AF}^{(d)})^{(1E)}_{njm}\bigg)\,,
\end{align*}
which should hold for each possible value of $d,j,m$. But this equation too can be further simplified:
\begin{align*}
0={}&{}\pm\sum\limits_{n}\frac{j(n+2)}{\sqrt{2j(j+1)}(d-3-n+j)}(\vok_{AF}^{(d)})^{(0B)}_{(n+1)jm}\\{}&{}\mp\sum\limits_{n}\frac{j(n+2)}{\sqrt{2j(j+1)}(d-3-n+j)}(\vok_{AF}^{(d)})^{(0B)}_{(n-1)jm}\\{}&{}+\sum\limits_{n}\frac{1}{\sqrt{2j(j+1)}}\left(\pm\frac{d}{n+4}(\vok_{AF}^{(d)})^{(1B)}_{njm}+i(\vok_{AF}^{(d)})^{(1E)}_{njm}\right)\,.
\end{align*}
Since the summations are over all possible values, we can shift the parameter $n$ as we desire. Then,
\begin{align*}
0={}&{}\pm\sum\limits_{n}\frac{j(n+1)}{\sqrt{2j(j+1)}(d-2-n+j)}(\vok_{AF}^{(d)})^{(0B)}_{njm}\mp\sum\limits_{n}\frac{j(n+3)}{\sqrt{2j(j+1)}(d-4-n+j)}\\{}&{}\times(\vok_{AF}^{(d)})^{(0B)}_{njm}+\sum\limits_{n}\frac{1}{\sqrt{2j(j+1)}}\left(\pm\frac{d}{n+4}(\vok_{AF}^{(d)})^{(1B)}_{njm}+i(\vok_{AF}^{(d)})^{(1E)}_{njm}\right)\,,
\end{align*}
which is
\begin{align}
0={}&{}\frac{1}{\sqrt{2j(j+1)}}\Bigg\{\pm\sum\limits_{n}\Bigg((\vok_{AF}^{(d)})^{(0B)}_{njm}\left(\frac{j(n+1)}{(d-2-n+j)}-\frac{j(n+3)}{(d-4-n+j)}\right)\nonumber\\
{}&{}+\frac{d}{n+4}(\vok_{AF}^{(d)})^{(1B)}_{njm}\Bigg)+i\sum\limits_{n}(\vok_{AF}^{(d)})^{(1E)}_{njm}\Bigg\}\label{Eq: k_pm zero}
\end{align}
in a more compact form.

Clearly, $\frac{1}{\sqrt{2j(j+1)}}$ does not have any root; hence, above equation can be reduced to
\begin{equation*}
\begin{aligned}
0={}&{}\pm\sum\limits_{n}\Bigg((\vok_{AF}^{(d)})^{(0B)}_{njm}\left(\frac{j(n+1)}{(d-2-n+j)}-\frac{j(n+3)}{(d-4-n+j)}\right)+\frac{d}{n+4}(\vok_{AF}^{(d)})^{(1B)}_{njm}\Bigg)\\{}&{}+i\sum\limits_{n}(\vok_{AF}^{(d)})^{(1E)}_{njm}\,,
\end{aligned}
\end{equation*}
where the finiteness of $\frac{1}{\sqrt{2j(j+1)}}$ is explicitly exploited. This is not that trivial at the first glance though, as $j=0$ value blows up this multiplier\footnotemark\footnotetext{Since $j$ is a nonnegative integer, $j+1$ is not a pole of this fraction.}. However, a careful examination reveals that \equref{Eq: k_pm zero} for $j=0$ is actually a trivial equation $0=0$, because there is no $(\vok_{AF}^{(d)})^{(1B)}_{njm}$ for $j=0$, and $(\vok_{AF}^{(d)})^{(1B)}_{njm}$ is multiplied by $\sqrt{j}$ in overall. Hence the finiteness of $\frac{1}{\sqrt{2j(j+1)}}$ is not ill-conditioned.

The form of the restrictions then reduces to
\begin{align*}
0={}&{}\sum\limits_{n}\Bigg((\vok_{AF}^{(d)})^{(0B)}_{njm}\left(\frac{j(n+1)}{(d-2-n+j)}-\frac{j(n+3)}{(d-4-n+j)}\right)+\frac{d}{n+4}(\vok_{AF}^{(d)})^{(1B)}_{njm}\Bigg)\\{}&{}+i\sum\limits_{n}(\vok_{AF}^{(d)})^{(1E)}_{njm}\,,\\
0={}&{}\sum\limits_{n}\Bigg((\vok_{AF}^{(d)})^{(0B)}_{njm}\left(\frac{j(n+1)}{(d-2-n+j)}-\frac{j(n+3)}{(d-4-n+j)}\right)+\frac{d}{n+4}(\vok_{AF}^{(d)})^{(1B)}_{njm}\Bigg)\\{}&{}-i\sum\limits_{n}(\vok_{AF}^{(d)})^{(1E)}_{njm}\,.\\
\end{align*}

These two restrictions can simultaneously hold only if both summations themselves are zero. Hence, with a simple algebraic manipulation, the second condition read as
\begin{equation}
\boxed{\begin{aligned}
\sum\limits_{n}\left(-\frac{2j(d-1+j)}{(d-2-n+j)(d-4-n+j)}(\vok_{AF}^{(d)})^{(0B)}_{njm}+\frac{d}{n+4}(\vok_{AF}^{(d)})^{(1B)}_{njm}\right)={}&{} 0\\
\sum\limits_{n}(\vok_{AF}^{(d)})^{(1E)}_{njm}={}&{} 0
\end{aligned}}\label{second condition for vacuum polarization}
\end{equation}

Therefore, any VOM has physical solutions only if \equref{first condition for vacuum polarization} and \equref{second condition for vacuum polarization} hold, and the resultant physical solutions are conventional transverse polarization vectors obeying the conventional dispersion relation $\omega=p$.

The results are summarized in \tabref{Table-vos}.

\begin{table}
\caption[Coefficient Space of Vacuum Orthogonal Model.]{Coefficient Space of Vacuum Orthogonal Model. The general situation of CPT-odd Model demonstrated in \tabref{Table-cscops} reduces to the following table when the focus is restricted to the vacuum orthogonal coefficients only. In this special model, birefringent solutions are no more physical as their dispersion relations are spurious, whereas $\hat{k}_{AF}^{(np)}$ is still nonphysical as it fails to produce non-gauge polarization vectors in VOM as well. The dispersion relations of $\hat{k}_{AF}^{(np)}$ and the polarization vectors of $\hat{k}_{AF}^{(bf)}$ are not calculated, simply because these coefficient subspaces are physically irrelevant. They are defined as the coefficient subspaces which do not obey either \equref{first condition for vacuum polarization} or \equref{second condition for vacuum polarization}, which is shown with a negation diacritic $\neg$ in front of the equation they fail to obey respectively. Finally, $\hat{k}_{AF}^{(cn)}$ remains the same in vacuum orthogonal models, with its conventional dispersion relation and conventional polarization vectors, that are one gauge plus two transverse solutions.}
\label{Table-vos}
\begin{tabularx}{\textwidth}{XXXl}
\hline\hline {\hskip -0.1in}\begin{tabular}{l}
Coefficient\\Subspace
\end{tabular} & {\hskip 0.1in}Conditions & {\hskip -0.1in}\begin{tabular}{l}
Dispersion\\Relation
\end{tabular}& \begin{tabular}{c}
Polarization Vectors $A^\mu$\end{tabular}\\\hline\\[-0.15in]
$\hat{k}_{AF}^{(cn)}$ & \begin{tabular}{l}
\equref{first condition for vacuum polarization}\\\equref{second condition for vacuum polarization}
\end{tabular}
& $\omega=p$ & $\small \left\{\begin{pmatrix}
1\\0\\1\\0
\end{pmatrix},\begin{pmatrix}
0\\1\\0\\0
\end{pmatrix},\begin{pmatrix}
0\\0\\0\\1
\end{pmatrix}\right\}$\\\\[-0.15in]
$\hat{k}_{AF}^{(bf)}$ & \begin{tabular}{l}
$\neg$\equref{first condition for vacuum polarization}\\\equref{second condition for vacuum polarization}
\end{tabular}& Spurious & Irrelevant\\\\[-0.15in]
$\hat{k}_{AF}^{(np)}$ & \begin{tabular}{l}
$\neg$\equref{second condition for vacuum polarization}
\end{tabular} & Irrelevant & Gauge only
\\\hline\hline
\end{tabularx}
\end{table}

\section{The Propagator}
\label{vom propagator}
The propagator of the general VOM can be extracted from the general CPT-odd photon propagator, \equref{eq: Propagator}, by imposing the appropriate restrictions. However, before we go into that, let us change our notation for mathematical easiness: Clearly, one can rewrite \equref{eq: Propagator} as
\begin{equation}
\hat{G}_\mu^{\;\nu}={}-\frac{\delta_\mu^{\;\nu}}{(p_\sigma p^\sigma)}+(\hat{G}_{AF})_\mu^{\;\nu} \label{general form of propagator}
\end{equation}
for brevity, where the total propagator is now the conventional propagator receiving a \emph{Lorentz violating propagator contribution}. This contribution is
\begin{equation}
\begin{aligned}
(\hat{G}_{AF})_\mu^{\;\nu}={}&{}\delta_\mu^{\;+}\delta_+^{\;\nu}\left(\frac{1}{(p_\sigma p^\sigma)}-\frac{1}{(p_\sigma p^\sigma)+2(\omega(\hat{k}_{AF})_r-p(\hat{k}_{AF})_0)}\right)\\{}&{}+\delta_\mu^{\;-}\delta_-^{\;\nu}\left(\frac{1}{(p_\sigma p^\sigma)}-\frac{1}{(p_\sigma p^\sigma)-2(\omega(\hat{k}_{AF})_r-p(\hat{k}_{AF})_0)}\right) 
\end{aligned}\label{LV contribution in the propagator}
\end{equation}
in the general model.

In \secref{vom dispersion}, it is shown that
\begin{equation*}
2(\omega(\hat{k}_{AF})_r-p(\hat{k}_{AF})_0)=(p_\mu p^\mu)\mathcal{R}(\omega,p)
\end{equation*}
in the VOM, as can be seen from \equref{eq: k_s in vom}, where $\mathcal{R}(\omega,p)$ is defined by \equref{definition of R}. Then,
\begin{align*}
(\hat{G}_{AF})_\mu^{\;\nu}={}&{}\delta_\mu^{\;+}\delta_+^{\;\nu}\left(\frac{1}{(p_\sigma p^\sigma)}-\frac{1}{(p_\sigma p^\sigma)\left(1+\mathcal{R}(\omega,p)\right)}\right)\\{}&{}+\delta_\mu^{\;-}\delta_-^{\;\nu}\left(\frac{1}{(p_\sigma p^\sigma)}-\frac{1}{(p_\sigma p^\sigma)\left(1-\mathcal{R}(\omega,p)\right)}\right)\,,
\end{align*}
hence,
\begin{equation*}
(\hat{G}_{AF})_\mu^{\;\nu}={}\frac{1}{(p_\sigma p^\sigma)}\left(\delta_\mu^{\;+}\delta_+^{\;\nu}\frac{\mathcal{R}(\omega,p)}{1+\mathcal{R}(\omega,p)}-\delta_\mu^{\;-}\delta_-^{\;\nu}\frac{\mathcal{R}(\omega,p)}{1-\mathcal{R}(\omega,p)}\right)\,.
\end{equation*}

This propagator contribution can be further simplified by noting that it contains redundant generality as the relevant coefficient space still contains $\hat{k}_{AF}^{(bf)}$. But this redundancy can be extracted by taking $(\hat{k}_{AF})_r$ to $(\hat{k}_{AF})_0$ in $\mathcal{R}(\omega,p)$. From the definition of $\mathcal{R}(\omega,p)$,
\begin{equation*}
\begin{aligned}
\mathcal{R}(\omega,p){}&{}=\frac{2p(\hat{k}_{AF})_s}{(p_\mu p^\mu)}=2\frac{p(\hat{k}_{AF})_0-\omega(\hat{k}_{AF})_r}{\omega^2-p^2}\,,\\
\lim\limits_{(\hat{k}_{AF})_r\rightarrow(\hat{k}_{AF})_0}\mathcal{R}(\omega,p){}&{}=-\frac{2(\hat{k}_{AF})_0}{\omega+p}\,.
\end{aligned}
\end{equation*}
Then,  the propagator contribution becomes
\begin{equation*}
(\hat{G}_{AF})_\mu^{\;\nu}={}-\frac{1}{(p_\sigma p^\sigma)}\left(\delta_\mu^{\;+}\delta_+^{\;\nu}\frac{2(\hat{k}_{AF})_0}{\omega+p-2(\hat{k}_{AF})_0}-\delta_\mu^{\;-}\delta_-^{\;\nu}\frac{2(\hat{k}_{AF})_0}{\omega+p+2(\hat{k}_{AF})_0}\right)\,,
\end{equation*}
from which \equref{general form of propagator} reads
\begin{equation}
\boxed{\hat{G}_\mu^{\;\nu}={}-\frac{\delta_\mu^{\;\nu}}{(p_\sigma p^\sigma)}-\frac{1}{(p_\sigma p^\sigma)}\left(\delta_\mu^{\;+}\delta_+^{\;\nu}\frac{2(\hat{k}_{AF})_0}{\omega+p-2(\hat{k}_{AF})_0}-\delta_\mu^{\;-}\delta_-^{\;\nu}\frac{2(\hat{k}_{AF})_0}{\omega+p+2(\hat{k}_{AF})_0}\right)}\label{eq: Physical vacuum orthogonal propagator}
\end{equation}
\chapter{Special Model Analysis}
\label{CHAPTER:SPECIAL MODEL}
In the earlier chapters, the model at hand contained a number of different LV coefficients at generic dimensions. In practice, however, it is quite formidable to work with infinitely many coefficients, and one usually examine a LV model with a handful of nonzero coefficients so as to avoid cumbersome calculations.

Special models, in which only some coefficients belonging to a particular category are taken to be nonzero, are introduced in \secref{Section: photon sector special models}. However, in any particular analysis, one usually goes further and simplify the LVT even more in the chosen special model. Actually, the drastic simplification of an analysis is the case in which only one coefficient at a time is taken to be nonzero and is analyzed by itself. This procedure, which is also called \emph{Kostelecký's Cutlass} \cite{Tasson:2014dfa}, is a general principle which both simplifies the analysis and enables extraction of corresponding properties of individual coefficients.

Even though the principle of Kostelecký's Cutlass is promoted above, we will not apply here in this chapter, but instead examine our general special model, that is the vacuum orthogonal model developed in \charef{vacuum orthogonal model}, with simplifications via different limits: Firstly, its isotropic limit will be considered where all non-isotropic LV coefficients are taken to be zero; and secondly, its leading order dimension will be analyzed in which only $d=5$ LV coefficients are allowed in the model.
\section{Vacuum Orthogonal and Isotropic Model at All Orders}
\label{Section: isotropic vom}
\subsection{Derivation of the General Model}
\label{Section: derivation of isotropic vom}
In \secref{Section: photon sector special models}, the isotropic model was introduced as a special model in which all LV coefficients that are not rotationally invariant are taken to be zero in a preferred frame. The model's mathematical structure is quite simplified compared to the general case, which is why it is also referred as ``fried-chicken" model, emphasizing that it is quite common yet not the whole story\footnotemark\footnotetext{The analogy is referring to the fact that one would not be able to get all necessary nutrients just by feeding on fried chicken even though it is easy and available.}. However, what is considered here is not the general isotropic model, but instead the isotropic limit of VOM, in other words, hybrid model of isotropic and vacuum orthogonal models.

As stated earlier, the isotropic nature of the coefficients are valid only in the preferred frame, as LVT are Lorentz tensors under OLT; hence, their nature will change under observer boosts and observer rotations\footnotemark\footnotetext{What is meant here is that there will be nonzero $j\ne0$ coefficients in the boosted frame even though all coefficients with $j\ne0$ are zero in the preferred frame.}. Therefore, one should always choose a definite observer frame before starting the specific analysis\footnotemark\footnotetext{Actually, this is true for all LV analyses. Since LVT mix up under OLT, it is always advisable to fix the observer frame, and then choose the LVT that will be analyzed in that frame, and finally search the effects of these coefficients only on that frame to avoid any possible confusion. In this thesis, there is no specific LVT analysis which is conducted to find some particular effects of some chosen coefficients and then to bound them by comparing with the experimental data. Instead, the generic coefficient spaces are analyzed, and general results such as that vacuum orthogonal models remain vacuum orthogonal at all orders are obtained. That is why no specific reference frame is, and will be, chosen although it is strictly stressed that the choice of reference frame is vital.}. Again, as stated earlier, \emph{Cosmic Microwave Background} frame is a natural choice for its theoretical relevance to the notion of an isotropic frame, yet it is much more practical for the purpose of bounding the coefficients if sun-centered canonical frame is chosen as the isotropic frame due to its direct relevance to experimental results.

Let us analyze the dispersion relation of the general VOM, \equref{General Vacuum-orthogonal dispersion relation2}, under the isotropic limit then. This is a wise limit for several reasons. Firstly, as mentioned above, isotropic models are somewhat popular, hence such an analysis would be beneficial for the literature. Secondly, the assertion that is made in \secref{vom dispersion}, which states that dispersion relation of VOM does not have a nonconventional root, can be more transparently seen in this limit, as we will se below. Finally, the transformation from the general case to the isotropic one is quite trivial, hence is mathematically advantageous. Actually, we simply impose $j=m=0$ in the coefficients and the dispersion relation. From \tabref{Table-vocir}, the only surviving coefficients are determined to be $(\vok_{AF}^{(d)})^{(0B)}_{n00}$; if we insert this to the $\mathcal{P}$ and $\mathcal{Q}$, given by \equref{P and Q of dispersion relation}, and also impose $j=0$, \equref{General Vacuum-orthogonal dispersion relation2} becomes
\begin{equation}
\boxed{0=\left(1-\frac{p^2}{\pi}\left(\sum\limits_{d=\text{odd}>3}\sum\limits_{n=\text{even}\ge 0}^{d-5}\omega^{d-5-n}p^n(\vok_{AF}^{(d)})^{(0B)}_{n00}\right)^2\right)(p_\mu p^\mu)^2} \label{Compact form}
\end{equation}
where we used
\begin{equation*}
\prescript{}{0}{Y}_{00}(\mathbf{\hat{p}})=\frac{1}{\sqrt{4\pi}}\,,
\end{equation*}
which is the well known $\mathbf{Y}_{0}^{0}$ as spin weighted spherical harmonics with the spin weight $s=0$ are simply the usual spherical harmonics.

The simple derivation above is addition of isotropy restriction on already derived general VOM dispersion relation. In general, however, one can derive the hybrid isotropic VOM from the generic CPT-odd dispersion relation \equref{General CPT-odd Spherical Dispersion Relation} as well.

To do this, one would need the spherical expansion of the helicity components $(\hat{k}_{AF})_i$ of the operator $\hat{k}_{AF}$ in terms of the only relevant LV coefficients, that are $(k_{AF}^{(d)})^{(0B)}_{n00}$. But this expansion can be obtained the general one, given by \equref{Equ: k_i}, with the imposition $j=m=0$, hence
\begin{equation*}
\begin{aligned}
(\hat{k}_{AF})_0 {}&{}=\frac{1}{\sqrt{4\pi}}\sum\limits_{dn}\omega^{d-3-n}p^n(k_{AF}^{(d)})^{(0B)}_{n00}\,,\\
(\hat{k}_{AF})_r {}&{}=\frac{1}{\sqrt{4\pi}}\sum\limits_{dn}\omega^{d-3-n}p^n\frac{2+n-d}{n+2}(k_{AF}^{(d)})^{(0B)}_{(n-1)00}\,,\\
(\hat{k}_{AF})_\pm {}&{}=0\,.
\end{aligned}
\end{equation*}

Moreover, one needs the prescription to convert $(k_{AF}^{(d)})^{(0B)}_{n00}$ to $(\vok_{AF}^{(d)})^{(0B)}_{n00}$. This prescription as well can be obtained from the general one, given by \equref{Eq:general to vacuum orthogonal}, with the same imposition, that is $j=m=0$, thus
\begin{equation*}
(k_{AF}^{(d)})^{(0B)}_{n00}=\frac{n+3}{d}\left((\vok_{AF}^{(d)})^{(0B)}_{n00}-(\vok_{AF}^{(d)})^{(0B)}_{(n-2)00}\right)\,.
\end{equation*}
Therefore
\begin{align*}
(\hat{k}_{AF})_0{}&{}=\frac{1}{\sqrt{4\pi}}\sum\limits_{dn}\omega^{d-3-n}p^n\frac{n+3}{d}\left((\vok_{AF}^{(d)})^{(0B)}_{n00}-(\vok_{AF}^{(d)})^{(0B)}_{(n-2)00}\right)\,,\\
(\hat{k}_{AF})_r{}&{}=\frac{1}{\sqrt{4\pi}}\sum\limits_{dn}\omega^{d-4-n}p^{n+1}\frac{n+3-d}{d}\left((\vok_{AF}^{(d)})^{(0B)}_{n00}-(\vok_{AF}^{(d)})^{(0B)}_{(n-2)00}\right)\,,
\end{align*}
which gives
\begin{equation*}
\begin{aligned}
p(\hat{k}_{AF})_0-\omega(\hat{k}_{AF})_r={}&{}\frac{1}{\sqrt{4\pi}}\sum\limits_{dn}\omega^{d-3-n}p^{n+1}\left((\vok_{AF}^{(d)})^{(0B)}_{n00}-(\vok_{AF}^{(d)})^{(0B)}_{(n-2)00}\right)\,.
\end{aligned}
\end{equation*}

Vacuum-orthogonality bounds the frequency $n$ as $n\le d-4$ whereas isotropy restricts it to even values only, both of which can be read from \tabref{Table-vocir}. Then
\begin{equation*}
\begin{aligned}
p(\hat{k}_{AF})_0-\omega(\hat{k}_{AF})_r=\frac{1}{\sqrt{4\pi}}\sum\limits_{d=\text{odd}}\bigg({}&{}\sum\limits_{n=\text{even}\ge 0}^{d-5}\omega^{d-3-n}p^{n+1}(\vok_{AF}^{(d)})^{(0B)}_{n00}\\{}&{}-\sum\limits_{n=\text{even}\ge 0}^{d-5}\omega^{d-5-n}p^{n+3}(\vok_{AF}^{(d)})^{(0B)}_{n00}\bigg)\,,
\end{aligned}
\end{equation*}
which is simply
\begin{equation}
p(\hat{k}_{AF})_0-\omega(\hat{k}_{AF})_r={}\frac{p\left(\omega^2-p^2\right)}{\sqrt{4\pi}}\sum\limits_{d=\text{odd}}\sum\limits_{n=\text{even}\ge 0}^{d-5}\omega^{d-5-n}p^{n}(\vok_{AF}^{(d)})^{(0B)}_{n00} \label{eq: pk0-omegakr}\,,
\end{equation}
thus
\begin{equation*}
\begin{aligned}
\left(p(\hat{k}_{AF})_0-\omega(\hat{k}_{AF})_r\right)^2={}(p_\mu p^\mu)^2\frac{p^2}{4\pi}\left(\sum\limits_{d=\text{odd}}\sum\limits_{n=\text{even}\ge 0}^{d-5}\omega^{d-5-n}p^{n}(\vok_{AF}^{(d)})^{(0B)}_{n00}\right)^2
\end{aligned}
\end{equation*}
for isotropic VOM. Once this is substituted into the general dispersion relation of CPT-odd modified photon, \equref{General CPT-odd Spherical Dispersion Relation}, we obtain the same dispersion relation, that is \equref{Compact form}.

Let us now check the claim in \secref{vom dispersion} about that the nonconventional dispersion relation roots are spurious in VOM. Clearly, in the notation introduced there, the dispersion relation above indicates that
\begin{equation*}
\mathcal{R}(\omega,p)=\frac{p}{\sqrt{\pi}}\sum\limits_{d=\text{odd}>3}\sum\limits_{n=\text{even}\ge 0}^{d-5}\omega^{d-5-n}p^n(\vok_{AF}^{(d)})^{(0B)}_{n00}\,,
\end{equation*}
where the dispersion relations associated with the birefringent solutions are
\begin{equation*}
\mathcal{R}(\omega,p)=\pm 1\,.
\end{equation*}
In the leading order, there even does not arise nonconventional dispersion relations, as the dispersion relation reduces to the form
\begin{equation}
0={}\left(1-\frac{((\vok_{AF}^{(5)})^{(0B)}_{000}p)^2}{\pi}\right)(p_\mu p^\mu)^2\,, \label{vod in 5}
\end{equation}
where the multiplicative term possesses no roots for $\omega$. This multiplicative term is practically irrelevant as SME is built on EFT approach, and it is expected to hold only for $\lvert\xi_{dn}p\rvert\ll1$.

The spurious nature of the nonconventional dispersion relation roots can be exploited only with next-to-leading order, and higher order terms. Particularly, in the next-to-leading order, $\mathcal{R}(\omega,p)$ reads
\begin{equation*}
\mathcal{R}(\omega,p)=\frac{p}{\sqrt{\pi}}\left(\omega^2(\vok_{AF}^{(7)})^{(0B)}_{000}+p^2(\vok_{AF}^{(7)})^{(0B)}_{200}\right)\,,
\end{equation*}
which simply means that
\begin{equation}
\omega^2=\pm\frac{\sqrt{\pi}}{(\vok_{AF}^{(7)})^{(0B)}_{000}p}-\frac{(\vok_{AF}^{(7)})^{(0B)}_{200}}{(\vok_{AF}^{(7)})^{(0B)}_{000}}p^2\,,\label{eq: isotropic solution}
\end{equation}
which blow up as Lorentz violation is turned off. The spurious nature is caused by the first term, which will remain in all orders as moving on to the next order will simply introduce extra perturbative terms without effecting the first term.

\subsection{Ring Coefficients Form}
\label{Section: ring coefficients}
In literature, it is somehow customary to work with so-called \emph{ring coefficients} when working with the isotropic models. The general derivation conducted above is carried out with usual coefficients as they are more transparent to infer conclusions; however, the ring-coefficient form will be derived here as well.

In \cite{0905.0031}, ring coefficients are defined by the expansion they satisfy; that is,
\begin{subequations}
\begin{align}
(\hat{k}_{AF})_0{}&{}=\sum\limits_{dn}\frac{\omega^{d-3-n}p^n}{\sqrt{4\pi}}(\ring{k}_{AF}^{(d)})_n\,,\\
(\hat{k}_{AF})_r{}&{}=-\sum\limits_{dn}\frac{\omega^{d-3-n}p^n}{\sqrt{4\pi}}\frac{d-2-n}{n+2}(\ring{k}_{AF}^{(d)})_{(n-1)}\,,
\end{align}\label{ring defining equation}
\end{subequations}
where they are named after the ring above the coefficient denoting that the coefficient is associated with an isotropic model only. Indeed, that there are no spin weighted spherical harmonics in the defining equations above clearly indicates the necessity of isotropy for the usage of ring coefficients.

For the VOM, the isotropy condition reduced the general dispersion relation to \\\equref{Compact form} as we derived in \secref{Section: derivation of isotropic vom}. Once we compare \equref{ring defining equation} with that, we can deduce the transformation from the usual coefficients to the ring coefficients:
\begin{equation}
(\ring{k}_{AF}^{(d)})_n=\frac{n+3}{d}\left((\vok_{AF}^{(d)})^{(0B)}_{n00}-(\vok_{AF}^{(d)})^{(0B)}_{(n-2)00}\right)\,.\label{ring-coefficient}
\end{equation}

Clearly, it is not possible to trivially replace the usual terms in \equref{Compact form} with their ring coefficient correspondents; one must instead derive the ring coefficient form from the general dispersion relation \equref{General CPT-odd Spherical Dispersion Relation}. However, this actually reduces to the calculation of $p(\hat{k}_{AF})_0-\omega(\hat{k}_{AF})_r$ in terms of ring coefficients.

For the isotropic VOM, $p(\hat{k}_{AF})_0-\omega(\hat{k}_{AF})_r$ is given by \equref{eq: pk0-omegakr}, which can be written as
\begin{equation*}
\begin{aligned}
p(\hat{k}_{AF})_0-\omega(\hat{k}_{AF})_r={}&{}\frac{1}{\sqrt{4\pi}}\sum\limits_{d=\text{odd}}\bigg(\omega^{d-3}p(\vok_{AF}^{(d)})^{(0B)}_{000}+\sum\limits_{n=2}^{d-5}\omega^{d-3-n}p^{n+1}\\{}&{}\times\left((\vok_{AF}^{(d)})^{(0B)}_{n00}-(\vok_{AF}^{(d)})^{(0B)}_{(n-2)00}\right)-p^{d-2}(\vok_{AF}^{(d)})^{(0B)}_{(d-5)00}\bigg)\,,
\end{aligned}
\end{equation*}
which then reads
\begin{equation*}
\begin{aligned}
p(\hat{k}_{AF})_0-\omega(\hat{k}_{AF})_r={}&{}\frac{1}{\sqrt{4\pi}}\sum\limits_{d=\text{odd}}\bigg(\omega^{d-3}p(\vok_{AF}^{(d)})^{(0B)}_{000}\\{}&{}+\sum\limits_{n=2}^{d-5}\omega^{d-3-n}p^{n+1}\frac{d}{n+3}(\ring{k}_{AF}^{(d)})_n-p^{d-2}(\vok_{AF}^{(d)})^{(0B)}_{(d-5)00}\bigg)\,.
\end{aligned}
\end{equation*}
But this is actually equivalent to
\begin{align}
&\begin{aligned}
p(\hat{k}_{AF})_0-\omega(\hat{k}_{AF})_r={}&{}\frac{1}{\sqrt{4\pi}}\sum\limits_{d=\text{odd}}\Bigg(\omega^{d-3}p(\vok_{AF}^{(d)})^{(0B)}_{000}+\sum\limits_{n=2}^{d-5}\omega^{d-3-n}p^{n+1}\frac{d}{n+3}(\ring{k}_{AF}^{(d)})_n\\{}&{}-p^{d-2}\left((\vok_{AF}^{(d)})^{(0B)}_{000}+\sum\limits_{n=2}^{d-5}\frac{d}{n+3}(\ring{k}_{AF}^{(d)})_n\right)\Bigg)\,,
\end{aligned}\nonumber\\
&\begin{aligned}
p(\hat{k}_{AF})_0-\omega(\hat{k}_{AF})_r={}&{}\frac{1}{\sqrt{4\pi}}\sum\limits_{d=\text{odd}}\bigg(\left(p(\vok_{AF}^{(d)})^{(0B)}_{000}(\omega^{d-3}-p^{d-3})\right)\\{}&{}+\sum\limits_{n=2}^{d-5}\left(\omega^{d-3-n}-p^{d-3-n}\right)\frac{d}{n+3}\left((\ring{k}_{AF}^{(d)})_np^{n+1}\right)\bigg)\,,\label{ringization}
\end{aligned}
\end{align}
as \equref{ring-coefficient} dictates
\begin{equation*}
(\vok_{AF}^{(d)})^{(0B)}_{(d-5)00}=(\vok_{AF}^{(d)})^{(0B)}_{000}+\sum\limits_{n=2}^{d-5}\frac{d}{n+3}(\ring{k}_{AF}^{(d)})_n\,.
\end{equation*}

Judging from \equref{ring-coefficient}, if we additionally take
\begin{equation*}
(\ring{k}_{AF}^{(d)})_0=\frac{3}{d}(\vok_{AF}^{(d)})^{(0B)}_{000}\,,
\end{equation*}
we can rewrite \equref{ringization} as
\begin{equation*}
p(\hat{k}_{AF})_0-\omega(\hat{k}_{AF})_r=\frac{1}{\sqrt{4\pi}}\sum\limits_{d=\text{odd}}\sum\limits_{n=0}^{d-5}\frac{d}{n+3}\left(\omega^{d-3-n}-p^{d-3-n}\right)\left((\ring{k}_{AF}^{(d)})_np^{n+1}\right)\,.
\end{equation*}

The apparent form of the equation is quite complicated at the first glance; yet, it can be put in a nicely compact form once the equality
\begin{equation*}
x^{n+1}-y^{n+1}=(x-y)\sum\limits_{i=0}^{n}x^{n-i}y^i
\end{equation*}
is invoked. Since $d-3-n$ is always greater or equal to 2, and is always even, we can use the above equality as
\begin{equation*}
\omega^{d-3-n}-p^{d-3-n}=(\omega^2-p^2)\sum\limits_{i=0}^{(d-5-n)/2}\omega^{d-5-n-2i}p^{2i}
\end{equation*}
for the case at hand. Therefore,
\begin{equation*}
p(\hat{k}_{AF})_0-\omega(\hat{k}_{AF})_r=\frac{(p_\mu p^\mu)}{\sqrt{4\pi}}\sum\limits_{d=\text{odd}}\sum\limits_{n=0}^{d-5}\sum\limits_{i=0}^{(d-5-n)/2}\frac{d}{n+3}\omega^{d-5-n-2i}p^{2i}\left((\ring{k}_{AF}^{(d)})_np^{n+1}\right)\,.
\end{equation*}

If this is inserted into the general CPT-odd modified photon dispersion relation, the isotropic VOM dispersion relation reads as 
\begin{equation}
\boxed{0=\left(1-\frac{p^2}{\pi}\left(\sum\limits_{d=\text{odd}}\sum\limits_{n=\text{even}\ge 0}^{d-5}\sum\limits_{i=\text{even}\ge 0}^{(d-5-n)}\frac{d}{n+3}\omega^{d-5-n-i}p^{i}\left((\ring{k}_{AF}^{(d)})_np^{n}\right)\right)^2\right)(p_\mu p^\mu)^2} \label{disperion relation in ring coefficients}
\end{equation}

Although it is provided for completeness, and because of its somewhat more popular usage, the general form given by \equref{Compact form} is more compact than the ring coefficient form above.

\section{Leading Order Vacuum Orthogonal Model}
\label{Section: finite order vom}
One of the main problems for any analysis in the \emph{nm}SME is that the coefficient space is infinite. Although generic examinations are possible to a degree, such as what is done throughout this study, the contact to the experimental results can be achieved only when a subset of the coefficient space is considered. This subset is actually a drastic reduction most of the time, as bounds on the coefficients are most effectively given only when they are considered one by one, the one-at-a-time approach that is sometimes called \emph{Kostelecký's Cutlass} as explained in the beginning of this chapter.

In this thesis, however, the main focus is on the generic analysis of the nonminimal Lorentz and CPT violating photon sector, without specifying in any particular coefficient. Henceforth, the one-at-a-time approach invoked to analyze the effects of a specific coefficient is not employed here; nonetheless, the restriction of the coefficient space to a finite subspace will be demonstrated in this section for different purposes: explicit analysis of the spurious roots and the relevant coefficients' determination.

\subsection{Analysis of Leading Order VOM}
\label{Section: leading vom}
The dispersion relation of VOM is readily given by \equref{General Vacuum-orthogonal dispersion relation3}, which is also valid for this special case. All that is needed, then, is simply the restriction of \equref{definition of R} to \mbox{$d=5$} and expansion of it in relevant LVT. Despite this would give the physical dispersion relation of the leading order VOM, the most general relevant dispersion relation, one that includes the nonphysical contributions of $(\hat{k}_{AF})_\pm$ hence is applicable to the case $\hat{k}_{AF}\in\hat{k}_{AF}^{(np)}$, can only be obtained from the more general form that is \equref{General Vacuum-orthogonal dispersion relation}. For the sake of completeness, we will start from the latter equation.

With $d=5$, \equref{General Vacuum-orthogonal dispersion relation} becomes
\begin{align*}
0={}&{}\left(p_\mu p^\mu\right)^2-4\Bigg\{\sum\limits_{njm}\omega^{2-n}p^n\prescript{}{0}{Y}_{jm}(\mathbf{\hat{p}})\Bigg((\vok_{AF}^{(5)})^{(0B)}_{njm}\frac{p}{\omega^2}\left(\frac{(3-n)\omega^2}{3-n+j}-\frac{(1-n)p^2}{1-n+j}\right)\\{}&{}-\frac{5p^2}{(n+4)(n+2)\omega}(\vok_{AF}^{(5)})^{(1B)}_{njm}+(\vok_{AF}^{(5)})^{(0B)}_{njm}\frac{j}{p}\left(\frac{\omega^2}{3-n+j}-\frac{p^2}{1-n+j}\right)\\{}&{}+\frac{5\omega}{(n+2)(n+4)}(\vok_{AF}^{(d)})^{(1B)}_{njm}\Bigg)\Bigg\}^2\\{}&{}+8p_\mu p^\mu\sum\limits_{n_1n_2j_1j_2m_1m_2}\omega^{4-n_1-n_2}p^{n_1+n_2}\prescript{}{+1}{Y}_{j_1m_1}(\mathbf{\hat{p}})\prescript{}{-1}{Y}_{j_2m_2}(\mathbf{\hat{p}})\\
{}&{}\times\frac{1}{\sqrt{4j_1j_2(j_1+1)(j_2+1)}}\Bigg\{\Bigg(\left(\frac{\omega j_1(n_1+1)}{p(3-n_1+j_1)}-\frac{pj_1(n_1+3)}{\omega(1-n_1+j_1)}\right)\\{}&{}\times(\vok_{AF}^{(5)})_{n_1j_1m_1}^{(0B)}+\frac{5}{n_1+4}(\vok_{AF}^{(5)})^{(1B)}_{n_1j_1m_1}\Bigg)\\{}&{}\times\Bigg(\left(\frac{\omega j_2(n_2+1)}{p(3-n_2+j_2)}-\frac{pj_2(n_2+3)}{\omega(1-n_2+j_2)}\right)(\vok_{AF}^{(5)})_{n_2j_2m_2}^{(0B)}+\frac{5}{n_2+4}(\vok_{AF}^{(5)})^{(1B)}_{n_2j_2m_2}\Bigg)\\{}&{}+(\vok_{AF}^{(5)})^{(1E)}_{n_1j_1m_1}(\vok_{AF}^{(5)})^{(1E)}_{n_2j_2m_2}\Bigg\}\,.
\end{align*}
Let us expand the summation over n for the first part.
\begin{align*}
0={}&{}\left(p_\mu p^\mu\right)^2-4\Bigg\{\sum\limits_{jm}\omega^{2}\prescript{}{0}{Y}_{jm}(\mathbf{\hat{p}})\Bigg((\vok_{AF}^{(5)})^{(0B)}_{0jm}\frac{p}{\omega^2}\left(\frac{3\omega^2}{3+j}-\frac{p^2}{1+j}\right)\\{}&{}-\frac{5p^2}{8\omega}(\vok_{AF}^{(5)})^{(1B)}_{0jm}+(\vok_{AF}^{(5)})^{(0B)}_{0jm}\frac{j}{p}\left(\frac{\omega^2}{3+j}-\frac{p^2}{1+j}\right)+\frac{5\omega}{8}(\vok_{AF}^{(d)})^{(1B)}_{0jm}\Bigg)\\{}&{}+\sum\limits_{jm}\omega p\prescript{}{0}{Y}_{jm}(\mathbf{\hat{p}})\Bigg((\vok_{AF}^{(5)})^{(0B)}_{1jm}\frac{p}{\omega^2}\left(\frac{2\omega^2}{2+j}\right)-\frac{5p^2}{15\omega}(\vok_{AF}^{(5)})^{(1B)}_{1jm}\nonumber+(\vok_{AF}^{(5)})^{(0B)}_{1jm}\frac{j}{p}\\{}&{}\times\left(\frac{\omega^2}{2+j}-\frac{p^2}{j}\right)+\frac{5\omega}{15}(\vok_{AF}^{(d)})^{(1B)}_{1jm}\Bigg)\Bigg\}^2+8p_\mu p^\mu\sum\limits_{n_1n_2j_1j_2m_1m_2}\omega^{4-n_1-n_2}p^{n_1+n_2}\\
{}&{}\times\prescript{}{+1}{Y}_{j_1m_1}(\mathbf{\hat{p}})\prescript{}{-1}{Y}_{j_2m_2}(\mathbf{\hat{p}})\frac{1}{\sqrt{4j_1j_2(j_1+1)(j_2+1)}}\Bigg\{\Bigg(\bigg(\frac{\omega j_1(n_1+1)}{p(3-n_1+j_1)}\\{}&{}-\frac{pj_1(n_1+3)}{\omega(1-n_1+j_1)}\bigg)(\vok_{AF}^{(5)})_{n_1j_1m_1}^{(0B)}+\frac{5}{n_1+4}(\vok_{AF}^{(5)})^{(1B)}_{n_1j_1m_1}\Bigg)\Bigg(\bigg(\frac{\omega j_2(n_2+1)}{p(3-n_2+j_2)}\\{}&{}-\frac{pj_2(n_2+3)}{\omega(1-n_2+j_2)}\bigg)(\vok_{AF}^{(5)})_{n_2j_2m_2}^{(0B)}+\frac{5}{n_2+4}(\vok_{AF}^{(5)})^{(1B)}_{n_2j_2m_2}\Bigg)\\{}&{}+(\vok_{AF}^{(5)})^{(1E)}_{n_1j_1m_1}(\vok_{AF}^{(5)})^{(1E)}_{n_2j_2m_2}\Bigg\}\,,
\end{align*}
then, due to the restriction on j,
\begin{align*}
0={}&{}\left(p_\mu p^\mu\right)^2-4\Bigg\{\sum\limits_{jm}\prescript{}{0}{Y}_{jm}(\mathbf{\hat{p}})\bigg((\vok_{AF}^{(5)})^{(0B)}_{0jm}p\left(\omega^2-p^2\right)-\frac{5p^2\omega}{8}(\vok_{AF}^{(5)})^{(1B)}_{0jm}\\{}&{}+\frac{5\omega^3}{8}(\vok_{AF}^{(d)})^{(1B)}_{0jm}\bigg)+\sum\limits_{jm}\omega p\prescript{}{0}{Y}_{jm}(\mathbf{\hat{p}})\Bigg((\vok_{AF}^{(5)})^{(0B)}_{1jm}\frac{p}{\omega^2}\left(\frac{2\omega^2}{3}\right)-\frac{5p^2}{15\omega}(\vok_{AF}^{(5)})^{(1B)}_{1jm}\\{}&{}+(\vok_{AF}^{(5)})^{(0B)}_{1jm}\frac{1}{p}\left(\frac{\omega^2}{3}-p^2\right)+\frac{5\omega}{15}(\vok_{AF}^{(d)})^{(1B)}_{1jm}\Bigg)\Bigg\}^2+8p_\mu p^\mu\sum\limits_{n_1n_2j_1j_2m_1m_2}\omega^{4-n_1-n_2}\\{}&{}p^{n_1+n_2}\prescript{}{+1}{Y}_{j_1m_1}(\mathbf{\hat{p}})\prescript{}{-1}{Y}_{j_2m_2}(\mathbf{\hat{p}})\frac{1}{\sqrt{4j_1j_2(j_1+1)(j_2+1)}}\Bigg\{\Bigg(\bigg(\frac{\omega j_1(n_1+1)}{p(3-n_1+j_1)}\\
{}&{}-\frac{pj_1(n_1+3)}{\omega(1-n_1+j_1)}\bigg)(\vok_{AF}^{(5)})_{n_1j_1m_1}^{(0B)}+\frac{5}{n_1+4}(\vok_{AF}^{(5)})^{(1B)}_{n_1j_1m_1}\Bigg)\\{}&{}\times\Bigg(\left(\frac{\omega j_2(n_2+1)}{p(3-n_2+j_2)}-\frac{pj_2(n_2+3)}{\omega(1-n_2+j_2)}\right)(\vok_{AF}^{(5)})_{n_2j_2m_2}^{(0B)}+\frac{5}{n_2+4}(\vok_{AF}^{(5)})^{(1B)}_{n_2j_2m_2}\Bigg)\\
{}&{}+(\vok_{AF}^{(5)})^{(1E)}_{n_1j_1m_1}(\vok_{AF}^{(5)})^{(1E)}_{n_2j_2m_2}\Bigg\}\,.
\end{align*}
Therefore,
\begin{align*}
0={}&{}\left(p_\mu p^\mu\right)^2-4(p_\mu p^\mu)^2\bigg(\sum_{jm}\prescript{}{0}{Y}_{jm}(\mathbf{\hat{p}})\Big((\vok_{AF}^{(5)})_{0jm}^{(0B)}p+\frac{\omega}{3}(\vok_{AF}^{(5)})_{1jm}^{(0B)}+\frac{5\omega}{8}(\vok_{AF}^{(5)})_{0jm}^{(1B)}\\
{}&{}+\frac{p}{3}(\vok_{AF}^{(5)})_{1jm}^{(1B)}\Big)\bigg)^2+8p_\mu p^\mu\sum\limits_{n_1n_2j_1j_2m_1m_2}\omega^{4-n_1-n_2}p^{n_1+n_2}\prescript{}{+1}{Y}_{j_1m_1}(\mathbf{\hat{p}})\prescript{}{-1}{Y}_{j_2m_2}(\mathbf{\hat{p}})\\{}&{}\times\frac{1}{\sqrt{4j_1j_2(j_1+1)(j_2+1)}}\Bigg\{\Bigg(\left(\frac{\omega j_1(n_1+1)}{p(3-n_1+j_1)}-\frac{pj_1(n_1+3)}{\omega(1-n_1+j_1)}\right)(\vok_{AF}^{(5)})_{n_1j_1m_1}^{(0B)}\\
{}&{}+\frac{5}{n_1+4}(\vok_{AF}^{(5)})^{(1B)}_{n_1j_1m_1}\Bigg)\Bigg(\left(\frac{\omega j_2(n_2+1)}{p(3-n_2+j_2)}-\frac{pj_2(n_2+3)}{\omega(1-n_2+j_2)}\right)(\vok_{AF}^{(5)})_{n_2j_2m_2}^{(0B)}\\{}&{}+\frac{5}{n_2+4}(\vok_{AF}^{(5)})^{(1B)}_{n_2j_2m_2}\Bigg)+(\vok_{AF}^{(5)})^{(1E)}_{n_1j_1m_1}(\vok_{AF}^{(5)})^{(1E)}_{n_2j_2m_2}\Bigg\}\,.
\end{align*}

Let us now expand for $(\vok_{AF}^{(5)})^{(1E)}_{njm}$. The above dispersion relation becomes
\begin{align*}
0={}&{}\left(p_\mu p^\mu\right)^2-4(p_\mu p^\mu)^2\bigg(\sum_{jm}\prescript{}{0}{Y}_{jm}(\mathbf{\hat{p}})\Big((\vok_{AF}^{(5)})_{0jm}^{(0B)}p+\frac{\omega}{3}(\vok_{AF}^{(5)})_{1jm}^{(0B)}+\frac{5\omega}{8}(\vok_{AF}^{(5)})_{0jm}^{(1B)}\\
{}&{}+\frac{p}{3}(\vok_{AF}^{(5)})_{1jm}^{(1B)}\Big)\bigg)^2+8p_\mu p^\mu\sum\limits_{n_1n_2j_1j_2m_1m_2}\omega^{4-n_1-n_2}p^{n_1+n_2}\prescript{}{+1}{Y}_{j_1m_1}(\mathbf{\hat{p}})\\{}&{}\times\prescript{}{-1}{Y}_{j_2m_2}(\mathbf{\hat{p}})\frac{1}{\sqrt{4j_1j_2(j_1+1)(j_2+1)}}(\vok_{AF}^{(5)})^{(1E)}_{n_1j_1m_1}(\vok_{AF}^{(5)})^{(1E)}_{n_2j_2m_2}+8p_\mu p^\mu\\{}&{}\times\sum\limits_{j_1j_2m_1m_2}\prescript{}{+1}{Y}_{j_1m_1}(\mathbf{\hat{p}})\prescript{}{-1}{Y}_{j_2m_2}(\mathbf{\hat{p}})\frac{1}{\sqrt{4j_1j_2(j_1+1)(j_2+1)}}\\{}&{}\times\sum\limits_{n_1n_2}\omega^{4-n_1-n_2}p^{n_1+n_2}\mathcal{S}(n_1,n_2)\,,
\end{align*}
which with the help of \tabref{Table-rocflovom} is then 
\begin{align}
0={}&{}\left(p_\mu p^\mu\right)^2-4(p_\mu p^\mu)^2\bigg(\sum_{jm}\prescript{}{0}{Y}_{jm}(\mathbf{\hat{p}})\Big((\vok_{AF}^{(5)})_{0jm}^{(0B)}p+\frac{\omega}{3}(\vok_{AF}^{(5)})_{1jm}^{(0B)}+\frac{5\omega}{8}(\vok_{AF}^{(5)})_{0jm}^{(1B)}\nonumber\\{}&{}+\frac{p}{3}(\vok_{AF}^{(5)})_{1jm}^{(1B)}\Big)\bigg)^2+8p_\mu p^\mu\sum\limits_{m_1m_2}\bigg(
(\vok_{AF}^{(5)})^{(1E)}_{11m_1}(\vok_{AF}^{(5)})^{(1E)}_{11m_2}\frac{1}{4}\omega^2p^2\prescript{}{+1}{Y}_{1m_1}(\mathbf{\hat{p}})\nonumber\\{}&{}\times\prescript{}{-1}{Y}_{1m_2}(\mathbf{\hat{p}})
+(\vok_{AF}^{(5)})^{(1E)}_{11m_1}(\vok_{AF}^{(5)})^{(1E)}_{22m_2}\frac{1}{4\sqrt{3}}\omega p^3\Big(\prescript{}{+1}{Y}_{1m_1}(\mathbf{\hat{p}})\prescript{}{-1}{Y}_{2m_2}(\mathbf{\hat{p}})\nonumber\\{}&{}+\prescript{}{+1}{Y}_{2m_1}(\mathbf{\hat{p}})\prescript{}{-1}{Y}_{1m_2}(\mathbf{\hat{p}})\Big)+(\vok_{AF}^{(5)})^{(1E)}_{22m_1}(\vok_{AF}^{(5)})^{(1E)}_{22m_2}\frac{1}{12} p^4\prescript{}{+1}{Y}_{2m_1}(\mathbf{\hat{p}})\prescript{}{-1}{Y}_{2m_2}(\mathbf{\hat{p}})\bigg)\nonumber\\{}&{}+8p_\mu p^\mu\sum\limits_{j_1j_2m_1m_2}\prescript{}{+1}{Y}_{j_1m_1}(\mathbf{\hat{p}})\prescript{}{-1}{Y}_{j_2m_2}(\mathbf{\hat{p}})\frac{1}{\sqrt{4j_1j_2(j_1+1)(j_2+1)}}\nonumber\\{}&{}\times\sum\limits_{n_1=0}^{1}\sum\limits_{n_2=0}^{1}\omega^{4-n_1-n_2}p^{n_1+n_2}\mathcal{S}(n_1,n_2)\,,\label{eq: leading order vom dispersion with S}
\end{align}
where
\begin{align*}
\mathcal{S}(n_1,n_2)={}&{}\Bigg(\left(\frac{\omega j_1(n_1+1)}{p(3-n_1+j_1)}-\frac{pj_1(n_1+3)}{\omega(1-n_1+j_1)}\right)(\vok_{AF}^{(5)})_{n_1j_1m_1}^{(0B)}\\{}&{}+\frac{5}{n_1+4}(\vok_{AF}^{(5)})^{(1B)}_{n_1j_1m_1}\Bigg)\Bigg(\left(\frac{\omega j_2(n_2+1)}{p(3-n_2+j_2)}-\frac{pj_2(n_2+3)}{\omega(1-n_2+j_2)}\right)\\{}&{}\times(\vok_{AF}^{(5)})_{n_2j_2m_2}^{(0B)}+\frac{5}{n_2+4}(\vok_{AF}^{(5)})^{(1B)}_{n_2j_2m_2}\Bigg)
\end{align*}
is defined for brevity.

Here, it may look problematic to have summation of $n$ inside summation of $j$ as the range of $j$ depends on the value of $n$. However, the summation of $j$ nd $m$ there simply refers to the summation of all their possible values, enabling performing the $n$ summation without any ambiguity.

The range of $n$ can be checked from \tabref{Table-rocflovom}, which states that the nonzero possible $\mathcal{S}(n_1,n_2)$ are as follows\footnotemark\footnotetext{In each calculation, the restriction of $n$ is applied in the first step, and that of $j$ in the second.}.
\begin{itemize}
\item[$\mathcal{S}(0,0)$]
\begin{align*}
={}&{}\Bigg(\left(\frac{\omega j_1}{p(3+j_1)}-\frac{3pj_1}{\omega(1+j_1)}\right)(\vok_{AF}^{(5)})_{0j_1m_1}^{(0B)}+\frac{5}{4}(\vok_{AF}^{(5)})^{(1B)}_{0j_1m_1}\Bigg)\\{}&{}\times\Bigg(\left(\frac{\omega j_2}{p(3+j_2)}-\frac{3pj_2}{\omega(1+j_2)}\right)(\vok_{AF}^{(5)})_{0j_2m_2}^{(0B)}+\frac{5}{4}(\vok_{AF}^{(5)})^{(1B)}_{0j_2m_2}\Bigg)\\
={}&{}\frac{25}{16}(\vok_{AF}^{(5)})^{(1B)}_{0j_1m_1}(\vok_{AF}^{(5)})^{(1B)}_{0j_2m_2}
\end{align*}

\item[$\mathcal{S}(0,1)$]
\begin{align*}
={}&{}\Bigg(\left(\frac{\omega j_1}{p(3+j_1)}-\frac{3pj_1}{\omega(1+j_1)}\right)(\vok_{AF}^{(5)})_{0j_1m_1}^{(0B)}+\frac{5}{4}(\vok_{AF}^{(5)})^{(1B)}_{0j_1m_1}\Bigg)\\{}&{}\times\Bigg(\left(\frac{2\omega j_2}{p(2+j_2)}-\frac{4pj_2}{\omega j_2}\right)(\vok_{AF}^{(5)})_{1j_2m_2}^{(0B)}+\frac{5}{5}(\vok_{AF}^{(5)})^{(1B)}_{1j_2m_2}\Bigg)\\
={}&{}\frac{5}{4}(\vok_{AF}^{(5)})^{(1B)}_{0j_1m_1}\left(\left(\frac{2\omega}{3p}-\frac{4p}{\omega}\right)(\vok_{AF}^{(5)})_{1j_2m_2}^{(0B)}+(\vok_{AF}^{(5)})^{(1B)}_{1j_2m_2}\right)
\end{align*}

\item[$\mathcal{S}(1,0)$]
\begin{align*}
={}&{}\Bigg(\left(\frac{2\omega j_1}{p(2+j_1)}-\frac{4pj_1}{\omega j_1}\right)(\vok_{AF}^{(5)})_{1j_1m_1}^{(0B)}+\frac{5}{5}(\vok_{AF}^{(5)})^{(1B)}_{1j_1m_1}\Bigg)\\{}&{}\times\Bigg(\left(\frac{\omega j_2}{p(3+j_2)}-\frac{3pj_2}{\omega(1+j_2)}\right)(\vok_{AF}^{(5)})_{0j_2m_2}^{(0B)}+\frac{5}{4}(\vok_{AF}^{(5)})^{(1B)}_{0j_2m_2}\Bigg)\\
={}&{}\frac{5}{4}(\vok_{AF}^{(5)})^{(1B)}_{0j_2m_2}\left(\left(\frac{2\omega}{3p}-\frac{4p}{\omega}\right)(\vok_{AF}^{(5)})_{1j_1m_1}^{(0B)}+(\vok_{AF}^{(5)})^{(1B)}_{1j_1m_1}\right)
\end{align*}

\item[$\mathcal{S}(1,1)$]
\begin{align*}
={}&{}\Bigg(\left(\frac{2\omega j_1}{p(2+j_1)}-\frac{4pj_1}{\omega j_1}\right)(\vok_{AF}^{(5)})_{1j_1m_1}^{(0B)}+\frac{5}{5}(\vok_{AF}^{(5)})^{(1B)}_{1j_1m_1}\Bigg)\\{}&{}\times\Bigg(\left(\frac{2\omega j_2}{p(2+j_2)}-\frac{4pj_2}{\omega j_2}\right)(\vok_{AF}^{(5)})_{1j_2m_2}^{(0B)}+\frac{5}{5}(\vok_{AF}^{(5)})^{(1B)}_{1j_2m_2}\Bigg)\\
={}&{}\left(\left(\frac{2\omega}{3p}-\frac{4p}{\omega}\right)(\vok_{AF}^{(5)})_{1j_1m_1}^{(0B)}+(\vok_{AF}^{(5)})^{(1B)}_{1j_1m_1}\right)\\{}&{}\times\left(\left(\frac{2\omega}{3p}-\frac{4p}{\omega}\right)(\vok_{AF}^{(5)})_{1j_2m_2}^{(0B)}+(\vok_{AF}^{(5)})^{(1B)}_{1j_2m_2}\right)
\end{align*}
\end{itemize}

With $\mathcal{S}(n_1,n_2)$ inserted back, \equref{eq: leading order vom dispersion with S} becomes
\begin{align*}
0={}&{}\left(p_\mu p^\mu\right)^2-4(p_\mu p^\mu)^2\Bigg(\sum_{jm}\prescript{}{0}{Y}_{jm}(\mathbf{\hat{p}})\bigg((\vok_{AF}^{(5)})_{0jm}^{(0B)}p+\frac{\omega}{3}(\vok_{AF}^{(5)})_{1jm}^{(0B)}+\frac{5\omega}{8}(\vok_{AF}^{(5)})_{0jm}^{(1B)}\\{}&{}+\frac{p}{3}(\vok_{AF}^{(5)})_{1jm}^{(1B)}\bigg)\Bigg)^2\nonumber+8p_\mu p^\mu\sum\limits_{m_1m_2}\bigg((\vok_{AF}^{(5)})^{(1E)}_{11m_1}(\vok_{AF}^{(5)})^{(1E)}_{11m_2}\frac{1}{4}\omega^2p^2\prescript{}{+1}{Y}_{1m_1}(\mathbf{\hat{p}})\\{}&{}\times\prescript{}{-1}{Y}_{1m_2}(\mathbf{\hat{p}})
+(\vok_{AF}^{(5)})^{(1E)}_{11m_1}(\vok_{AF}^{(5)})^{(1E)}_{22m_2}\frac{1}{4\sqrt{3}}\omega p^3\Big(\prescript{}{+1}{Y}_{1m_1}(\mathbf{\hat{p}})\prescript{}{-1}{Y}_{2m_2}(\mathbf{\hat{p}})\\{}&{}+\prescript{}{+1}{Y}_{2m_1}(\mathbf{\hat{p}})\prescript{}{-1}{Y}_{1m_2}(\mathbf{\hat{p}})\Big)+(\vok_{AF}^{(5)})^{(1E)}_{22m_1}(\vok_{AF}^{(5)})^{(1E)}_{22m_2}\frac{1}{12} p^4\prescript{}{+1}{Y}_{2m_1}(\mathbf{\hat{p}})\prescript{}{-1}{Y}_{2m_2}(\mathbf{\hat{p}})
\bigg)\\{}&{}+8p_\mu p^\mu\sum\limits_{j_1j_2m_1m_2}\prescript{}{+1}{Y}_{j_1m_1}(\mathbf{\hat{p}})\prescript{}{-1}{Y}_{j_2m_2}(\mathbf{\hat{p}})\frac{1}{\sqrt{4j_1j_2(j_1+1)(j_2+1)}}\\{}&{}\Bigg(\frac{25p^4}{16}(\vok_{AF}^{(5)})^{(1B)}_{0j_1m_1}(\vok_{AF}^{(5)})^{(1B)}_{0j_2m_2}+\frac{5p^3\omega}{4}(\vok_{AF}^{(5)})^{(1B)}_{0j_1m_1}\bigg(\left(\frac{2\omega}{3p}-\frac{4p}{\omega}\right)(\vok_{AF}^{(5)})_{1j_2m_2}^{(0B)}\\{}&{}+(\vok_{AF}^{(5)})^{(1B)}_{1j_2m_2}\bigg)\nonumber+\frac{5p^3\omega}{4}(\vok_{AF}^{(5)})^{(1B)}_{0j_2m_2}\left(\left(\frac{2\omega}{3p}-\frac{4p}{\omega}\right)(\vok_{AF}^{(5)})_{1j_1m_1}^{(0B)}+(\vok_{AF}^{(5)})^{(1B)}_{1j_1m_1}\right)\\{}&{}+\omega^2p^2\left(\left(\frac{2\omega}{3p}-\frac{4p}{\omega}\right)(\vok_{AF}^{(5)})_{1j_1m_1}^{(0B)}+(\vok_{AF}^{(5)})^{(1B)}_{1j_1m_1}\right)\\{}&{}\times\left(\left(\frac{2\omega}{3p}-\frac{4p}{\omega}\right)(\vok_{AF}^{(5)})_{1j_2m_2}^{(0B)}+(\vok_{AF}^{(5)})^{(1B)}_{1j_2m_2}\right)\Bigg)\,,
\end{align*}
which can be written as 
\begin{align*}
0={}&{}\left(p_\mu p^\mu\right)^2-4(p_\mu p^\mu)^2\Bigg(\sum_{jm}\prescript{}{0}{Y}_{jm}(\mathbf{\hat{p}})\bigg((\vok_{AF}^{(5)})_{0jm}^{(0B)}p+\frac{\omega}{3}(\vok_{AF}^{(5)})_{1jm}^{(0B)}+\frac{5\omega}{8}(\vok_{AF}^{(5)})_{0jm}^{(1B)}\\{}&{}+\frac{p}{3}(\vok_{AF}^{(5)})_{1jm}^{(1B)}\bigg)\Bigg)^2+8p_\mu p^\mu\sum\limits_{m_1m_2}\bigg(
(\vok_{AF}^{(5)})^{(1E)}_{11m_1}(\vok_{AF}^{(5)})^{(1E)}_{11m_2}\frac{1}{4}\omega^2p^2\prescript{}{+1}{Y}_{1m_1}(\mathbf{\hat{p}})\nonumber\\{}&{}\times\prescript{}{-1}{Y}_{1m_2}(\mathbf{\hat{p}})
+(\vok_{AF}^{(5)})^{(1E)}_{11m_1}(\vok_{AF}^{(5)})^{(1E)}_{22m_2}\frac{1}{4\sqrt{3}}\omega p^3\Big(\prescript{}{+1}{Y}_{1m_1}(\mathbf{\hat{p}})\prescript{}{-1}{Y}_{2m_2}(\mathbf{\hat{p}})\\{}&{}+\prescript{}{+1}{Y}_{2m_1}(\mathbf{\hat{p}})\prescript{}{-1}{Y}_{1m_2}(\mathbf{\hat{p}})\Big)+(\vok_{AF}^{(5)})^{(1E)}_{22m_1}(\vok_{AF}^{(5)})^{(1E)}_{22m_2}\frac{1}{12} p^4\prescript{}{+1}{Y}_{2m_1}(\mathbf{\hat{p}})\prescript{}{-1}{Y}_{2m_2}(\mathbf{\hat{p}})\bigg)\\
{}&{}+8p_\mu p^\mu\sum\limits_{j_1j_2m_1m_2}\Bigg(
(\vok_{AF}^{(5)})^{(0B)}_{1j_1m_1}(\vok_{AF}^{(5)})^{(0B)}_{1j_2m_2}\left(\frac{2}{3}\omega^2-4p^2\right)^2
+\Big((\vok_{AF}^{(5)})^{(0B)}_{1j_1m_1}\\{}&{}\times(\vok_{AF}^{(5)})^{(1B)}_{0j_2m_2}+(\vok_{AF}^{(5)})^{(0B)}_{1j_2m_2}(\vok_{AF}^{(5)})^{(1B)}_{0j_1m_1}\Big) \frac{5}{4}p^2\left(\frac{2}{3}\omega^2-4p^2\right)
+\Big((\vok_{AF}^{(5)})^{(0B)}_{1j_1m_1}\\{}&{}\times(\vok_{AF}^{(5)})^{(1B)}_{1j_2m_2}+(\vok_{AF}^{(5)})^{(0B)}_{1j_2m_2}(\vok_{AF}^{(5)})^{(1B)}_{1j_1m_1}\Big)\omega p\left(\frac{2}{3}\omega^2-4p^2\right)
+(\vok_{AF}^{(5)})^{(1B)}_{0j_1m_1}\\{}&{}\times(\vok_{AF}^{(5)})^{(1B)}_{0j_2m_2}\frac{25}{16}p^4
+\left((\vok_{AF}^{(5)})^{(1B)}_{0j_1m_1}(\vok_{AF}^{(5)})^{(1B)}_{1j_2m_2}+(\vok_{AF}^{(5)})^{(1B)}_{0j_2m_2}(\vok_{AF}^{(5)})^{(1B)}_{1j_1m_1}\right)\frac{5}{4}p^3\omega\\{}&{}
+(\vok_{AF}^{(5)})^{(1B)}_{1j_1m_1}(\vok_{AF}^{(5)})^{(1B)}_{1j_2m_2}\omega^2p^2\Bigg) \prescript{}{+1}{Y}_{j_1m_1}(\mathbf{\hat{p}})\prescript{}{-1}{Y}_{j_2m_2}(\mathbf{\hat{p}})\frac{1}{\sqrt{4j_1j_2(j_1+1)(j_2+1)}}\,,
\end{align*}
which is arranged so as to expose the coefficient combinations.

Now, the restriction on $j$ can be applied straightforwardly. Hence, one can show that

\begin{align*}
0={}&{}\left(p_\mu p^\mu\right)^2-4(p_\mu p^\mu)^2\Bigg(\sum_{m}\bigg(\prescript{}{0}{Y}_{0m}(\mathbf{\hat{p}})(\vok_{AF}^{(5)})_{00m}^{(0B)}p+\prescript{}{0}{Y}_{1m}(\mathbf{\hat{p}})\bigg(\frac{\omega}{3}(\vok_{AF}^{(5)})_{11m}^{(0B)}\\{}&{}+\frac{5\omega}{8}(\vok_{AF}^{(5)})_{01m}^{(1B)}\bigg)+\prescript{}{0}{Y}_{2m}(\mathbf{\hat{p}})\frac{p}{3}(\vok_{AF}^{(5)})_{12m}^{(1B)}\bigg)\Bigg)^2+8p_\mu p^\mu\\{}&{}\times\sum\limits_{m_1m_2}\left(\begin{array}{r l}\small
(\vok_{AF}^{(5)})^{(1E)}_{11m_1}(\vok_{AF}^{(5)})^{(1E)}_{11m_2} {}&{}\times\frac{1}{4}\omega^2p^2\prescript{}{+1}{Y}_{1m_1}(\mathbf{\hat{p}})\prescript{}{-1}{Y}_{1m_2}(\mathbf{\hat{p}})\\
+(\vok_{AF}^{(5)})^{(1E)}_{11m_1}(\vok_{AF}^{(5)})^{(1E)}_{22m_2} {}&{}\times\frac{1}{4\sqrt{3}}\omega p^3\bigg(\prescript{}{+1}{Y}_{1m_1}(\mathbf{\hat{p}})\prescript{}{-1}{Y}_{2m_2}(\mathbf{\hat{p}})\\&\qquad\quad\qquad+\prescript{}{+1}{Y}_{2m_1}(\mathbf{\hat{p}})\prescript{}{-1}{Y}_{1m_2}(\mathbf{\hat{p}})\bigg)\\+(\vok_{AF}^{(5)})^{(1E)}_{22m_1}(\vok_{AF}^{(5)})^{(1E)}_{22m_2} {}&{}\times\frac{1}{12} p^4\prescript{}{+1}{Y}_{2m_1}(\mathbf{\hat{p}})\prescript{}{-1}{Y}_{2m_2}(\mathbf{\hat{p}})
\end{array}\right)\nonumber\\
{}&{}+8p_\mu p^\mu
\end{align*}

\begin{align*}
{}&{}\times\sum\limits_{m_1m_2}\left(\begin{array}{l}\small
(\vok_{AF}^{(5)})^{(0B)}_{11m_1}(\vok_{AF}^{(5)})^{(0B)}_{11m_2} \times\frac{1}{4}\left(\frac{2}{3}\omega^2-4p^2\right)^2\prescript{}{+1}{Y}_{1m_1}(\mathbf{\hat{p}})\prescript{}{-1}{Y}_{1m_2}(\mathbf{\hat{p}})\\
+(\vok_{AF}^{(5)})^{(0B)}_{11m_1}(\vok_{AF}^{(5)})^{(1B)}_{01m_2} \times\frac{5}{16}p^2\left(\frac{2}{3}\omega^2-4p^2\right)\prescript{}{+1}{Y}_{1m_1}(\mathbf{\hat{p}})\prescript{}{-1}{Y}_{1m_2}(\mathbf{\hat{p}})\\
+(\vok_{AF}^{(5)})^{(0B)}_{11m_2}(\vok_{AF}^{(5)})^{(1B)}_{01m_1} \times\frac{5}{16}p^2\left(\frac{2}{3}\omega^2-4p^2\right)\prescript{}{+1}{Y}_{1m_1}(\mathbf{\hat{p}})\prescript{}{-1}{Y}_{1m_2}(\mathbf{\hat{p}})\\
+(\vok_{AF}^{(5)})^{(0B)}_{11m_1}(\vok_{AF}^{(5)})^{(1B)}_{12m_2} \times\frac{1}{4\sqrt{3}}\omega p\left(\frac{2}{3}\omega^2-4p^2\right)\prescript{}{+1}{Y}_{1m_1}(\mathbf{\hat{p}})\prescript{}{-1}{Y}_{2m_2}(\mathbf{\hat{p}})\\
+(\vok_{AF}^{(5)})^{(0B)}_{11m_2}(\vok_{AF}^{(5)})^{(1B)}_{12m_1} \times\frac{1}{4\sqrt{3}}\omega p\left(\frac{2}{3}\omega^2-4p^2\right)\prescript{}{+1}{Y}_{1m_1}(\mathbf{\hat{p}})\prescript{}{-1}{Y}_{2m_2}(\mathbf{\hat{p}})\\
+(\vok_{AF}^{(5)})^{(1B)}_{01m_1}(\vok_{AF}^{(5)})^{(1B)}_{01m_2} \times\frac{25}{64}p^4\prescript{}{+1}{Y}_{1m_1}(\mathbf{\hat{p}})\prescript{}{-1}{Y}_{1m_2}(\mathbf{\hat{p}})\\
+(\vok_{AF}^{(5)})^{(1B)}_{01m_1}(\vok_{AF}^{(5)})^{(1B)}_{12m_2} \times\frac{5}{16\sqrt{3}}p^3\omega\prescript{}{+1}{Y}_{1m_1}(\mathbf{\hat{p}})\prescript{}{-1}{Y}_{2m_2}(\mathbf{\hat{p}})\\
+(\vok_{AF}^{(5)})^{(1B)}_{01m_2}(\vok_{AF}^{(5)})^{(1B)}_{12m_1} \times\frac{5}{16\sqrt{3}}p^3\omega\prescript{}{+1}{Y}_{1m_1}(\mathbf{\hat{p}})\prescript{}{-1}{Y}_{2m_2}(\mathbf{\hat{p}})\\
+(\vok_{AF}^{(5)})^{(1B)}_{12m_1}(\vok_{AF}^{(5)})^{(1B)}_{12m_2} \times\frac{1}{12}\omega^2p^2\prescript{}{+1}{Y}_{2m_1}(\mathbf{\hat{p}})\prescript{}{-1}{Y}_{2m_2}(\mathbf{\hat{p}})	
\end{array}\right)\,,
\end{align*}
which is written such for notational brevity. In compact form, it becomes
\begin{equation}\small
\boxed{\begin{aligned}
0={}&{}\left(p_\mu p^\mu\right)^2-4(p_\mu p^\mu)^2\Bigg(\sum_{m}\bigg(\prescript{}{0}{Y}_{0m}(\mathbf{\hat{p}})(\vok_{AF}^{(5)})_{00m}^{(0B)}p+\prescript{}{0}{Y}_{1m}(\mathbf{\hat{p}})\\{}&{}\times\left(\frac{\omega}{3}(\vok_{AF}^{(5)})_{11m}^{(0B)}+\frac{5\omega}{8}(\vok_{AF}^{(5)})_{01m}^{(1B)}\right)+\prescript{}{0}{Y}_{2m}(\mathbf{\hat{p}})\frac{p}{3}(\vok_{AF}^{(5)})_{12m}^{(1B)}\bigg)\Bigg)^2+8p_\mu p^\mu\\
{}&{}\times\sum\limits_{m_1m_2}\left(\begin{array}{r l}
(\vok_{AF}^{(5)})^{(0B)}_{11m_1}(\vok_{AF}^{(5)})^{(0B)}_{11m_2} {}&{}\times\frac{1}{4}\left(\frac{2}{3}\omega^2-4p^2\right)^2\prescript{}{+1}{Y}_{1m_1}(\mathbf{\hat{p}})\prescript{}{-1}{Y}_{1m_2}(\mathbf{\hat{p}})\\
+(\vok_{AF}^{(5)})^{(0B)}_{11m_1}(\vok_{AF}^{(5)})^{(1B)}_{01m_2} {}&{}\times\frac{5}{16}p^2\left(\frac{2}{3}\omega^2-4p^2\right)\prescript{}{+1}{Y}_{1m_1}(\mathbf{\hat{p}})\prescript{}{-1}{Y}_{1m_2}(\mathbf{\hat{p}})\\
+(\vok_{AF}^{(5)})^{(0B)}_{11m_2}(\vok_{AF}^{(5)})^{(1B)}_{01m_1} {}&{}\times\frac{5}{16}p^2\left(\frac{2}{3}\omega^2-4p^2\right)\prescript{}{+1}{Y}_{1m_1}(\mathbf{\hat{p}})\prescript{}{-1}{Y}_{1m_2}(\mathbf{\hat{p}})\\
+(\vok_{AF}^{(5)})^{(0B)}_{11m_1}(\vok_{AF}^{(5)})^{(1B)}_{12m_2} {}&{}\times\frac{1}{4\sqrt{3}}\omega p\left(\frac{2}{3}\omega^2-4p^2\right)\prescript{}{+1}{Y}_{1m_1}(\mathbf{\hat{p}})\prescript{}{-1}{Y}_{2m_2}(\mathbf{\hat{p}})\\
+(\vok_{AF}^{(5)})^{(0B)}_{11m_2}(\vok_{AF}^{(5)})^{(1B)}_{12m_1} {}&{}\times\frac{1}{4\sqrt{3}}\omega p\left(\frac{2}{3}\omega^2-4p^2\right)\prescript{}{+1}{Y}_{1m_1}(\mathbf{\hat{p}})\prescript{}{-1}{Y}_{2m_2}(\mathbf{\hat{p}})\\
+(\vok_{AF}^{(5)})^{(1B)}_{01m_1}(\vok_{AF}^{(5)})^{(1B)}_{01m_2} {}&{}\times\frac{25}{64}p^4\prescript{}{+1}{Y}_{1m_1}(\mathbf{\hat{p}})\prescript{}{-1}{Y}_{1m_2}(\mathbf{\hat{p}})\\
+(\vok_{AF}^{(5)})^{(1B)}_{01m_1}(\vok_{AF}^{(5)})^{(1B)}_{12m_2} {}&{}\times\frac{5}{16\sqrt{3}}p^3\omega\prescript{}{+1}{Y}_{1m_1}(\mathbf{\hat{p}})\prescript{}{-1}{Y}_{2m_2}(\mathbf{\hat{p}})\\
+(\vok_{AF}^{(5)})^{(1B)}_{01m_2}(\vok_{AF}^{(5)})^{(1B)}_{12m_1} {}&{}\times\frac{5}{16\sqrt{3}}p^3\omega\prescript{}{+1}{Y}_{1m_1}(\mathbf{\hat{p}})\prescript{}{-1}{Y}_{2m_2}(\mathbf{\hat{p}})\\
+(\vok_{AF}^{(5)})^{(1B)}_{12m_1}(\vok_{AF}^{(5)})^{(1B)}_{12m_2} {}&{}\times\frac{1}{12}\omega^2p^2\prescript{}{+1}{Y}_{2m_1}(\mathbf{\hat{p}})\prescript{}{-1}{Y}_{2m_2}(\mathbf{\hat{p}})\\
+(\vok_{AF}^{(5)})^{(1E)}_{11m_1}(\vok_{AF}^{(5)})^{(1E)}_{11m_2} {}&{}\times\frac{1}{4}\omega^2p^2\prescript{}{+1}{Y}_{1m_1}(\mathbf{\hat{p}})\prescript{}{-1}{Y}_{1m_2}(\mathbf{\hat{p}})\\
+(\vok_{AF}^{(5)})^{(1E)}_{11m_1}(\vok_{AF}^{(5)})^{(1E)}_{22m_2} {}&{}\times\frac{1}{4\sqrt{3}}\omega p^3\bigg(\prescript{}{+1}{Y}_{1m_1}(\mathbf{\hat{p}})\prescript{}{-1}{Y}_{2m_2}(\mathbf{\hat{p}})\\&\qquad\qquad\quad+\prescript{}{+1}{Y}_{2m_1}(\mathbf{\hat{p}})\prescript{}{-1}{Y}_{1m_2}(\mathbf{\hat{p}})\bigg)\\+(\vok_{AF}^{(5)})^{(1E)}_{22m_1}(\vok_{AF}^{(5)})^{(1E)}_{22m_2} {}&{}\times\frac{1}{12} p^4\prescript{}{+1}{Y}_{2m_1}(\mathbf{\hat{p}})\prescript{}{-1}{Y}_{2m_2}(\mathbf{\hat{p}})
\end{array}\right)
\end{aligned}}\normalsize\label{vo dispersion relation in component form for d5}
\end{equation}

This boxed equation is the most general dispersion relation for VOM at the leading order in $d$. As a consistency check, one can easily show that the only non-vanishing component in the isotropic limit, that is $(\vok_{AF}^{(5)})^{(0B)}_{00m}$, gives rise to the earlier worked out dispersion relation, \equref{vod in 5}.

As stated out in the beginning of the section, the main purpose for the derivation of the most general form of the dispersion relation of VOM at the leading order was the sake of completeness; as a matter of fact, the physically relevant dispersion relation is the one which can be extracted out of the general formula above by restricting the attention to the coefficient subspace set $\{\hat{k}_{AF}^{(cn)}, \hat{k}_{AF}^{(bf)}\}$. That amounts to the drop of the last term in \equref{vo dispersion relation in component form for d5} as a result of which the simplified dispersion relation reads as
\begin{equation*}
\begin{aligned}
0={}&{}\left(p_\mu p^\mu\right)^2-4(p_\mu p^\mu)^2\Bigg(\sum_{m}\bigg(\prescript{}{0}{Y}_{0m}(\mathbf{\hat{p}})(\vok_{AF}^{(5)})_{00m}^{(0B)}p\\{}&{}+\prescript{}{0}{Y}_{1m}(\mathbf{\hat{p}})\left(\frac{\omega}{3}(\vok_{AF}^{(5)})_{11m}^{(0B)}+\frac{5\omega}{8}(\vok_{AF}^{(5)})_{01m}^{(1B)}\right) +\prescript{}{0}{Y}_{2m}(\mathbf{\hat{p}})\frac{p}{3}(\vok_{AF}^{(5)})_{12m}^{(1B)}\bigg)\Bigg)^2\,.
\end{aligned}
\end{equation*}

Referring to the form of \equref{General Vacuum-orthogonal dispersion relation3}, we can reorganize this equation as
\begin{equation*}
\begin{aligned}
0={}&{}(p_\mu p^\mu)^2\Bigg\{1-2\sum_{m}\Bigg(\prescript{}{0}{Y}_{0m}(\mathbf{\hat{p}})(\vok_{AF}^{(5)})_{00m}^{(0B)}p+\prescript{}{0}{Y}_{1m}(\mathbf{\hat{p}})\bigg(\frac{\omega}{3}(\vok_{AF}^{(5)})_{11m}^{(0B)}\\{}&{}+\frac{5\omega}{8}(\vok_{AF}^{(5)})_{01m}^{(1B)}\bigg)+\prescript{}{0}{Y}_{2m}(\mathbf{\hat{p}})\frac{p}{3}(\vok_{AF}^{(5)})_{12m}^{(1B)}\Bigg)\Bigg\}\Bigg\{1+2\sum_{m}\Bigg(\prescript{}{0}{Y}_{0m}(\mathbf{\hat{p}})(\vok_{AF}^{(5)})_{00m}^{(0B)}p\\
{}&{}+\prescript{}{0}{Y}_{1m}(\mathbf{\hat{p}})\left(\frac{\omega}{3}(\vok_{AF}^{(5)})_{11m}^{(0B)}+\frac{5\omega}{8}(\vok_{AF}^{(5)})_{01m}^{(1B)}\right)+\prescript{}{0}{Y}_{2m}(\mathbf{\hat{p}})\frac{p}{3}(\vok_{AF}^{(5)})_{12m}^{(1B)}\Bigg)\Bigg\}\,,
\end{aligned}
\end{equation*}
from which the roots can be extracted as 
\begin{equation*}
\begin{aligned}
\omega={}&{} p\,,\\
\omega={}&{}\pm\frac{1}{2a}-\frac{b}{a}p\,,
\end{aligned}
\end{equation*}
where
\begin{equation*}
\begin{aligned}
a:={}&{}\sum\limits_m\prescript{}{0}{Y}_{1m}(\mathbf{\hat{p}})\left(\frac{1}{3}(\vok_{AF}^{(5)})_{11m}^{(0B)}+\frac{5}{8}(\vok_{AF}^{(5)})_{01m}^{(1B)}\right)\,,\\{}
b:={}&{}\sum\limits_m\left(\prescript{}{0}{Y}_{0m}(\mathbf{\hat{p}})(\vok_{AF}^{(5)})_{00m}^{(0B)}+\prescript{}{0}{Y}_{2m}(\mathbf{\hat{p}})\frac{1}{3}(\vok_{AF}^{(5)})_{12m}^{(1B)}\right)\,.
\end{aligned}
\end{equation*}

Not unlike \equref{eq: isotropic solution}, this solution is of the form of a spurious term and a perturbation addition. Indeed, it can be shown that the first term giving rise to the spurious nature is present at higher orders as well, and what is actually modified at higher orders is simply the addition of new perturbative terms.
\subsection{Coefficient Subspace of Leading Order VOM}
\label{Section: coefficients of leading vom}

Throughout the thesis, so far, some generic analysis have been studied and their effects on the coefficient space have been investigated. These include the derivation of the dispersion relations for various cases, examination of polarization vectors and the construction of the propagator; additionally, coefficient space has been examined and it was shown that it splits into different subspaces. It was stated that for any particular analysis of Lorentz and CPT violating nonminimal photon sector, the relations that components of a priori chosen $\hat{k}_{AF}$ satisfy determine the nature of solutions, if they exist.

For the vacuum orthogonal models, the examination yielded one step more stringent bound and dictated that birefringent solutions need to be spurious. In \secref{Section: leading vom}, we indeed explicitly showed that $\hat{k}_{AF}\in\hat{k}_{AF}^{(bf)}$ is spurious\footnotemark\footnotetext{Although the explicit calculation is done only for the leading order case, that dispersion relation is shown to be spurious at the leading order is sufficient to deduce that it is spurious at any order, because addition of higher order contributions to the dispersion relation will only bring more perturbative terms to the solution without affecting the divergent term, though this is not proved in this thesis as it is beyond the scope of this study.}, with which that \mbox{$\hat{k}_{AF}\in\hat{k}_{AF}^{(np)}$} is nonphysical enforces the only solution to be the conventional one. However, the question of whether there actually is a nontrivial coefficient subspace $\hat{k}_{AF}^{(cn)}$ has not been addressed yet. What is meant here is that, it may very well turn out to be the case that $\hat{k}_{AF}^{(cn)}$ is simply the trivial null space, which indicates that there is no LV whatsoever in the first place, and hence detracts from the earlier result that VOM remain conventional despite the presence of LV.

The relevant constraints for VOM are given by \equref{first condition for vacuum polarization} and \equref{second condition for vacuum polarization}. In the leading order, they become
\begin{subequations}
\begin{align}
0={}&{}\sum\limits_{n}(\vok_{AF}^{(5)})^{(0B)}_{njm}\left(-\frac{4}{5}+\frac{4j(6+j)}{5(3-n+j)(1-n+j)}\right)+\sum\limits_{n}(\vok_{AF}^{(5)})^{(1B)}_{njm}\nonumber\\
{}&{}\times\left(\frac{1}{n+2}-\frac{5}{(n+4)(n+2)}-\frac{(2-n)(n+4)}{5(2-n+j)(n+2)}\right)\,,\\
0={}&{}\sum\limits_{n}\left(-\frac{2j(4+j)}{(3-n+j)(1-n+j)}(\vok_{AF}^{(5)})^{(0B)}_{njm}+\frac{5}{n+4}(\vok_{AF}^{(5)})^{(1B)}_{njm}\right)\,,\\
0={}&{}\sum\limits_{n}(\vok_{AF}^{(5)})^{(1E)}_{njm}\,,
\end{align}\label{d=5 conditions}
\end{subequations}
where merely $d=5$ is imposed.

\begin{table}
\caption[Index Ranges for Leading Order VOM Coefficients]{Index Ranges for Leading Order VOM Coefficients. The index range for the arbitrary order is given by \tabref{Table-vocir}, from which the leading order case can be extracted as below. As explained there, the frequency dependence $n$ and the total angular momentum $j$ are restricted as below due to the way the coefficients are constructed.}
\label{Table-rocflovom}
\begin{tabularx}{\textwidth}{XXX}
\hline\hline Coefficient & $n$ & $j$\\\hline\\[-0.15in]
$(\vok_{AF}^{(5)})^{(0B)}_{njm}$ & 0, 1, 2 & $n$, $n-2$ $\ge 0$\\\\[-0.15in]
$(\vok_{AF}^{(5)})^{(1B)}_{njm}$ & 0, 1, 2 & $n+1$, $n-1$ $\ge 1$\\\\[-0.15in]
$(\vok_{AF}^{(5)})^{(1E)}_{njm}$ & 1, 2 & $n$\\\hline\hline
\end{tabularx}
\end{table}

It is clear from \tabref{Table-rocflovom} that the only nonvanishing relevant coefficients for the leading order VOM are \mbox{$(\vok_{AF}^{(5)})^{(0B)}_{00m}$}, \mbox{$(\vok_{AF}^{(5)})^{(0B)}_{11m}$}, \mbox{$(\vok_{AF}^{(5)})^{(0B)}_{22m}$}, \mbox{$(\vok_{AF}^{(5)})^{(0B)}_{20m}$}, \mbox{$(\vok_{AF}^{(5)})^{(1B)}_{01m}$}, \mbox{$(\vok_{AF}^{(5)})^{(1B)}_{12m}$}, \mbox{$(\vok_{AF}^{(5)})^{(1B)}_{23m}$}, \mbox{$(\vok_{AF}^{(5)})^{(1B)}_{21m}$}, \mbox{$(\vok_{AF}^{(5)})^{(1E)}_{11m}$} and \mbox{$(\vok_{AF}^{(5)})^{(1E)}_{22m}$}\footnotemark\footnotetext{The $m$ values are irrelevant for practical purposes as the conditions that they must satisfy do not explicitly depend on the value of $m$.}. These coefficients should satisfy the conditions \equref{d=5 conditions} for each possible $j$ value. Let us examine them one by one.
\begin{itemize}
\item For $j=0$:
\begin{equation*}
\begin{aligned}
0={}&{}\frac{-4}{5}\sum\limits_{n}(\vok_{AF}^{(5)})^{(0B)}_{n0m}+\sum\limits_{n}(\vok_{AF}^{(5)})^{(1B)}_{n0m}\left(\frac{1}{n+2}-\frac{5}{(n+4)(n+2)}-\frac{(n+4)}{5(n+2)}\right)\\
0={}&{}\sum\limits_{n}\frac{5}{n+4}(\vok_{AF}^{(5)})^{(1B)}_{n0m}\\
0={}&{}\sum\limits_{n}(\vok_{AF}^{(5)})^{(1E)}_{n0m}
\end{aligned}
\end{equation*}
As the only coefficients with $j=0$ are $(\vok_{AF}^{(5)})^{(0B)}_{00m}$ and $(\vok_{AF}^{(5)})^{(0B)}_{20m}$, above conditions amounts to
\begin{equation}
0=(\vok_{AF}^{(5)})^{(0B)}_{00m}+(\vok_{AF}^{(5)})^{(0B)}_{20m}\,.\label{d=5, j=0 conditions}
\end{equation}

\item For $j=1$:
\begin{equation*}
\begin{aligned}
0={}&{}\sum\limits_{n}(\vok_{AF}^{(5)})^{(0B)}_{n1m}\left(-\frac{4}{5}+\frac{28}{5(4-n)(2-n)}\right)\nonumber\\
{}&{}+\sum\limits_{n}(\vok_{AF}^{(5)})^{(1B)}_{n1m}\left(\frac{1}{n+2}-\frac{5}{(n+4)(n+2)}-\frac{(2-n)(n+4)}{5(3-n)(n+2)}\right)\\
0={}&{}\sum\limits_{n}\left(-\frac{10}{(4-n)(2-n)}(\vok_{AF}^{(5)})^{(0B)}_{n1m}+\frac{5}{n+4}(\vok_{AF}^{(5)})^{(1B)}_{n1m}\right)\\
0={}&{}\sum\limits_{n}(\vok_{AF}^{(5)})^{(1E)}_{n1m}
\end{aligned}
\end{equation*}
As the only coefficients with $j=1$ are \mbox{$(\vok_{AF}^{(5)})^{(0B)}_{11m}$}, \mbox{$(\vok_{AF}^{(5)})^{(1B)}_{01m}$}, \mbox{$(\vok_{AF}^{(5)})^{(1B)}_{21m}$} and \mbox{$(\vok_{AF}^{(5)})^{(1E)}_{11m}$}, the above conditions amounts to
\begin{subequations}
\begin{align}
0={}&{}\frac{16}{15}(\vok_{AF}^{(5)})^{(0B)}_{11m}-\frac{47}{120}(\vok_{AF}^{(5)})^{(1B)}_{01m}+\frac{1}{24}(\vok_{AF}^{(5)})^{(1B)}_{21m}\,,\\
0={}&{}-\frac{10}{3}(\vok_{AF}^{(5)})^{(0B)}_{11m}+\frac{5}{4}(\vok_{AF}^{(5)})^{(1B)}_{01m}+\frac{5}{6}(\vok_{AF}^{(5)})^{(1B)}_{21m}\,,\\
0={}&{}(\vok_{AF}^{(5)})^{(1E)}_{11m}\,.
\end{align}\label{d=5, j=1 conditions}
\end{subequations}

\item For $j=2$:
\begin{equation*}
\begin{aligned}
0={}&{}\sum\limits_{n}(\vok_{AF}^{(5)})^{(0B)}_{n2m}\left(-\frac{4}{5}+\frac{64}{5(5-n)(3-n)}\right)\nonumber\\
{}&{}+\sum\limits_{n}(\vok_{AF}^{(5)})^{(1B)}_{n2m}\left(\frac{1}{n+2}-\frac{5}{(n+4)(n+2)}-\frac{(2-n)(n+4)}{5(4-n)(n+2)}\right)\\
0={}&{}\sum\limits_{n}\left(-\frac{24}{(5-n)(3-n)}(\vok_{AF}^{(5)})^{(0B)}_{n2m}+\frac{5}{n+4}(\vok_{AF}^{(5)})^{(1B)}_{n2m}\right)\\
0={}&{}\sum\limits_{n}(\vok_{AF}^{(5)})^{(1E)}_{n2m}
\end{aligned}
\end{equation*}
As the only coefficients with $j=2$ are \mbox{$(\vok_{AF}^{(5)})^{(0B)}_{22m}$}, \mbox{$(\vok_{AF}^{(5)})^{(1B)}_{12m}$} and \mbox{$(\vok_{AF}^{(5)})^{(1E)}_{22m}$}, above conditions amounts to
\begin{subequations}
\begin{align}
0={}&{}\frac{52}{15}(\vok_{AF}^{(5)})^{(0B)}_{22m}-\frac{1}{9}(\vok_{AF}^{(5)})^{(1B)}_{12m}\,,\\
0={}&{}-8(\vok_{AF}^{(5)})^{(0B)}_{22m}+(\vok_{AF}^{(5)})^{(1B)}_{12m}\,,\\
0={}&{}(\vok_{AF}^{(5)})^{(1E)}_{22m}\,.
\end{align}\label{d=5, j=2 conditions}
\end{subequations}

\item For $j=3$:
The only coefficient with $j=3$ is $(\vok_{AF}^{(5)})^{(1B)}_{23m}$. Then, from \equref{d=5 conditions}, we simply have
\begin{subequations}
\begin{align}
0={}&{}\frac{1}{24}(\vok_{AF}^{(5)})^{(1B)}_{23m}\,,\\
0={}&{}\frac{5}{6}(\vok_{AF}^{(5)})^{(1B)}_{23m}\,.
\end{align}\label{d=5, j=3 conditions}
\end{subequations}
\end{itemize}

Therefore, the coefficient subspace $\hat{k}_{AF}^{(cn)}$ is spanned by the \emph{the only nonzero coefficients} \mbox{$(\vok_{AF}^{(5)})^{(0B)}_{00m}$}, \mbox{$(\vok_{AF}^{(5)})^{(0B)}_{11m}$}, \mbox{$(\vok_{AF}^{(5)})^{(0B)}_{20m}$}, \mbox{$(\vok_{AF}^{(5)})^{(1B)}_{01m}$} and \mbox{$(\vok_{AF}^{(5)})^{(1B)}_{11m}$}, where they obey the constraints
\begin{subequations}
\begin{align}
(\vok_{AF}^{(5)})^{(0B)}_{20m}={}&{}-(\vok_{AF}^{(5)})^{(0B)}_{00m}\,,\\
(\vok_{AF}^{(5)})^{(1B)}_{01m}={}&{}\frac{296}{109}(\vok_{AF}^{(5)})^{(0B)}_{11m}\,,\\
(\vok_{AF}^{(5)})^{(1B)}_{21m}={}&{}-\frac{8}{109}(\vok_{AF}^{(5)})^{(0B)}_{11m}\,,
\end{align}\label{eq: leading order VOM coefficient subspace}
\end{subequations}
as can be seen from \equref{d=5, j=0 conditions}, \equref{d=5, j=1 conditions}, \equref{d=5, j=2 conditions} and \equref{d=5, j=3 conditions}.

The above result indicates that there indeed exists a nontrivial coefficient subspace spanned by two free parameters, say $(\vok_{AF}^{(5)})^{(0B)}_{00m}$ and $(\vok_{AF}^{(5)})^{(0B)}_{11m}$, that constitute $\hat{k}_{AF}^{(cn)}$ in the leading order in $d$. Thus, the assertion that VOM remain VOM at all orders, which was made in \charef{vacuum orthogonal model}, is indeed a nontrivial assertion: The dispersion relations and the polarization vectors of the model can still remain conventional despite the presence of two LV and CPTV coefficients in the theory.

Here, there is one question that remains unanswered: Is there indeed a nonzero LV in the model even though there are nonzero LV coefficients? In other words, do the LV effects generated by these LV coefficients cancel each other or not?

This question is critical because it is futile to say that the solutions of the model are conventional if the overall Lorentz violation generated by each coefficient adds up to zero: That model would be equivalent to the conventional theory with some redundant coefficients in it.

The most straightforward way to check this is to examine if there is a nonzero $\hat{k}_{AF}$ given rise by these coefficients. From \equref{eq: leading order VOM coefficient subspace} and \equref{Equ: k_i}, it can be shown that $(\hat{k}_{AF})_0$ takes the form
\begin{equation}
(\hat{k}_{AF})_0=p^2\sum\limits_{m}\prescript{}{0}{Y}_{jm}(\mathbf{\hat{p}})(k_{AF}^{(5)})^{(0B)}_{11m}\,,\label{eq: k_0 in leading order VOM}
\end{equation}
where contributions of $(\vok_{AF}^{(5)})^{(0B)}_{20m}$ and $(\vok_{AF}^{(5)})^{(0B)}_{00m}$ cancel one another. This indicates that although we have a coefficient subspace spanned by two free parameters, only one of them actually generates a nonzero LV in total. Nonetheless, since this proves that there can be nonzero LV, that is non-null $\hat{k}_{AF}^{(cn)}$, the claims and conclusions made throughout the thesis have been justified.

The results are summarized in \tabref{Table-cslovom}.
\begin{table}[h]
\caption[Coefficient Space of Leading Order VOM]{Coefficient Space of Leading Order VOM. It is shown that there is indeed a nontrivial coefficient subspace for $\hat{k}_{AF}$ for which the LV inserted into the theory does not produce any alteration in the dispersion relations and the polarization vectors. In other words, for the LV whose coefficients and the constraints that they satisfy are given below, the solutions for the photon field remain conventional, although the propagator does get a modification. Below, the properties of this coefficient subspace, $\hat{k}_{AF}^{(cn)}$, is given for the leading order case, that is $d=5$. The coefficients are left compact in their $m$ values, which run from $-j$ to $j$.}
\label{Table-cslovom}
\begin{tabularx}{\textwidth}{X@{\hskip -2cm}X}
\hline\hline\\[-0.20in] Free Coefficients: & $(\vok_{AF}^{(5)})^{(0B)}_{00m}$ \& $(\vok_{AF}^{(5)})^{(0B)}_{11m}$\\\\
Nonzero Coefficients:& $(\vok_{AF}^{(5)})^{(0B)}_{00m}$, $(\vok_{AF}^{(5)})^{(0B)}_{11m}$, $(\vok_{AF}^{(5)})^{(0B)}_{20m}$, $(\vok_{AF}^{(5)})^{(1B)}_{01m}$, $(\vok_{AF}^{(5)})^{(1B)}_{11m}$\\\\
Constraint Relations:& $\begin{aligned}
(\vok_{AF}^{(5)})^{(0B)}_{20m}={}&{}-(\vok_{AF}^{(5)})^{(0B)}_{00m}\,,\\
(\vok_{AF}^{(5)})^{(1B)}_{01m}={}&{}\frac{296}{109}(\vok_{AF}^{(5)})^{(0B)}_{11m}\,,\\
(\vok_{AF}^{(5)})^{(1B)}_{21m}={}&{}-\frac{8}{109}(\vok_{AF}^{(5)})^{(0B)}_{11m}
\end{aligned}$\\\\
Field Theoretical Properties:& \begin{tabular}{l}
Conventional Dispersion Relation\\
Conventional Polarization Vectors\\
Nonconventional Propagator\footnotemark\end{tabular}\\\hline\hline
\end{tabularx}
\end{table}
\footnotetext{The propagator can be obtained from inserting \equref{eq: k_0 in leading order VOM} into \equref{eq: Physical vacuum orthogonal propagator}.}
\chapter{Discussion and Conclusion}
\label{CHAPTER:CONCLUSION}
The quest for Quantum Theory of Gravity has engaged the physics community almost for a century now. The abundance of candidate theories, however, is not sufficient to solve the mystery as discussed in the Introduction: We are unable to probe into the Planck realm, hence are unable to use the beautiful --despite sometimes scary for a theoretician-- results of the experiments to decide which theories will survive.

In this thesis, one of the important alternatives to access to the Planck physics has been studied, that is the Lorentz and CPT violation. Indeed, as indicated in the Introduction, it is much more practical to seek the exotic effects of the Quantum Theory of Gravity --whatever it is-- in the attainable energies, than to search for direct effects in the Planck energies; and, Lorentz and CPT violations are perfect candidates for such an exotic effect, both because Lorentz symmetry already needs a modification for Planck level physics, and because they are already violated in most of the candidate Planck level theories.

In \appref{CHAPTER:LORENTZ}, some of the various Lorentz violating models in the literature are briefly introduced; yet, an action level effective field theoretical approach to the Lorentz violation has been shown to be the most general formalism for inclusion of LV effects. In this thesis, we have worked with this formalism, which is called \emph{Standard Model Extension}, and explicitly analyzed the photon sector. The details of this framework including its philosophy, along with the brief explanation of the \emph{Spontaneous Symmetry Breaking} mechanism, are left to the appendices as well as the details of the so-called \emph{helicity basis}, which is the basis most suited for an expansion regarding to the direct relevance to observations and experiments, and which is extensively used throughout this thesis.

In \charef{CHAPTER:PRELIMINARIES}, the Lorentz and CPT symmetries are reviewed and how they should be broken is elaborated. Indeed, it is a nontrivial issue to break the Lorentz symmetry, and yet to have a theory still independent of the observer. This is a fundamental issue as physics, with or without Lorentz symmetry, should always be independent of the observer, which is the core of Relativity Principle. In SME, this is achieved via the discrimination between the so-called observer Lorentz transformations and particle Lorentz transformations, as explained there. The EFT approach and the renormalizability issue are also covered in that chapter where current bounds are presented, indicating the necessity of consideration of nonrenormalizable photon sector, also called nonminimal photon sector which is the main content of this study. The Lagrangian of this model is constructed and spherical decomposition of LVT in the introduced helicity basis are provided there as well. 

The framework of SME allows all Lorentz violating action level terms, some of which violate the CPT invariance and some of which do not; and, they are all considered in the \charef{CHAPTER:PRELIMINARIES}. In \charef{CHAPTER:CPT ODD PHOTON} however, the model is restricted to CPT violating part only, as that is the focus of this thesis and is analyzed throughly from a quantum field theoretical point of view. This analysis comprised of the derivation of dispersion relation, polarization vectors and the propagator.

The dispersion relation for the general Lorentz and CPT violating photon is straightforwardly calculated both in terms of the general LVT and in terms of the spherically decomposed coefficients; however, the straightforward calculation of polarization vectors are shown to be quite formidable. An alternative procedure called rank-nullity is employed, which enabled the extraction of not only polarization vectors but also the conditions on the components of LVT for each possible solution: the birefringent, conventional, and gauge only namely. Judging from this result, corresponding coefficient subspaces are introduced and their resultant properties are summarized in \tabref{Table-cscops}. This means that the LV effects of any Lorentz and CPT violating nonminimal photon model can be readily looked up from this table by simply using the relations that the components of LVT satisfy in the helicity basis.

The calculation of propagator is shown to be even more formidable. Different approaches are carried on to approximate the propagator where two of them --namely an ansatz and the perturbation expansion-- yielded covariant but not analytical results, and where the last method --explicit expansion in the helicity basis-- yielded an analytic result which obviously fails to be in a covariant form. It was argued why the last method is superior to the other ones as the nonphysical possibilities of LVT, that is the LVT which would yield gauge solution only, can easily be removed from the propagator if it is in explicitly helicity basis. However, how one would use this form of the propagator in an actual Feynman calculus is not addressed as it is beyond the scope of this thesis.

A general problem with any analysis of nonminimal SME is the enormous number of LVT that need to be considered in the model, which complicates the comparison of the theory with the experiments and makes bounding the individual coefficients more nontrivial. That's why most of the analysis in the literature are done addressing only some subsets of the whole coefficient set. These subsets are called special models, and some of the most known ones are introduced and discussed at the end of \charef{CHAPTER:CPT ODD PHOTON}.

In \charef{vacuum orthogonal model}, the analysis is further restricted to one of these special models, that is the so-called vacuum orthogonal model. The dispersion relations, polarization vectors and the propagator for this special model are extracted. It is found that the birefringent solution possibility which is valid for the general case becomes unphysical as the birefringent dispersion relation in this model turns out to be a spurious solution; in other words, it becomes a high energy effect to be ignored in the low energies, indicating that the only physically possible polarization vectors and the dispersion relation they satisfy are the usual conventional ones.

That the only possibility for the VOM is the conventional case is stated as \emph{vacuum orthogonal model remains vacuum orthogonal at all orders}. This refers to the fact that the vacuum orthogonal models in general should not have leading order LV effects, but what is shown in this thesis is that they do not have LV effects at any order if they are restricted to the CPTV LVT. This simply means that observing no deviation from the Lorentz symmetry in the vacuum properties of the photon does not indicate that the underlying model is Lorentz invariant as LV VOM also gives rise to such a photon. The coefficient subspace resulting in such a photon is calculated at \secref{vom coefficient}.

Although it was deduced that the birefringent solutions are spurious and that there is a coefficient subspace of LVT which does not lead to any LV effect on vacuum, the spurious nature is not explicitly demonstrated and whether the coefficient subspace is not a null one is not addressed in the general model. In order to avoid unnecessary cumbersome calculations, the VOM is analyzed in special limits in \charef{CHAPTER:SPECIAL MODEL}. In the first case, the VOM is restricted to isotropic coefficients only, where the birefringent solutions are explicitly shown to be spurious. In addition, although it is not useful for this thesis, the so-called ring coefficient form of the dispersion relation is derived for the isotropic VOM so as to provide a complete and convenient source for any possible future analysis. In the second case, the leading order VOM is analyzed --that is VOM with LVT of $d=5$ only-- for which the birefringent dispersion relations are shown to be spurious as well. In addition, the coefficient subspace is analyzed and is shown to be a nontrivial one: It is concluded that there indeed exists a form of LV which does not produce any observable effect on the vacuum whatsoever, although these LVT can be tested by other means such as laboratory tests which are beyond the scope of this thesis.

%
\bibliography{thesis}
\bibliographystyle{unsrt}

\appendix
\chapter{Lorentz Violating Models}
\label{CHAPTER:LORENTZ}
The study of LV is an actively growing area essentially because of two reasons. The first one is that there has never been an experiment which proved that Lorentz symmetry should be absolutely correct; hence, we would at worst strengthen the bounds on the possible deviations from Lorentz symmetry even though we do not observe a violation. The second one is that it would be a profound breakthrough should LV be discovered, as that would invalidate, at least diminish the validity of, almost all modern theories of physics.

The truth is that the validity of Einstein's relativity has been questioned ever since he built it in 1905. There has been scattered experiments throughout the history; however, a systematic examination of how the Lorentz symmetry would be violated had not been undergone up until the end of 1990s. And it is actually this systematic investigation that inspired the community for the search on LV, as it turned out that there are quite many possible forms of LV among which some were never thought of before.

The systematic analysis of LV is conducted through the model called \emph{Standard Model Extension} (SME), which is based on effective field theory, and whose details will be extensively discussed in the next section. The other formalisms dealing with LV will be briefly introduced in \secref{Section: other formalisms}, but will not be elaborated as they are shown to be contained within SME.
\section{Standard Model Extension}
\label{Section: sme in general}

\subsection{Philosophy of SME}
\label{Section: philosophy of sme}
During the investigation of LV, it turned out that most classic tests of rotation and boost invariance are insufficient to determine whether Lorentz symmetry is broken or not, as there are some ways of LV whose effects are not distinguishable in such experiments. It is realized in the end of 20th century that theoretical consideration of all possibilities of LV is required before determining the reach of an experiment into the ways that Lorentz symmetry is violated; and is also realized that such a systematic framework is possible only with a suitably general EFT, as EFT is the only known framework that allows a theory of exotic physics to be able to explain every already known things about elementary particle dynamics. This dynamical approach is in contrast to the kinematic approaches, such as Robertson's framework and its Mansouri-Sexl extension, where kinematic approaches are severely restricted in scope, and cannot analyze all possible LV effects\cite{hep-ph/0611177}. Such an approach is briefly discussed in \secref{Section: robertson}.

Being an EFT, SME contains local operators that can be built up from SM fields. Then, all possible forms of local operators under some specific restrictions are included into the SME Lagrangian. The form of Lagrangian will be examined in \secref{Section: msme in general}, and the mentioned restrictions will be explained below; however, we should stress an important point before those: Not every different operator in SME needs to correspond to a distinct physical Lorentz violation; that is, the forms of physically distinguishable LV are over-counted in SME, where the effort of extracting different operators resulting in the same physical LV is actively ongoing. We will not be analyzing such efforts in this thesis, but concisely, the over-countings are corrected either by eliminating some coefficients via field-redefinitions, coordinate-redefinitions, or gamma-matrix redefinitions, and by absorbing some coefficients into others\cite{hep-ph/9703464,hep-ph/9809521}.

The most important restriction in the construction of the SME Lagrangian is that all terms should be properly contracted, hence are coordinate Lorentz scalars, so that coordinate independence is guaranteed. This is strictly important because observer Lorentz invariance should not be broken as explained in \secref{Section: olt and plt}. In addition to this, the desirable properties like usual gauge structure, conventional quantization, energy-momentum conservation, hermiticity resulting in microcausality and positivity of energy are all expected to hold due to the construction of the framework and that the breaking is spontaneous\cite{hep-ph/9809521}. Additional conditions --such as spacetime translational invariance, rotational invariance, power-counting renormalizability-- can be assumed in the model at will, albeit such assumptions would reduce the coefficient space of LVT.

\subsection{SME Lagrangians}
\label{Section: msme in general}
The general form of SME Lagrangian is
\begin{equation*}
\mathcal{L}_\text{SME}=\mathcal{L}_\text{SM}+\delta\mathcal{L}\,,
\end{equation*}
where the correction term  $\delta\mathcal{L}$ includes all possible forms of combinations of SM operators contracted with LV tensor coefficients, where these coefficients are assumed to be the vacuum expectation values of some fields in the fundamental Planck scale theory as explained in \appref{CHAPTER:SPONTANEOUS}. These combinations are called \emph{Lorentz Violating Terms} (LVT).

In general, all possible combinations of LVT are included into $\mathcal{L}_\text{SME}$; however, only those with the power-counting renormalizability are considered under \emph{m}SME\cite{hep-ph/9703464}, to be defined in the next section. The combinations typically consist of SM fields and a coefficient contracted with these fields, introducing the LV. These coefficients are assumed to be vacuum expectation value of some Lorentz tensors in the Planck level theory.

Regarding the fact that the overall number of LVT can be quite huge, it is preferred to work in sector by sector; in addition, this bypasses any ambiguity regarding the uniqueness of the coefficients of different sectors as some coefficients in different sectors may actually correspond to the same physical effect. Then,
\begin{equation*}
\delta\mathcal{L}=\mathcal{L}_{\text{lepton}}+\mathcal{L}_{\text{quark}}+\mathcal{L}_{\text{Yukawa}}+\mathcal{L}_{\text{Higgs}}+\mathcal{L}_{\text{gauge}}\,,
\end{equation*}
where each sector can also be devided according to their CPT properties, such as
\begin{equation*}
\mathcal{L}_{\text{Higgs}}=\mathcal{L}^{\text{CPT-odd}}_{\text{Higgs}}+\mathcal{L}_{\text{Higgs}}^{\text{CPT-even}}\,.
\end{equation*}

The general form of all LVT in each sector can be found in \cite{hep-ph/9809521}, an interested reader is advised to consult there. A typical form is as follows:
\begin{equation*}
\mathcal{L}^{\text{CPT-odd}}_{\text{Higgs}}=i(k_\phi)^\mu\phi^\dagger D_\mu\phi+\text{H.c.}\,,
\end{equation*}
where $D_\mu$ is the covariant derivative and $(k_\phi)^\mu$ is the CPT-odd coefficient of Lorentz violation in Higgs Sector. 

The extended QED can be obtained from $\delta\mathcal{L}$ by initially breaking the SU(2)$\times$U(1) symmetry, eliminating the gluon and weak boson fields ($G_\mu$ and $W^\pm_\mu$, $Z^0_\mu$ respectively), and finally letting physical Higgs field vanish without touching the expectation value of the Higgs doublet\footnotemark\footnotetext{This expectation value should remain to generate fermion masses.}. According to the last reference, only the photon remains as a boson after this process, hence we have the extended QED.

The photon sector of this extended QED can be written as
\begin{equation}
\mathcal{L}^{\text{QED}}_{\text{photon}}=-\frac{1}{4}F_{\mu\nu}F^{\mu\nu}-\frac{1}{4}(k_F)_{\kappa\lambda\mu\nu}F^{\kappa\lambda}F^{\mu\nu}+\frac{1}{2}(k_{AF})^\kappa\epsilon_{\kappa\lambda\mu\nu}A^\lambda F^{\mu\nu}\,, \label{Eq: Minimal QED Lagrangian}
\end{equation}
where the first term is the conventional one, and the second and third ones are CPT-even and CPT-odd LVT respectively.
\subsection{Expansion of SME Beyond Renormalizability}
\label{Section: nmsme in general}

The SME in its original form comprised of renormalizable LVT, as renormalizability is regarded a desirable asset of a field theoretical model. However, it is also generally assumed in high energy physics community that GR and SM are simply low energy limits of a fundamental theory in Planck level; and since the GR is nonrenormalizable, the low energy description of this fundamental theory can very well be expected to be nonrenormalizable. From the experience, we can expect a smooth transition between the low-energy description and the underlying theory: With this approach, we can expend the low energy description of this unknown theory over a mass scale in energy; naturally, the zeroth order term would be SM itself. Then, we can rename the renormalizable SME introduced earlier as \emph{Minimal Standard Model Extension} (\emph{m}SME), and take it as the first order correction in the expansion. Then, all possible terms restricted by the SME conditions except renormalizability constitute the higher order corrections, and are called \emph{non-Minimal Standard Model Extension} (\emph{m}SME).

The methodology to go from minimal to nonminimal is quite trivial: To preserve observer Lorentz invariance, the newly introduced elements should be Lorentz tensors to be properly contracted. Remaining consistent with current observational data, this can be achieved only by introducing higher derivatives in the Lagrangian \cite{0905.0031}. Hence, the form of LVT remains the same if the derivatives are combined with the coefficients.

To clarify that a coefficient includes the derivatives, it is re-denoted with a hat on top of it, meaning that it is now an operator. For example, the photon sector Lagrangian of \emph{nm}SME can be obtained from that of \emph{m}SME, \equref{Eq: Minimal QED Lagrangian}, by simply introducing the hat and by arranging the correct order\footnotemark\footnotetext{Unlike \emph{m}SME, the place of coefficients in \emph{nm}SME is not arbitrary as they are operators now.}:
\begin{equation*}
\mathcal{L}^{\text{QED}}_{\text{photon}}=-\frac{1}{4}F_{\mu\nu}F^{\mu\nu}-\frac{1}{4}F_{\kappa\lambda}(\hat{k}_F)^{\kappa\lambda\mu\nu}F_{\mu\nu}+\frac{1}{2}\epsilon^{\kappa\lambda\mu\nu}A_\lambda(\hat{k}_{AF})_\kappa F_{\mu\nu}\,.
\end{equation*}
where the hatted coefficients are defined as
\begin{equation*}
\begin{aligned}
(\hat{k}_{AF})_\kappa={}&{}\sum\limits_{d=\text{odd}}(k_{AF}^{(d)})_\kappa^{\;\;\alpha_1\alpha_2...\alpha_{(d-3)}}\partial_{\alpha_1}\partial_{\alpha_2}...\partial_{\alpha_{(d-3)}}\,,\\
(\hat{k}_{F})^{\kappa\lambda\mu\nu}={}&{} \sum\limits_{d=\text{even}}(k_{F}^{(d)})^{\kappa\lambda\mu\nu\alpha_1\alpha_2...\alpha_{(d-4)}}\partial_{\alpha_1}\partial_{\alpha_2}...\partial_{\alpha_{(d-4)}}\,.
\end{aligned}
\end{equation*}

\section{Other Formalisms for Photon Sector Lorentz Violation}
\label{Section: other formalisms}
\subsection{Gambini and Pullin Model}
\label{Section: gambini}

In their EFT approach, Rodolfo Gambini \& Jorge Pullin introduce a parameter $\chi l_P$\cite{gr-qc/9809038,89.231301}. This parameter controls an isotropic and non-renormalizable LV, and contains both \mbox{CPT-odd} and CPT-even parts. In the framework of SME, the model of Gambini and Pullin is equivalent to all LVT being zero except
\begin{equation*}
\begin{aligned}
(\ring{k}_{AF}^{(5)})_2 {}&{}=-2\sqrt{4\pi}\chi l_P\,,\\
(\ring{c}_{F}^{(6)})_2 {}&{}=(\ring{k}_{F}^{(6)})_2=-\frac{\sqrt{4\pi}}{5}\left(\chi l_P\right)^2\,,
\end{aligned}
\end{equation*}
where the first one is exactly one of the LVT examined in this study, whereas the second one is CPT-even and beyond the scope of this thesis. From the CPT-odd point of view, both this model and its nonlinear generalization\cite{65.103509} can be shown to be contained in SME\cite{0905.0031}.

\subsection{Myers and Pospelov Model}
\label{Section: myers}
Using an EFT approach like Gambini and Pullin, Robert Myers \& Maxim Pospelov specify a background vector $n^\mu$, and control the LV with it\cite{90.211601,78.125011}. They restrict this vector to be timelike, which then results in a unique parameter $\xi/M_P$ that controls the LV in the preferred frame. It can be then shown that only the CPT-odd contribution affects the photon propagation\cite{0905.0031}, and this LVT can be written as
\begin{equation*}
(k_{AF}^{(5)})^{\kappa\mu\nu}=-\frac{\xi}{M_P}(n^\kappa n^\mu n^\nu-\frac{1}{5}n^2n^{(\mu}\eta^{\nu)\kappa})
\end{equation*}
in the framework of SME. Again, since in the preferred frame the LVT takes the form $n^\mu=(1,0,0,0)$, which is isotropic, whole LV can be written in terms of ring coefficients for that particular frame. Thus, it can be shown that the model of Myers and Pospelov is equivalent to SME with all LVT being zero except
\begin{equation*}
(\ring{k}_{AF}^{(5)})_0=\frac{3\xi\sqrt{4\pi}}{5M_P}\,.
\end{equation*}
\subsection{Bolokhov and Pospelov Model}
\label{Section: bolokhov}
Another model of Maxim Pospelov is one that he constructs with Pavel Bolokhov\cite{77.025022}. In this model is introduced a totally symmetric and traceless parameter $C^{\mu\nu\rho}$, which induces a non-renormalizable and CPT-odd LV. The property of this parameter is that it results in leading order vacuum propagation alteration, hence corresponds to the components of $k_V$ in SME. By construction, $C^{\mu\nu\rho}$ is equivalent to $d=5$ LV in SME, and the equivalence
\begin{equation*}
\left.k^{(5)}_{(V)jm}\sim(k_{AF}^{(5)})^{\mu\nu\rho}\right\rvert_{\stackunder{\text{\scriptsize symmetric}}{\text{\scriptsize traceless}}
}=-2C^{\mu\nu\rho}
\end{equation*}
can be shown\cite{0905.0031}.

As this model deals with leading order vacuum propagation, 16 of 36 coefficients of $(k_{AF}^{(5)})$ are considered only, where all absent coefficients are those of vacuum-orthogonal nature. 

\subsection{Very Special Relativity}
\label{Section: vsr}
\emph{Very Special Relativity} (VSR) is an excellent example of a model with nonspontaneous breaking of Lorentz symmetry. In this model, developed by Andrew Cohen \& Sheldon Glashow, the Lorentz group is explicitly broken into the 4-parametric subgroup SIM(2)\cite{97.021601}.

VSR is still a model of EFT approach, but its Lagrangian dynamics are nontrivial as the dependence on the field operators of VSR is of non-polynomial form. However, it is possible to match it to Lagrange of SME via Taylor expansion; yet, different analytical continuations results in different matchings as there are singularities in VSR. Although a direct matching is not expected as LV methods are different, it is still unusual for VSR to get to be represented with different limits of SME\cite{0905.0031}.

There is a nice explanation of this phenomena in the last reference; nonetheless, we will not go into that here.

\subsection{Robertson-Mansouri-Sexl Model}
\label{Section: robertson}

Unlike the earlier examples, this model studied by \emph{Howard Robertson, Reza Mansouri and Roman Sexl}, (RMS), does not employ EFT, but instead, uses a kinematic approach such that LV is introduced in the transformation laws instead of the action level\cite{21.378,8.497}.

In this model, a universal preferred frame is assumed such that the propagation of light remains conventional in that frame. Since the coordinate transformations between inertial frames are dictated by a modified tensor $T^\mu_\nu$, which is a function of 3 different deformed boosts denoted as $a(\nu)$, $b(\nu)$, and $d(\nu)$ in the model, light can behave anisotropically in other inertial frames.

That transformation tensor is modified in RMS translates into that coordinate system used is different than that of SME; hence, a redefinition of lengths and time intervals would match RMS to SME\cite{0905.0031}. Thus, RMS can indeed be shown to be contained in SME, but we will not go any further in this thesis.

\subsection{Deformed Special Relativities}
\label{Section: dsr}

Like RMS, \emph{Deformed Special Relativity} (DSR) --also called doubly special relativity or kappa-deformed relativity-- is also a kinematic approach to LV, in which Lorentz transformations are modified in a nonlinear manner\cite{hep-th/0405273,0806.0339,hep-th/0207022,gr-qc/0602075}. 

Generically, the modification induces the replacement of the four momentum $p_\mu$ with a modified one denoted by $\pi_\mu$\cite{68.045001} which acts as unmodified under momentum space Lorentz transformations. As this replacement respects the rotation invariance, DSR should be recovered from the isotropic subset of SME if it can be.

Indeed, it can be shown\cite{0905.0031} that DSR emerges within SME under the restrictions
\begin{equation*}
\begin{aligned}
(\hat{k}_{AF})_\mu {}&{}=0\,,\\
(\hat{k}_F)^{\mu\nu\rho\sigma} {}&{}=\sqrt{g_{DSR}}(g_{DSR}^{-1})^{\mu\rho}(g_{DSR}^{-1})^{\nu\sigma}-\frac{1}{2}\left(\eta^{\mu\rho}\eta^{\nu\sigma}-\eta^{\nu\rho}\eta^{\mu\sigma}\right)
\end{aligned}
\end{equation*}
for the effective metric $g_{DSR}$ of DSR, where first equation simply indicates that there is no CPT violating Lorentz violation in DSR.
\chapter{Spontaneous Symmetry Breaking}
\label{CHAPTER:SPONTANEOUS}
The spontaneous symmetry breaking mechanism is a well-studied and well understood symmetry breaking tool which has applications in a variety of branches in physics, including elastic media, condensed matter physics, and elementary particle theory\cite{hep-ph/0611177}. The popularity of this mechanism lies with its theoretical appeal, as it allows the theory to have a certain symmetry irregardless of the fact that this symmetry is unobserved. More technically, spontaneous symmetry breaking allows the Hamiltonian to have a certain symmetry, and the ground state solution not to have it.

The mechanism is actually very straightforward. From the quantum field theoretical point of view, we know that fields should be quantized around their vacuum values; in other words, as long as we are sufficiently far from very high energy values --where the relativity of high is dependent on the physical potential at hand--, the solutions of the theory that we observe remain around the vacuo. Thus, if the potential of the field does not have the vacuum at the symmetry point, then the symmetry will be broken for the vacuum solution even though the potential, hence the Hamiltonian, has that symmetry.

The traditional potential form, also actually a little bit mandated by the power-counting renormalizability, for spontaneous symmetry breaking is \emph{the Mexican hat}, which can be seen in \figref{spontaneous_sb}. The form of potential satisfies the requirements: It is symmetric, hence the underlying theory has the symmetry; however, the symmetry point does not belong to the vacua. As the power-counting renormalizability vetoes field operators of mass dimension higher than 4, such a fourth order polynomial is the only choice for spontaneous symmetry breaking in a renormalizable theory with a scalar or a vector field --as both fields are of mass dimension one--, yet \emph{nm}SME deals with nonrenormalizable operators, meaning that this requirement would be loosened. In any case, we do not know, nor interested in, the fundamental theory in which this breaking happens, hence the breaking mechanism as well as the form of potential is irrelevant for our purposes in this thesis.

The spontaneous breaking of the Lorentz symmetry is hence the mechanism of a tensor type field getting a nonzero \emph{vacuum expectation value}. As that nonzero vacuum expectation values indicate the preferred directions, the symmetry of the spacetime is broken. The only exception, which can also be seen in \figref{spontaneous_sb}, is the case for the scalar fields: The vacuum getting a nonzero constant scalar value everywhere is in-detrimental to Lorentz symmetry. Another point of view why scalar fields do not break Lorentz symmetry is explained in \secref{Section: olt and plt} as well.
\begin{figure}
\sfigure{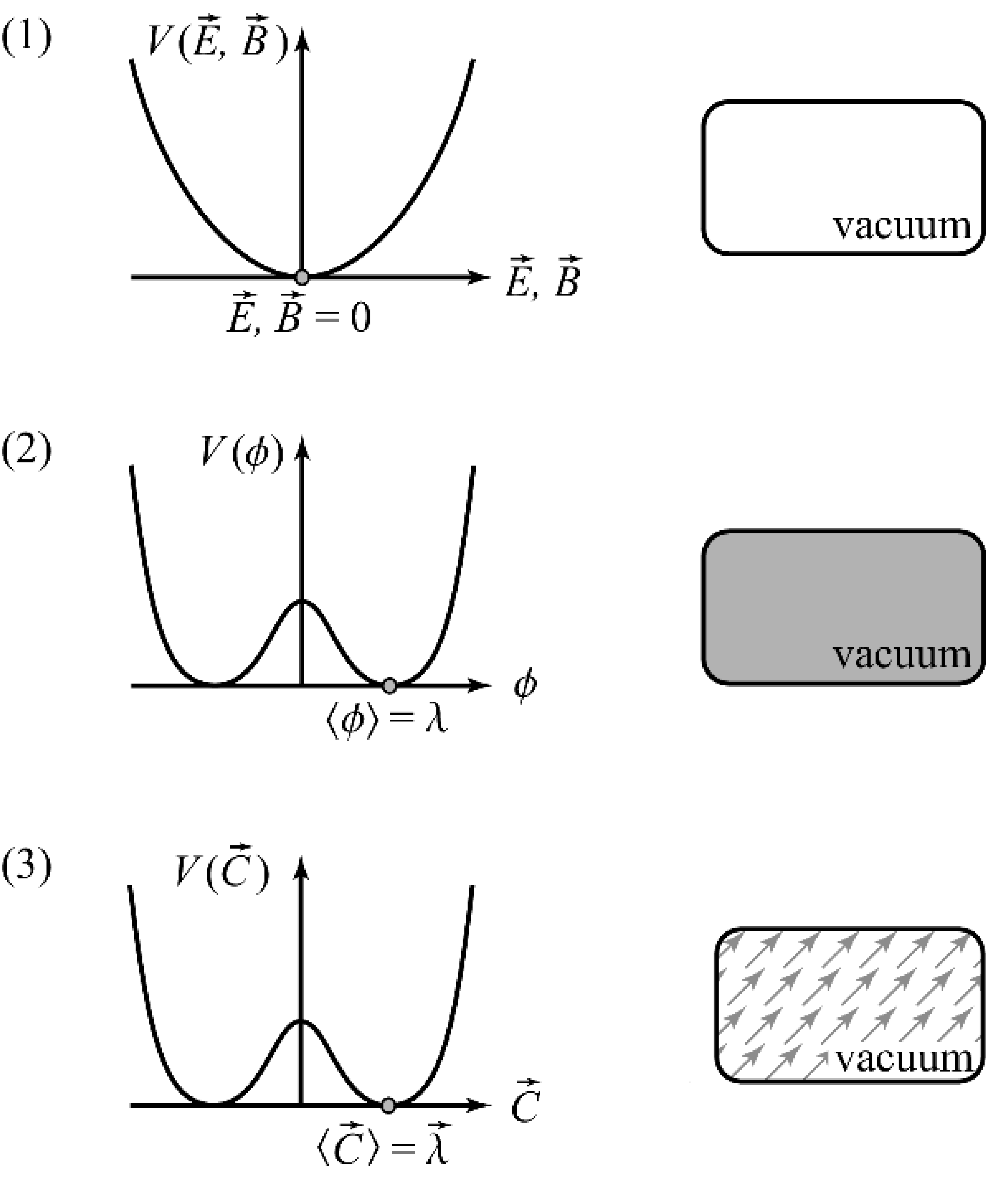}{0.4}{Spontaneous Symmetry Breaking}{Spontaneous Symmetry Breaking\cite{hep-ph/0611177}. The first diagram (1) represents the conventional electromagnetism in which the fields take the value zero for the vacuum, hence the vacuum preserves the symmetry of the potential. In the second diagram (2), the spontaneous symmetry breaking of a scalar field is shown: Although the potential, hence the Hamiltonian, has the symmetry, the vacuum solution is not symmetric; in more technical terms, vacuum expectation value of the field is not zero which is the symmetric value of the potential. This kind of a symmetry breaking is exactly the case for the celebrated Higgs mechanism. Finally, in the third diagram (3), the same mechanism is illustrated for the potential of a vector field. As shown in the boxes next to the diagrams, the third type of the symmetry breaking induces LV with possible CPTV, whereas second type may break an internal symmetry at most, breaking of which has nothing to do with the spacetime symmetries; and clearly, the first type does not induce a breaking whatsoever.}
\end{figure}
\chapter{Helicity Basis}
\label{CHAPTER:HELICITY}
There are various different applications of helicity basis, along which come lots of different notations and definitions. However, in this section, the helicity basis in 3 dimensional space with a positive signature is explored briefly, except the very end in which the 4 dimensional spacetime metric with its space part in helicity basis with negative signature is introduced, as that is the metric used throughout this thesis.

In a nutshell, helicity basis is the complex spherical basis in which rotations about the radial direction can be represented by a diagonal rotation matrix. Mathematically, the radial direction is arbitrary; yet, in this study, and in similar physical applications\cite{0905.0031}, the radial direction $\mathbf{\hat{r}}$ will be taken as the momentum direction $\mathbf{\hat{p}}$. Therefore, strictly speaking, we are dealing with \emph{helicity basis with respect to the momentum direction}.

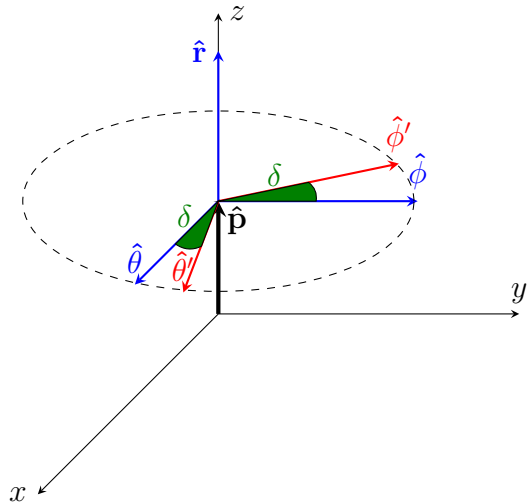
\begin{figure}
\centering
\caption[Rotations About the Radial Direction in Spherical Coordinates]{Rotations About the Radial Direction in Spherical Coordinates. For the helicity basis employed extensively throughout this thesis, the radial direction $\mathbf{\hat{r}}$ is chosen to be aligned with the momentum direction $\mathbf{\hat{p}}$. Exploiting the freedom to choose the coordinate system, $z$ axis can then be aligned with these vectors, ensuring \equref{eq:helicity p}. Then, a rotation of the basis vectors of amount $\delta$ about the radial direction $\mathbf{\hat{r}}$ can be shown as in the figure, where primed quantities are the rotated basis vectors. Then, this illustrated rotation can be written as \equref{eq:spherical transformation} in matrix notation.}
\label{fig-helicity rotation}
\begin{tikzpicture}[x=0.5cm,y=0.5cm,z=0.3cm,>=stealth]
\draw[ultra thick,->] (xyz cs:x=0) -- (xyz cs:y=3) node at (xyz cs:y=2.5,x=0.5) {$\mathbf{\hat{p}}$};
\draw[->] (xyz cs:x=0) -- (xyz cs:x=8) node[above] {$y$};
\draw[->] (xyz cs:y=0) -- (xyz cs:y=8) node[right] {$z$};
\draw[<-] (xyz cs:z=-8)node[left] {$x$} -- (xyz cs:z=0) ;

\draw[blue,thick,->] (xyz cs:x=0,y=3,z=0) -- (xyz cs:x=5.3,y=3)node[above] {$\mathbf{\hat{\phi}}$};
\draw[blue,thick,->] (xyz cs:x=0,y=3,z=0) -- (xyz cs:y=7)node[left] {$\mathbf{\hat{r}}$};
\draw[blue,thick,->] (xyz cs:x=0,y=3,z=0) -- (xyz cs:z=-3.7,y=3)node[above] {$\mathbf{\hat{\theta}}$};
\draw[red,thick,->] (xyz cs:x=0,y=3,z=0) -- (xyz cs:x=1.5,z=-4.05,y=3)node[above] {$\mathbf{\hat{\theta}'}$};
\draw[red,thick,->] (xyz cs:x=0,y=3,z=0) -- (xyz cs:x=4.8,y=4)node[above] {$\mathbf{\hat{\phi}'}$};
\draw[dashed] (0,3) ellipse (2.6cm and 1.2cm);
\node[green!50!black] at (xyz cs:z=-1.5,y=3.5) {$\mathbf{\delta}$};
\node[green!50!black] at (xyz cs:x=1.5,y=3.7) {$\mathbf{\delta}$};

\draw[fill=green!50!black]  (xyz cs:z=0,y=3) -- (xyz cs:y=3,z=-1.9) to[bend right] (xyz cs:x=0.75,z=-2,y=3) -- cycle;
 
\draw[fill=green!50!black]  (xyz cs:z=0,y=3) -- (xyz cs:x=2.6,y=3) to[bend right] (xyz cs:x=2.4,y=3.5) -- cycle;
\end{tikzpicture}
\vskip 0.3cm
\end{figure}

The freedom of choosing the $z$ direction of the coordinate system can be best exploit if $\hat{z}=\hat{r}$ is chosen. Hence, the standard angles in spherical coordinate system yield
\begin{equation}
\mathbf{\hat{p}}=\sin\theta\cos\phi\,\mathbf{\hat{e}_x}+\sin\theta\sin\phi\,\mathbf{\hat{e}_y}+\cos\theta\,\mathbf{\hat{e}_z}\label{eq:helicity p}\,.
\end{equation}

It is clear from the \figref{fig-helicity rotation} that any rotation by an angle $\delta$ about the momentum axis changes the spherical coordinate basis as
\begin{equation}
\begin{pmatrix}
\mathbf{\hat{e}_\theta}\\\mathbf{\hat{e}_\phi}
\end{pmatrix}'=
\begin{pmatrix}
\cos\delta {}&{} \sin\delta\\-\sin\delta{}&{}\cos\delta
\end{pmatrix}\begin{pmatrix}
\mathbf{\hat{e}_\theta}\\\mathbf{\hat{e}_\phi}
\end{pmatrix}\,.\label{eq:spherical transformation}
\end{equation}

As promised, the transformation matrix in \equref{eq:spherical transformation} can be diagonalized as
\begin{equation*}
\begin{pmatrix}
\mathbf{\hat{e}_+}\\\mathbf{\hat{e}_-}
\end{pmatrix}'=
\begin{pmatrix}
e^{-i\delta} {}&{} 0\\0{}&{} e^{i\delta}
\end{pmatrix}\begin{pmatrix}
\mathbf{\hat{e}_+}\\\mathbf{\hat{e}_-}
\end{pmatrix}
\end{equation*}
for
\begin{equation}
\mathbf{\hat{e}_\pm}=\mathbf{\hat{e}^\mp}=\frac{1}{\sqrt{2}}\left(\mathbf{\hat{e}_\theta}\pm i\mathbf{\hat{e}_\phi}\right)\,,\label{eq:helicity basis}
\end{equation}
where the overall phase freedom is used explicitly.

Here, the base vectors of the complex helicity basis, $\mathbf{\hat{e}_\pm}$, denote the positive and negative helicity respectively. The third base vector of the helicity basis is the radial $\mathbf{\hat{e}_r}$, which directly caries over from spherical coordinate system. From the relationship in \equref{eq:helicity basis}, it is clear that the basis of order $\{\mathbf{\hat{e}_+},\mathbf{\hat{e}_r},\mathbf{\hat{e}_-}\}$ employs the non-diagonal metric
\begin{equation}
g_{ab}=g^{ab}=\begin{pmatrix}
0{}&{} 0{}&{} 1\\0{}&{} 1{}&{} 0\\1 {}&{} 0 {}&{} 0
\end{pmatrix}\,.\label{eq:Helicity metric}
\end{equation}

Additionally, it can be derived from the spherical coordinate system that the totally antisymmetric tensor $\epsilon^{abc}$ in the helicity basis satisfies
\begin{equation*}
\epsilon_{+r-}=-\epsilon^{+r-}=i\,.
\end{equation*}

Many other quantities like the nonzero elements of Christoffel connection $\Gamma^c_{\,ab}$ or angular momentum commutation relations can be investigated in helicity basis; however, these details are glossed over as they are not required in this thesis, hence any interested reader is advised to look up to more comprehensive sources\cite{0905.0031}.

Lastly, there is a subtlety in the usage of helicity basis. Although the helicity basis is derived from the spherical coordinate system in 3 dimensions, hence have a positive signature, it is to be used as the space-part of the 4-dimensional Lorentzian metric, hence should have a negative signature. Therefore, in 4 dimensions of the basis $\{\mathbf{\hat{e}_t},\mathbf{\hat{e}_+},\mathbf{\hat{e}_r},\mathbf{\hat{e}_-}\}$, \equref{eq:Helicity metric} becomes
\begin{equation*}
\eta_{\mu\nu}=\eta^{\mu\nu}=\begin{pmatrix}
1{}&{} 0{}&{} 0{}&{} 0\\ 0{}&{} 0{}&{} 0{}&{} -1\\0{}&{} 0{}&{} -1{}&{} 0\\0{}&{} -1 {}&{} 0 {}&{} 0
\end{pmatrix}\,.
\end{equation*}
%
\end{document}